\documentclass[reprint,aps,amsmath,amssymb,twoside,prd,showkeys,superscriptaddress]{revtex4-1}
\pdfoutput=1
\usepackage{verbatim}
\usepackage{comment}
\usepackage{graphicx}
\usepackage[T1]{fontenc}
\usepackage{graphicx}
\usepackage{epsfig}
\usepackage{bm}
\usepackage{tensor}
\usepackage{amsfonts}
\usepackage{lipsum, babel}
\usepackage{amssymb}
\usepackage{float}
\usepackage{amsmath}
\usepackage{subfigure}
\usepackage{dcolumn}
\usepackage{cancel}
\usepackage[colorlinks]{hyperref}
\usepackage[utf8]{inputenc}

\usepackage[usenames,dvipsnames]{color}
\hypersetup{
     breaklinks=true,
    pdfstartview={FitH},    
    colorlinks=true,       
    linkcolor=blue,          
    citecolor=red,        
    filecolor=magenta,      
    urlcolor=blue,           
    anchorcolor=green,      
    linktocpage=true
}

\def\h{\mathcal{H}}

\def\doi{http://doi.org}

\def\h{\mathrm{h}}

\newcommand{\be}{\begin{equation}}
\newcommand{\ee}{\end{equation}}
\usepackage{amsmath}
\usepackage{amssymb}

\usepackage{multirow}
\usepackage{graphicx}
\usepackage{tikz}

\newlength\imagewidth
\newlength\imagescale

\newcommand{\HCd}{\mathcal{H}}

\def\HCdt0{\tilde{\HCd}_{0}}

\newcommand{\affcam}{DAMTP, Centre for Mathematical Sciences, University of Cambridge, Wilberforce Road, Cambridge CB3 0WA, United Kingdom}
\newcommand{\affcamast}{Kavli Institute of Cosmology (KICC), University of Cambridge, Madingley Road, Cambridge, CB3 0HA, United Kingdom}

\newcommand{\affFran}{Institut de Physique Theorique, Universite Paris-Saclay, CEA, CNRS, F-91191 Gif-sur-Yvette Cedex, France}
\newcommand{\cern}{CERN, Theoretical Physics Department, Geneva, Switzerland.}

\begin{document}

\rightline{CERN-TH-2022-207}

\title{Stringent Pulsar Timing Bounds on Light Scalar Couplings to Matter}
\author{David Benisty}
\email{db888@cam.ac.uk}
\affiliation{\affcam}\affiliation{\affcamast}
\author{Philippe Brax}
\email{philippe.brax@ipht.fr}
\affiliation{\affFran}
\affiliation{\cern}
\author{Anne-Christine Davis}
\email{ad107@cam.ac.uk}
\affiliation{\affcam}\affiliation{\affcamast}

\begin{abstract}
Pulsar Timing constraints on scalar-tensor theories with conformal and disformal couplings to matter are discussed. Reducing  the dynamics to the motion in the centre of mass frame and using the mean anomaly parametrisation, we find the first post-Newtonian corrections induced by the conformal and disformal interactions in the form of a generalized quasi-Keplerian solution. We also derive the radiation reaction force due to scalar radiation and the corresponding Post-Keplerian Parameters (PKP). We use different pulsar time of arrival (TOA) data sets to probe  the  scalar corrections to the PKP. In particular, we focus on systems with large orbital frequencies as the contributions to the  PKP terms induced by the disformal coupling  are sensitive to higher frequencies. We find that the most constraining  { {pulsar timings}}  are  PSR B1913+16 and the double pulsar PSR J0737-3039A/B,  being {{of the order of}} the Cassini bound on the conformal coupling obtained from the Shapiro effect in the solar system. { {The combined constraints using other pulsar timings give an upper bound on the conformal coupling $\beta^2 < 2.33 \cdot 10^{-5}$ and a lower bound on the disformal coupling scale of $\Lambda \geq 1.12 \ {\rm MeV}$ which is comparable to the Cassini bound and to the GW-170817 constraints respectively}}. Future measurements for pulsar timing with black hole companions are also discussed.
\end{abstract}
\keywords{Pulsar Timing; Modified Gravity; Dark Energy; Post Keplerian Parameters; Cassini Spacecraft;}
\maketitle
\section{Introduction}

The discovery of the accelerated expansion of the late Universe, requires a modification of
General Relativity (GR) as originally presented in 1915. This change could be as minimal as the addition of a
cosmological constant \cite{SupernovaCosmologyProject:1998vns,Weinberg:1988cp,Lombriser:2019jia,Copeland:2006wr,Frieman:2008sn,Riess:2019cxk}, { {which is so far the most likely explanation to the cosmological observations}}. { {On the theoretical side, the smallness of the cosmological constant could be  considered to be  fine tuned.}} { {Recent string theoretic conjectures such as  the swampland ones \cite{Ooguri:2018wrx,Garg:2018reu} would favour a more dynamical approach and   prescribe that   the Universe should be driven by quintessence in its late time phase }}\cite{Starobinsky:1979ty,Starobinsky:1980te,Guth:1980zm,Albrecht:1982wi,Mukhanov:1981xt,Guth:1982ec,Linde:1981mu,Barrow:1988xh,Barrow:1988xi,Elizalde:2008yf,Ratra:1987rm,Caldwell:1997ii,Zlatev:1998tr,Caldwell:1999ew,Chiba:1999ka,Bento:2002ps,Tsujikawa:2013fta,Caldwell:1997ii,Ratra:1987rm,Peebles:1987ek,Barreiro:1999zs,Carroll:1998zi,Chiba:1999wt,Sahni:1999qe}. Such a scalar field would  slow roll and eventually  mimic a cosmological constant. In these models the scalar field does not couple to ordinary matter. Other possibilities include a modification of general relativity itself where a scalar, which could be the scalar polarisation of a massive gravity model for instance, couples to matter \cite{Koivisto:2008ak,Zumalacarregui:2010wj,Koivisto:2012za,vandeBruck:2013yxa,Brax:2013nsa,Neveu:2014vua,Sakstein:2014isa,Sakstein:2014aca,Desmond:2019ygn}.

One very popular  model of light scalar coupled to matter is obtained by modifying the Einstein-Hilbert action into  a function of the Ricci scalar, the so-called $f(R)$ theories~\cite{Brax:2008hh,Sotiriou:2008rp}. This can be seen as  adding  a coupled scalar field to GR, i.e. becoming a scalar-tensor theory, with a specific coupling to matter equal to $\beta=1/\sqrt 6$, { {i.e. the Jordan  $g_{\mu\nu}^J$ and the Einstein  $g^E_{\mu\nu}$ metrics are related by a conformal rescaling.}}
\be
g_{\mu\nu}^J= e^{2\beta \phi/m_{\rm Pl}} g_{\mu\nu}^E.
\ee
In more general cases, the coupling between the scalar field and matter depends on the transformation between  the Jordan, where matter couples minimally, to the Einstein frames, where the Einstein-Hilbert term is canonical. Bekenstein gave the most general coupling of a scalar field to matter, which involves  both conformal and disformal transformations \cite{Bekenstein:1992pj,Sakstein:2014isa,vandeBruck:2015ida,Koivisto:2012za}
\begin{equation}
g_{\mu\nu}^J= A^2(\phi,X) \, g_{\mu\nu}^{E}+ B^2(\phi,X)\, \partial_\mu\phi \partial_\nu \phi,
\label{eq:jordan_frame_metric}    
\end{equation}
where this metric $g_{\mu\nu}^J$ defines the Jordan frame and $g_{\mu\nu}^{E}$ is the Einstein frame metric. Here we denote by $X= -\frac{1}{2} (\partial \phi)^2$ the standard kinetic terms. Such modifications may help resolve some issues in cosmology such as the Hubble tension \cite{Karwal:2021vpk}. 
Recently light scalar fields have also been suggested as possible candidates for dark matter \cite{Hui:2016ltb}. The coupling of such dark matter fields to matter is also crucial for their dynamics and their eventual detection \cite{Brax:2017xho,Trojanowski:2020xza,Brax:2020gqg}.

These theories can be tested using gravitational methods as shown by earlier studies which focused on two bodies in an orbital motion  \cite{Maheshwari:1980gmr,Damour:1991rd,Buonanno:1998gg,Damour:1988mr,Damour:1999cr,Memmesheimer:2004cv,Iorio:2014yga,Benisty:2021cmq,Zhang:2016njn,Zhang:2018prg,Zhang:2019ufz,Benisty:2022txp,Benisty:2022idt,Benisty:2022ive}, a well studied example in GR, from which similar properties can be inferred for modified gravity with conformal and disformal couplings \cite{Brax:2012ie,Brax:2013uh,Zhang:2017srh,Brax:2018bow,Davis:2019ltc,Brax:2019tcy,Liu:2017xef,Shibata:2022gec}. 

\begin{figure}[t!]
    \centering
    \pgfmathsetlength{\imagewidth}{\linewidth}%
    \pgfmathsetlength{\imagescale}{\imagewidth/524}%
    \begin{tikzpicture}[x=\imagescale,y=-\imagescale]
        \node[anchor=north west] at (0,0) {\includegraphics[width=\imagewidth]{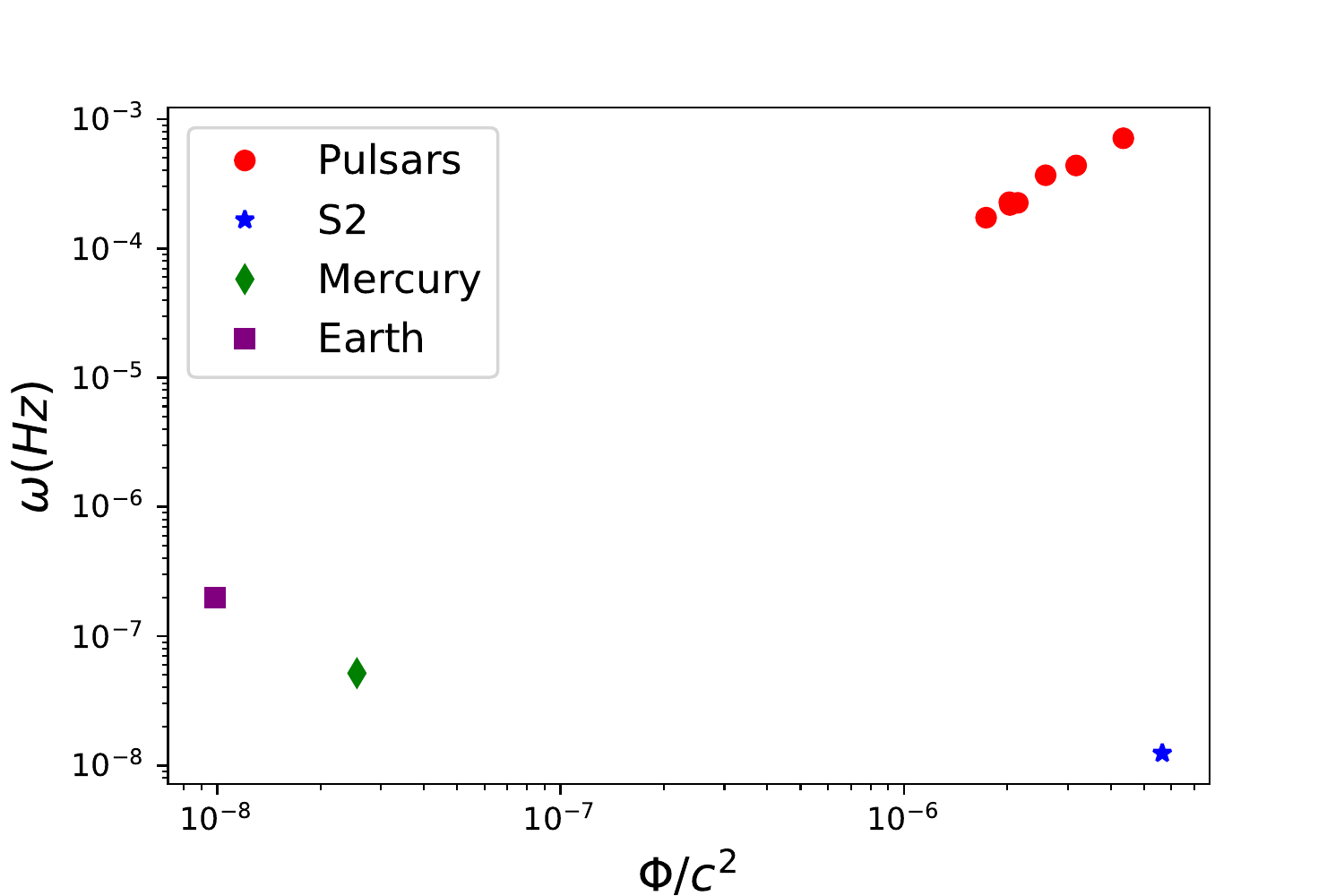}};
        \node[anchor=north west] at (180,120) {\includegraphics[width=0.45\imagewidth]{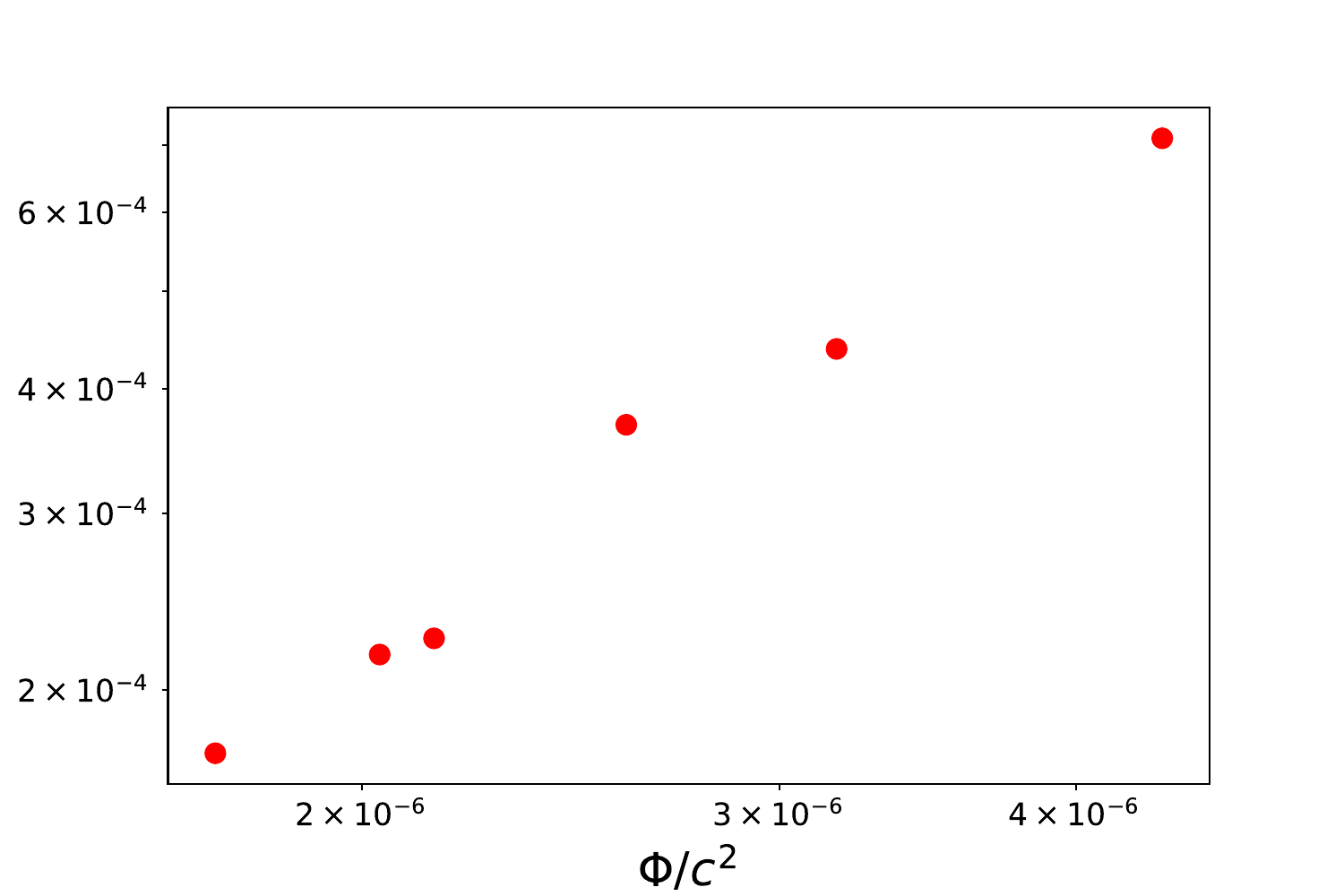}};
    \end{tikzpicture}
    \caption{\it{Comparison of selected pulsars  that are analysed in this paper vs. mercury and the S2 star from the galactic centre. In the middle of the figure, we  focus on the pulsar timings (in red).  Since the orbital periods of the pulsars is higher then the other systems, the constraints on the disformal interaction is stronger. { {The S2 stars orbit around the central supermassive black hole  of the Milky Way. Far away from the black hole, the two-body system can be approximated by the dynamics of two point-particles \cite{Benisty:2021cmq, Wong:2019yoc}.}}}}
\label{fig:com}
\end{figure}

In this paper, we shall work with the most general scalar-tensor theory associated with a Bekenstein coupling involving generic conformal and disformal couplings. These  theories potentially give rise to fifth forces which are subject to strict limits from solar system tests of general relativity \cite{Bertotti:2003rm}, and, at face value,   the archetypical $f(R)$ models would then appear to be ruled out. However, the fifth force effects  can be screened in the solar system, giving rise to screened modified gravity  with a phenomenology which depends on the  environment. Screening can take place in different ways either via the chameleon \cite{Khoury:2003aq, Brax:2004qh},  the  K-mouflage and  Vainshtein \cite{Babichev:2009ee,Vainshtein:1972sx} or Damour-Polyakov mechanisms \cite{Damour:1994zq,Olive:2007aj,Brax:2010gi,Hinterbichler:2010es}. All rely on the environment such that the fifth force becomes screened in the solar system and as a result the theory can evade all the local tests of gravity. On larger scales, these models can give rise to modifications to GR on cosmological scales \cite{Damour:1992we,Julie:2017pkb,Bertotti:2003rm,Williams:2004qba,Khoury:2003aq,Damour:1994zq,Vainshtein:1972sx,Babichev:2009ee}. 

{{ Screenings of the Chameleon and Damour-Polyakov types, see \cite{Brax:2013ida,Brax:2021wcv} for reviews, can be usefully compared to scalar models where scalarisation takes place \cite{Damour:1993hw}. Scalarised models are commonly used in analysing pulsar timing when looking for  new physics effects. In a nutshell, chameleon type screening relies on the stabilisation effects of matter, i.e. the effective mass of the scalar field in matter increases as the density increases. On the other hand, scalarisation relies on the destabilisation of the scalar field in the presence of matter \cite{Freire:2012mg,Doneva:2016xmf,Ramazanoglu:2016kul,Shao:2017gwu,Zhao:2022vig}. We will give more details in section \ref{sec:conDisInter}. }}

Here we consider the scalar interaction between moving bodies when the coupling between matter and the scalar field is mediated by the coupling functions
\be
\begin{split}
 A(\phi,X)= e^{\beta \phi/m_{\rm Pl}}, \quad B^2(\phi,X) = \frac{2}{m^2_{\rm Pl} \Lambda^2},   
 \label{choice}
\end{split}
\ee
which gives rise to a Yukawa interaction of coupling strength $\beta$ with matter and the disformal term is taken to be constant at leading order. {{These terms are the leading order contributions from an effective field theory point of view. Indeed the contributions depending on powers of $X$ only matter at short distances.}} We work on large distances where the terms in $X$ in the coupling functions can be neglected as they would lead to higher order corrections to the dynamics of the moving bodies and for most purposes it is enough to consider a small field expansion $A(\phi) \approx 1+ \beta \phi/m_{\rm Pl}$ and $B(\phi,X)$ as constant. Higher order terms would lead to effects which could be taken into account in perturbation theory and are neglected here. 
Our description would nonetheless apply to the effective interaction of screened bodies when screening takes place. In this case, the small values of the couplings that we will infer from Pulsar Timing  should be seen as resulting from the screening of dense objects such as pulsars.

The study of gravitational physics benefits from
a number of experimental advances which provide excellent chances for constraining these interactions. The first evidence for gravitational waves was provided by the binary Hulse-Taylor pulsar  PSR B1913+16 \cite{1981SciAm.245d..74W,Weisberg:2016jye}. Pulsars are extremely useful tools for testing gravity due to the extreme precision of the radio pulses they emit. Pulsars have a short spinning period. The monitoring of the times of arrival of the pulsar's radio pulses allows one to infer the properties of the orbit. Observations over long periods of time provide a unique way of obtaining  experimental constraints on the parameters of the orbits. In pulsar timing systems, observed pulse arrival times are sensitive to  relativistic effects that can be modeled in a theory-independent way using the  "post-Keplerian parameters" (PKP) \cite{1986AIHS...44..263D,Stairs:2003eg}.  They  are phenomenological corrections and additions to the Keplerian description of the binary motion. The PKP take different forms in different theories of gravity and so their measurement can be used to test these gravity theories \cite{Liu:2017xef,DeLaurentis:2011tp,DeLaurentis:2013zv,DeLaurentis:2013fra,DeLaurentis:2013zv,Dyadina:2016bzb,Narang:2022jkv}. Earlier studies such as  \cite{Brax:2019tcy,Davis:2019ltc,Brax:2019tcy,Brax:2020vgg,Brax:2021qqo,Benisty:2021cmq} show conformal and the disformal effects on the two body motion. In this paper we derive the whole PKP and compare it with the latest measurements of pulsars.

As we will see, the disformal strength is affected by the frequency of the orbital motion, where higher frequencies give larger disformal contributions. This follows from the higher derivative nature of the disformal interaction. The following dimensionless quantity $\epsilon_\Lambda$ relates the disformal coupling interaction to the frequency
\be
\epsilon_\Lambda = \frac{\left(\beta n_b / \Lambda\right)^2}{\left(1-e^2\right)^3},
\ee
where $e$ is the eccentricity of the motion and $n_b = 2\pi/P_b$ is the frequency of the motion. $\epsilon_\Lambda$ describes the contribution of the disformal interaction to the PKP. {Fig~\ref{fig:com} compares pulsar timings that are analysed in this paper vs. the precessions of mercury and of the S2 star around the galactic centre, as analysed in~\cite{Benisty:2021cmq}.} In the middle of this plot we focus on the pulsar {{timings}} (in red). Since the periods of the pulsar timings are higher than for other systems, the constraints on the disformal interaction is stronger and therefore motivates us to constrain the conformal and the disformal interactions with these systems.

The plan of this work is  as follows: Section \ref{sec:conDisInter} describes the action and the equations of motion. Section \ref{sec:meanTrueAnom} solves the system using the mean anomaly approach. In section \ref{sec:RadEm} we include the effects of the scalar radiated power on the trajectories of the reduced two-body system in the centre of mass frame. This leads to a radiation reaction force on the orbital motion, allowing us to compute the change in the period and the eccentricity due to this effect. Section \ref{sec:timedelays} derives the time delays PKP. Section \ref{sec:constraints} compares the PKP with different pulsar timings. Section \ref{sec:Dis} discusses the results with possible observational constraints on the conformal and disformal couplings. There are five technical appendices.

\section{Conformal and Disformal Interactions}
\label{sec:conDisInter}

\subsection{Screening and effective dynamics}
We will focus on models where screening of the Damour-Polyakov or the chameleon types take place. Such models are defined by the scalar tensor action 
\begin{eqnarray}
S&&=\int d^4x \sqrt{-g}\left ( \frac{R}{16 \pi G}-\frac{1}{2} g^{\mu\nu}\phi_{,\mu}\phi_{,\nu} -V(\phi)\right )\nonumber \\  &&+S_m(\psi_i, g^J_{\mu\nu}),\nonumber \\
\end{eqnarray}
where the Jordan metric is related to the Einstein metric via (\ref{eq:jordan_frame_metric}). In the spirit of effective field theories, we will consider that the dependence on the kinetic terms of the $B(\phi,X)$ function can be expanded as
\be 
B^2(\phi,X)= \sum_{n\ge 0} B_n(\phi)\frac{X^n}{(\Lambda^2 m_{\rm Pl}^2)^n}
\ee
where the powers of the Planck scale appear as $\phi$ is normalised in Planck units and derivatives appear suppressed by powers of $\Lambda$. This expansion is valid in the regime where $\partial/\Lambda \lesssim 1$ corresponding to distances larger than $1/\Lambda$. In practice, the length scale $1/\Lambda$ of interest is much shorter than the typical scales probed by gravitational physics observations and experiments implying that the leading term $B_0(\phi)$ dominates. Similarly as we can expand $B_0$ in powers of $\phi$
\be 
B_0(\phi)= \sum_{n\ge 0} B_{0,n} \frac{\phi^n}{m_{\rm Pl}^n}
\ee
and in the small $\phi$ regime corresponding to the screened theories that we will consider, the leading contribution comes from the constant term $B_{0,n}=2$ normalised as in (\ref{choice}). In the regime where $\partial/\Lambda \ll 1$ as will be the case for coupling scales $\Lambda$ in the MeV range, the effect of the disformal coupling can be treated in perturbation theory using for instance a ladder expansion  for the solutions of the Klein-Gordon equation \cite{Brax:2018bow}. The zeroth order in the disformal interaction corresponds to a scalar-tensor theory with the conformal coupling $A(\phi)$ only. In the presence of non-relativistic matter the Klein-Gordon equation reads \cite{Khoury:2003aq,Brax:2013ida,Brax:2017idh,Brax:2021wcv}
\be 
\Box \phi= \frac{\partial V_{\rm eff}}{\partial \phi}
\ee
where the effective potential is given by 
\be 
V_{\rm eff}(\phi)= V(\phi) + (A(\phi)-1) \rho_m
\ee
where $\rho_m$ is the conserved matter density in the Einstein frame. It is related to the Einstein matter density by $\rho_E= A(\phi) \rho_m$. The chameleon and Damour-Polyakov mechanisms operate when the effective potential has a unique minimum $\phi(\rho_m)$ which depends on the matter density. 

A typical example is given by the inverse power-law chameleon where \cite{Khoury:2003aq}
\be 
V(\phi)= \frac{M^{4+n}}{\phi^n}+\dots
\ee
and the coupling function  $A(\phi)=e^{\beta \phi/m_{\rm Pl}}$. The runaway potential of the quintessence type \cite{Ratra:1987rm} $V(\phi)$ is stabilised by the matter contributions with 
\be 
\phi(\rho_m)= (\frac{nM^{4+n} m_{\rm Pl}}{\beta \rho_m})^{1/n+1}.
\ee
Notice that $\phi(\rho_m)\ll m_{\rm Pl}$ as soon as $M\ll m_{\rm Pl}$. This implies that $A(\phi(\rho_m)\simeq 1$ and  that the Einstein and conserved matter densities are essentially equal. 
The mass of the chameleon is given by the second derivative of the effective potential
\be 
m^2(\rho_m)= \frac{d^2 V_{\rm eff}(\phi)}{d\phi^2}\vert_{\phi(\rho_m)}
\ee
which grows with the matter density in $\rho_m^{(n+2)/2(n+1)}$. 
This is the defining property of screened models, i.e. the mass of the stabilised scalar field grows rapidly with the matter density.

This has important consequences for compact objects that we will treat as composed of non-relativistic matter in the first place \cite{Brax:2012gr}. In this case, the scalar field profiles takes two different characteristic shapes. In the screened regime, the scalar field is nearly constant inside the body, that we consider as spherical for simplification's sake, with a value given by $\phi_{\rm in}=\phi(\rho_{\rm in})$. Here $\rho_{\rm in}$ is the density inside the object. This is guaranteed as long as $m_{\rm in}R\gg 1$ where $m_{\rm in}$ is the mass of the scalar field inside the body.  Outside the field extrapolates to the outside value $\phi_{\rm out}= \phi(\rho_{\rm out})$ where $\rho_{\rm out}$ is the matter density in the environment surrounding the compact object. For typical models such as the inverse chameleons $\phi_{\rm in}\ll \phi_{\rm out}$. Outside the objects the field behaves as
\be 
\phi= \phi_{\rm out} - \frac{\beta_{\rm eff}}{4\pi m_{\rm Pl}} \frac{ M}{r}
\ee
where the effective coupling of the scalar field to the compact object is given by 
\be 
\beta_{\rm eff}= \frac{ \phi_{\rm out}-\phi_{\rm in}}{2 m_{\rm Pl}\Phi_N(R)}\simeq \frac{ \phi_{\rm out}}{2 m_{\rm Pl}\Phi_N(R)}.
\label{coup}
\ee
Here  $\Phi_N(R)= GM/R$ is the absolute value of Newton's potential at the surface of the object. The screening criterion is simply that the compact object couples less strongly than a point particle embedded in the surrounding environment
\be 
\beta_{\rm eff}\le \beta(\phi_{\rm out})
\label{scr}
\ee
where 
\be 
\beta (\phi)= m_{\rm Pl} \frac{d\ln A}{d\phi}
\ee
is the coupling of a point particle to the scalar field.
One important point is that from the point of view of an outside observer, the compact object behaves like a point particle coupled to the scalar field with a strength $\beta_{\rm eff}$. As the field is nearly constant inside the body and $\phi_{\rm in}/m_{\rm Pl}\ll 1$ implying that $A(\phi_{\rm in})\simeq 1$, the mass of the field in the Einstein frame $M_E$ corresponding to the source for the gravitational field in the Poisson equation coincides with the conserved mass $M$ as
\begin{eqnarray} 
&& M_E= \int d^3 r A(\phi(r)) \rho_m \simeq \int_0^R d^3 r A(\phi_{\rm in})\rho_m \nonumber \\ && \simeq \int_0^R d^3r \rho_m= M. \nonumber \\
\end{eqnarray}
For screened models, the mass of the object can be identified with the mass in GR in the absence of scalar field.

On the other hand when screening does not operate, i.e. when (\ref{scr}) is not satisfied, the field inside the object is nearly constant and equal to $\phi_{\rm out}$. In this case, the object behaves for an outside observer as a point-particle with the coupling $\beta(\phi_{\rm out})$ and a mass $M$ which does not differ from the mass in GR. 

These results have been deduced in the case of non-relativistic matter. They can be extended to the case of matter where the pressure does not vanish. In this case the generalised Tolman-Oppenheimer-Volkov equations including the scalar constributions must be solved \cite{Brax:2017wcj}. In fact for screened models satisfying the solar system constraints the  field profile is still sharp and the previous results apply \cite{Brax:2013uh}. In particular, the mass of the object is still given by its GR value and the body behaves like a point-like particle with an effective charge depending on the environment. In the unscreened case, the scalar field is not perturbed by the presence of the objects and the previous results also apply. For objects where the screening is rather weak and the field profile is not sharp, numerical methods are necessary in order to solve the generalised TOV equations. This goes beyond the present paper.

The screened and unscreened cases behave very differently when it comes to the gravitational tests in the solar system and the radiation of scalar waves. In the screened case, the coupling to scalars must be small enough to evade tests such as the Cassini bound \cite{Bertotti:2003rm} when the effective coupling of the Sun is small enough. This gives a constraint on the physics of the scalar field in the solar system, i.e. $\phi_{\rm out}$ and $\beta(\phi_{\rm out})$ must be such that (\ref{scr}) is satisfied for $\Phi_\odot \sim 10^{-6}$. For pulsars, we have seen that they would behave like point-particles for the scalar field with a coupling (\ref{coup}) which would depend on the pulsar's environment via $\phi_{\rm out}$ and the pulsar's Newtonian potential. For binary system composed of two screened stars, the coupling would differ if their Newtonian potentials are not equal. In this case, scalar radiation in the form of dipolar radiation must be taken into account \cite{1975ApJ19659E}. Constraints from pulsar timing would result in bounds on the scalar value $\phi_{\rm out}$ in the pulsar's environment. 
On the other hand for the unscreened case, the coupling $\beta(\phi_{\rm out})$ must be tuned to be small enough in the solar system to pass the Cassini bound. In the case of pulsars, the constraints are then on $\beta(\phi_{\rm out})$ in the pulsar's environment. In particular, as the objects are universally coupled to the scalar field, the dipolar radiation vanishes and only the monopole and quadrupole radiations matter.

\subsection{Screening vs Scalarisation}

The screening mechanisms that we have reviewed in the previous section are inspired by the physics of the acceleration of the Universe. On large scale where the density is small, the scalar should have large effects on the dynamics of the Universe and therefore should not couple weakly. On the other hand, locally in the solar system the scalar should be screened to pass the gravitational tests. This is what has been achieved with models for which the effective mass in dense objects is large, i.e. preventing any strong interaction between the scalar field and the object. Only a thin shell at the surface of the object interacts with the scalar field, hence drastically reducing its coupling to the scalar.

Scalarisation \cite{Damour:1993hw,Doneva:2022ewd} appears in a different context whereby scalar effects are enhanced in the strong field regime of objects such as neutron stars. In sparse environments, the scalar field essentially decouples from matter whilst its coupling is driven to larger values in dense matter. This is in effect an anti-screening behaviour. This can be simply exemplified in the non-relativistic case with a simple scalar-tensor theory defined by a massive scalar field of potential \cite{Doneva:2016xmf,Ramazanoglu:2016kul}
\be 
V(\phi)= \frac{1}{2}m^2 \phi^2
\ee
and a coupling function 
\be 
A(\phi)= e^{-a\phi^2/2m_{\rm Pl}^2}
\ee
where $a$ is a constant.
In the absence of surrounding matter, point particles couples to the scalar with a strength
\be 
\beta (\phi)= -a \frac{\phi}{m_{\rm Pl}}.
\ee
In vacuum, the effective potential reduces to $V(\phi)$ whose minimum is for $\bar \phi=0$ implying that the scalar decouples from point-particles $\beta(\bar\phi)=0$. Obviously this is the opposite effect that one would like to achieve to generate  modifications of gravity on the largest scales of the Universe. On the other hand, a remarkable effect takes place in dense matter from which   compact objects potentially couple strongly to the scalar field.

This can be understood by studying the non-relativistic limit where the background space-time metric is taken to be flat and matter is pressure-less. When matter is present the coupling function add a potential term in
\be 
(A(\phi)-1)\rho_m \simeq -a \frac{\rho_m}{m_{\rm Pl}^2 }\phi^2 + \dots
\ee
corresponding to a destabilisation of the scalar field by a negative contribution to the scalar mass. The vanishing value of the scalar field is not stable for densities
\be 
\rho_m \ge \frac{m^2 m^2_{\rm Pl}}{a}.
\ee
This is an anti-symmetron effect \cite{Hinterbichler:2010es} whilst the symmetron contribution to the mass is positive in matter. This instability is eventually stabilised for a non-vanishing value of the field $\phi(\rho_m)$ satisfying $dV_{\rm eff}/d\phi=0$ implying that
\be 
e^{-a\phi(\rho_m)^2/2m_{\rm Pl}^2}= \frac{m^2 m^2_{\rm Pl}}{a \rho_m}
\ee
and therefore
\be 
A(\phi(\rho_m))= \frac{m^2 m^2_{\rm Pl}}{a \rho_m}.
\label{ain}
\ee
The coupling of the scalar to a point particle $\beta(\phi_m)$ increases like $\ln^{1/2}(\rho_m)$ implying that the scalar field couples stronger to point particles in matter than in vacuum.

When it comes to the field profile created by an object of radius $R$ and density $\rho_m$, the field interpolates between a vanishing value in vacuum at infinity and a value which would converge to $\phi(\rho_m)$ inside very large objects. Scalarisation takes place when $mR$ is bounded and the field evolves significantly inside the object. As $A(\phi)$ interpolates between a value of unity at infinity and a very different value which could be  (\ref{ain}) deep inside the objects for large bodies, we see that the Einstein frame density $\rho_E$ varies substantially with the scalar field and is not equal to the conserved density $\rho_m$  as in the screened case. This implies that the Einstein mass $M_E$ which sources the Poisson equation for the Newtonian potential does not coincide with the conserved mass $M$. The Einstein mass depends on the details of the scalar field profile in the object. Moreover the scalar coupling of the object depends on the profiles and becomes non-universal triggering the possibility of dipolar emission in binary system \cite{1975ApJ19659E}. We give a simplified treatment of the scalar profile in the non-relativistic approximation in the appendix \ref{app:sca}.

In the case of relativistic objects like neutron stars where pressure must be taken into account, the simple description given by the non-relativistic case must be complemented with a numerical integration of the TOV equations. Still the effects of the scalar field on the mass of the objects must be taken into account. This is very different from the case of screened scalars where such an effect of the scalar field is absent. 

\subsection{The effective point-particle Lagrangian}
In this paper we will focus on screened models in their  unscreened regime where the coupling of the scalar field to matter is universal and depends on the environment. The case for which compact objects are screened is left future works. 

The dynamics of gravity  interacting with a massless scalar field are described by
\be
S=\int d^4x \sqrt{-g}\left ( \frac{R}{16 \pi G}-\frac{1}{2} g^{\mu\nu}\phi_{,\mu}\phi_{,\nu}\right) +S_m(\psi_i, g^J_{\mu\nu}),
\ee
where matter fields are denoted by $\psi_i$ and  their action by $S_m$. In the following we will take the matter action to be the one of point-like particles and the scalar potential to be vanishing. This will provide  an appropriate description of the dynamics of macroscopic objects like neutron stars as long as finite size effects can be neglected. This setting  applies to screened models where  the scalar field between massive objects is assumed to be very light and the coupling to matter depends on the environment and is universal for different objects. In a different environment, the coupling would differ as the background density and distribution of the surrounding objects would be  different. Our model can therefore be seen as an effective description of the long range interaction between massive objects mediated by such  an unscreened scalar with the coupling $\beta \equiv \beta(\phi_{\rm out})$ which depends on the environment. In practice, we will require that the scalar field is massless on the size of the solar system corresponding to an estimated $100$ a.u. and the mass $m_\phi \lesssim 10^{-20}$ eV. Larger masses by two orders of magnitude would still be considered as massless for double pulsars but not in the solar system where the time delay effects would have to be modified, i.e. the enhancement of Newton's constant for time delays would not be present for a massive field. To simplify the analysis we will also assume that the coupling in the pulsar's environment is the same as the one in the solar system. If this is not the case then  two couplings $\beta_P$ and $\beta_{SS}$ would have to be introduced for the pulsars and the solar system.


In \cite{Brax:2018bow,Brax:2019tcy} the Lagrangian for the reduced action for two isolated bodies was derived and  reads
\begin{equation}
\mathcal{L} = \mathcal{L}_0 +  c^{-2} \mathcal{L}_{Dis} + c^{-2}\mathcal{L}_{1}+ \mathcal{O}(c^{-4}),
\end{equation}
with:
\begin{eqnarray}
&&
\mathcal{L}_0 = \frac{1}{2}v^2 + (1+2\beta^2)\frac{G M}{r}, \nonumber \\
&&\mathcal{L}_{Dis} = \xi\left(r\right) \left(v^2 - 2 (v\cdot \hat{n})^2 \right).
\nonumber \\
&&
\mathcal{L}_1 = \frac{1-3 \nu}{8}v^4 + \frac{G M}{2 r}[(3-2 \beta^2 +
\nonumber \\
&&\nu (1+2\beta^2))v^2 +\nu (v\cdot \hat{n})^2 -(1+2 \beta^2)\frac{G M }{r}],  \nonumber \\
\end{eqnarray}
and the dimensionless function $\xi\left(r\right)$ is defined as:
\begin{equation}
\xi\left(r\right) = \frac{4 \beta^2 G M^2}{\Lambda ^4 r^4}.
\end{equation}
The $\beta$ parameter characterises  the conformal interaction and the dimensionless function $\xi(r)$ the disformal coupling. When $\beta$ and $1/\Lambda$ go to zero the action reduces to the standard Einstein-Infeld-Hoffmann (EIH) action \cite{AIHPA_1985__43_1_107_0}. $\mu$ is the reduced mass and $M$ is the total mass of the system, $r$ is the separation and $v$ is the relative velocity. $\nu$ is the ratio between the reduced mass and the total mass $\nu = \mu/M$, with $0 \leq \nu \leq 1/4$. Since $v\cdot \hat{n} = \dot{r}$ we get that the energy per units of  reduced mass $\epsilon = E/\mu$ is
\begin{equation}
\epsilon = \frac{1}{2}v^2 - (1+2\beta^2) \frac{G M }{r} + \xi\left(r\right) \left(v^2 - 2 \dot{r}^2 \right) + c^{-2}\epsilon_1,
\end{equation}
with the $1^{st}$ PPN correction
\begin{eqnarray}
&&\epsilon_1 = \frac{3}{8}(1-3 \nu)v^4 + \frac{G M}{2 r}[(3-2 \beta^2 + \nonumber \\ &&\nu (1+2\beta^2))v^2 +\nu \dot{r}^2 +(1+2 \beta^2)\frac{G M }{r}],   \nonumber \\
\end{eqnarray}
and the angular momentum per units of  reduced mass $j = J/\mu$ is
\begin{equation}
\vec{j} = \left[1 + 2 \xi(r) + c^{-2}j_1 \right] \vec{r} \times \vec{v}
\end{equation}
where
\begin{equation}
j_1 = \frac{v^2}{2}(1-3 \nu)  + (3-2 \beta^2 + \nu (1+2\beta^2))\frac{G M}{r}.
\end{equation}
Although $\vec j$ is conserved, the vector $\vec r \times \vec v$ is not. While $\vec r \times \vec v$  no longer has a constant  magnitude, it has a constant direction. This is sufficient to establish that the orbital motion takes place  within a fixed orbital plane, just as in the Newtonian case. 

\begin{figure}
 	\centering
\includegraphics[width=0.35\textwidth]{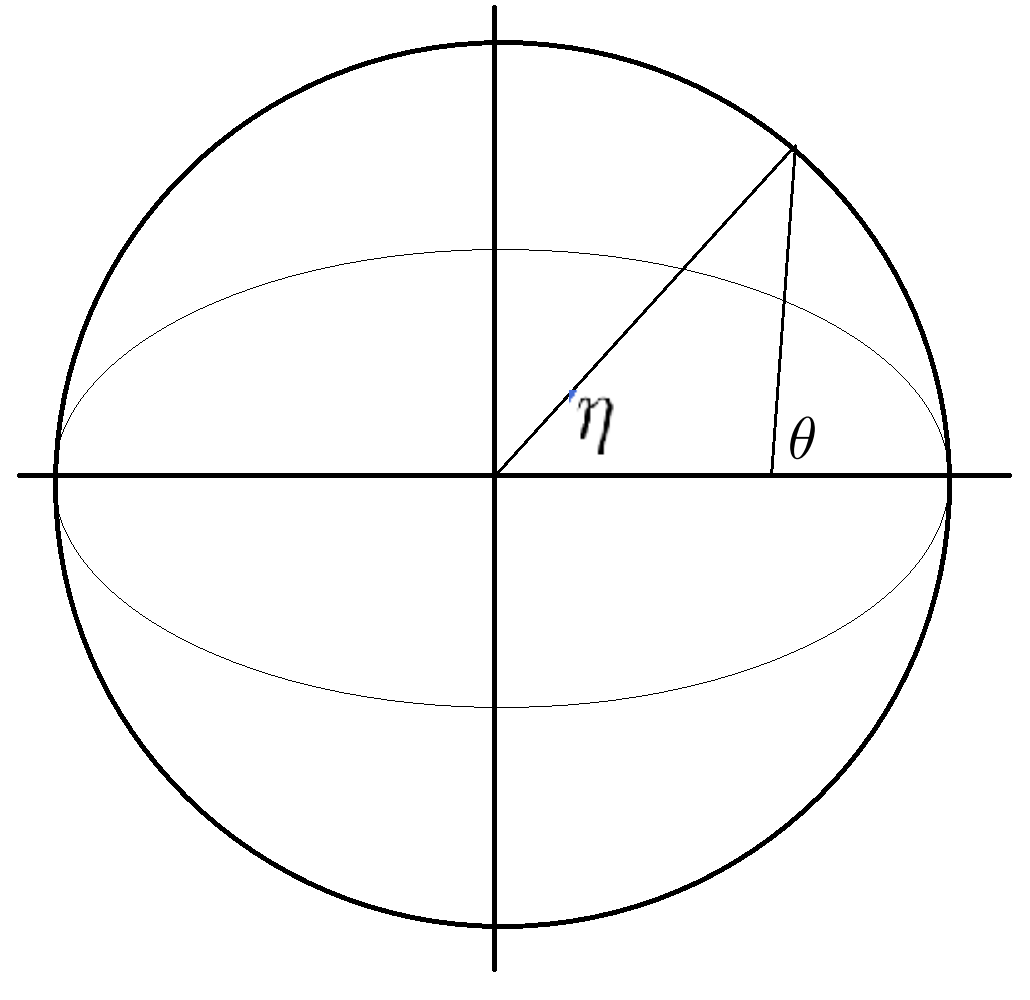}
\caption{\it{Illustration of the mean ($\eta$) vs. the true ($\theta$) anomaly. Their relation depends on the eccentricity of the trajectory. } } 
 	\label{fig:trueAnomMeanAnom}
\end{figure}

\section{Quasi-Keplerian Solution}
\label{sec:meanTrueAnom}
\subsection{Newtonian case}
In order to study the dynamics following from the above Lagrangian, we start by reviewing  the mean and true anomaly formalism  together with the Keplerian parameterisation for the Newtonian motion \cite{brouwer1961methods}. The conservation of  energy and  angular momentum reads
\begin{equation}
\epsilon = \frac{1}{2}\dot{r}^2 + \frac{j^2}{2r^2} -\frac{G M }{r}, \quad j = r^2 \dot{\theta},
\label{eq:NewEqu}
\end{equation}
where $\epsilon$ is the total energy per unit of  reduced mass, $j$ is the total angular momentum per unit of reduced mass, $r$ is the separation between the bodies and the dot is the derivative with respect to time. In order to solve the two-body problem, it is useful to parameterize the separation as
\begin{equation}
r/a = 1 - e \cos \eta .
 \label{eq:meanAnomDef}
\end{equation}
$e$ is the eccentricity, $a$ is the semi-major axis. The exact solution of Eq.~(\ref{eq:NewEqu}) is described by the well-known relations
\begin{equation}
n_b \left(t - t_0 \right) = \eta - e \sin \eta, \quad \theta = \tilde{\nu}_e.
\label{eq:meanAnomNew}
\end{equation}
$n_b = 2 \pi/  P_b$ is the frequency whilst $\tilde \nu$ is defined via
\begin{equation}
\tilde{\nu}_e \equiv 2\arctan \left[  \sqrt{\frac{1+e}{1-e}} \, \tan \frac{\eta}{2} \right].
\label{eq:nutildedef}
\end{equation}
This provides the Newtonian relationship between the mean anomaly $\eta$ and the true anomaly $\theta$. In the Newtonian case,  two  angles are used to describe the instantaneous position of the reduced body of mass $\mu$ on the ellipse, namely $\theta$, i.e. the true anomaly and $\eta$, i.e. the mean anomaly. Fig.~\ref{fig:trueAnomMeanAnom} shows the two anomalies and the relation between them.

The orbital parameters  are related to the energy and the angular momentum via
\begin{equation}
\begin{split}
a = \sqrt{-\frac{G M}{2\epsilon}}, \quad e^2 = 1 + 2 l^2 \epsilon , \quad n_b = \frac{\left(-2\epsilon\right)^{3/2}}{G M},    
\end{split}
\end{equation}
which corresponds to the Keplerian $3^{rd}$ law $n=\sqrt{G M/a^3}$.

\subsection{Conformal Interaction}
\begin{figure}[t!]
    \centering
\includegraphics[width=0.44\textwidth]{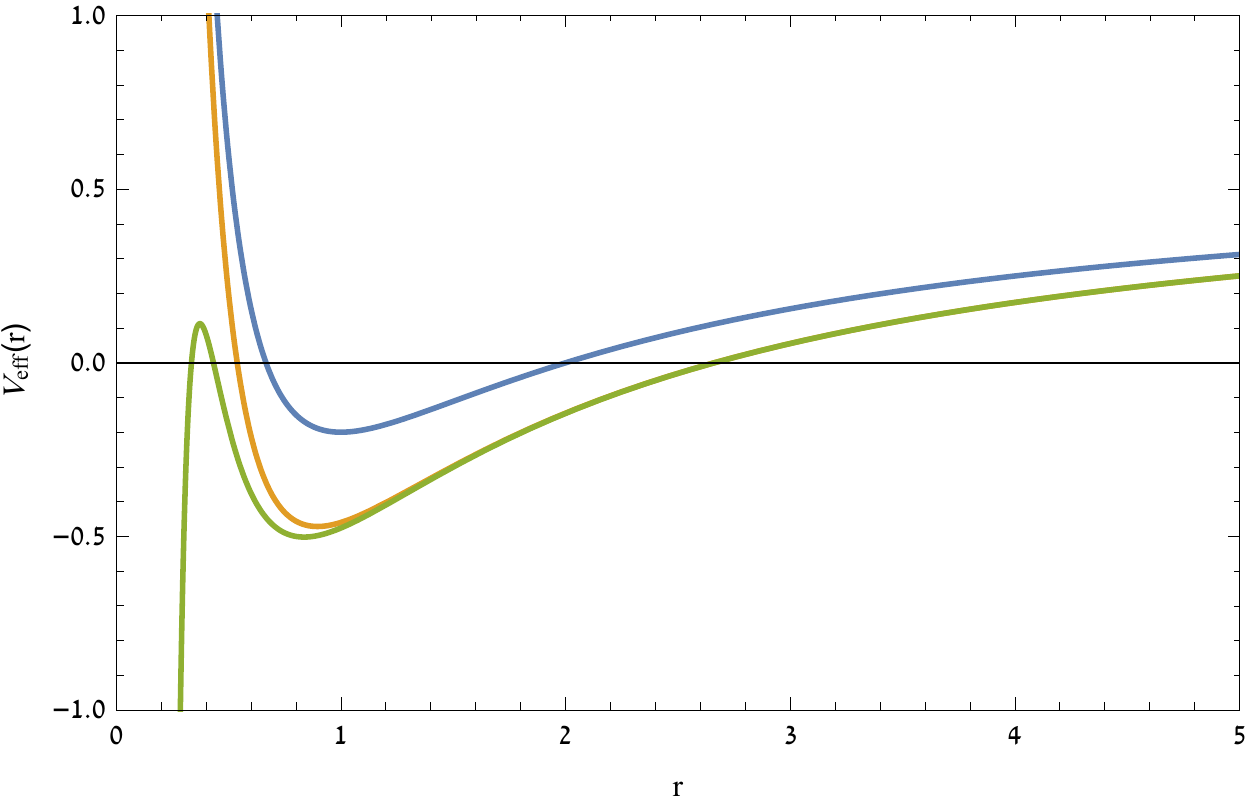}
    \caption{The effective potential for the $1^{st}$ Post Newtonian correction. The blue curve shows the PN case, the orange curve shows the conformal modification (with $\beta^2 $) and the green curve shows the modification with  the disformal interaction (with $\Lambda $).}
    \label{fig:effPotPN}
\end{figure}

We  now consider the general case and reinstate the conformal and the disformal interactions up the $1^{st}$ PN correction. In this case we find that the equations of motion obtained from the Lagrangian with the conformal and disformal corrections  become
\begin{equation}
\dot{r}^2 = \sum_{i = 1}^{5}{\frac{\alpha_i}{r^i}},\quad \dot{\theta}r^2 = \sum_{i = 1}^{4}{\frac{\gamma_i}{r^i}} ,
\label{eq:fullEoM}
\end{equation}
with the couplings listed below for completeness
\begin{eqnarray}
&&
\alpha_0 = 2\epsilon \left(1 + \frac{3}{2}(3 \nu -1)\frac{\epsilon}{c^2}\right)
\nonumber \\
&&\alpha_1 = 2 G M \left(1+\frac{\epsilon}{c^2}(7 \nu -6) +2 \beta^2(1 + \frac{2\epsilon}{c^2})\right) 
\nonumber \\
&&
\alpha_2 = -j^2 \left(1 - 2 (1-3 \nu ) \frac{\epsilon}{c^2}\right)\nonumber \\
&&+\left(5\nu - 10 +2\beta^2 (12\nu -11)\right)\left(\frac{G M}{c}\right)^2    
\nonumber \\
&&\alpha_3 = \left(8-( 3+8\beta^2) \nu\right) \frac{G M j^2}{c^2}
\nonumber\\
&&\alpha_4 = \epsilon \frac{16 \beta^2 G M^2}{ \Lambda ^4},  \quad \nonumber \\
&& \alpha_5 = \frac{16 \beta^2 G^2 M^3}{\Lambda ^4},
\nonumber \\
&&\gamma_0 =  j \left( 1 + (3 \nu -1)\frac{\epsilon}{c^2}\right),
\nonumber \\
&&\gamma_1 =  \left((2 + 4 \beta^2)\nu - 4 \right) \frac{G  M j }{c^2},
\nonumber \\
&&\gamma_2 = \gamma_3 = 0,\quad \gamma_4 =  -\frac{8 \beta^2 G j M^2}{ \Lambda^4}.
\nonumber \\
\end{eqnarray}
The GR case gives only contributions to the orders between $0$ and $3$  for the $\alpha$ parameters and $0$ to $1$ for the $\gamma$ parameters. The higher order terms  emerge from the disformal interaction as they depend on  the coupling $\Lambda$. The solution for the system is similar to the $3PN$ solution in the GR case with modified couplings. We will review how to describe the orbits below. 

For the limit $\Lambda \rightarrow \infty$ there are only modifications from the conformal coupling. Based on identities from \cite{Damour:1988mr} (see appendix \ref{appA}) the solution for this case reads
\begin{equation}
n (t-t_0) = \eta - e_t \sin \eta , \quad \frac{2 \pi}{\Phi} \left(\theta - \theta_0\right)   = \tilde{\nu}_{e_\theta},
\end{equation}
where the relation between the mean and true anomalies now involves the angular eccentricity $e_\theta$ 
whilst the orbital radius evolves according to $r/a_R=1-e_R \cos \eta $. All in all, this requires the definition of four parameters $(a_R,e_R,e_t,e_\theta)$. The parameters are given by
\begin{eqnarray}
&&
n = \frac{\left(-2\epsilon\right)^{3/2}}{G M}\left(1-2 \beta^2-\frac{\epsilon}{4 c^2}(\beta^2 (6 \nu +62)+\nu -15)\right),
\nonumber \\
&& a_R = -\frac{G M }{4 \epsilon}\left(2+4 \beta^2-\frac{\epsilon }{c^2}(2 \beta^2 (\nu +1)+\nu -7)\right),
\nonumber \\
&&
e_R^2 - 1 = 2 (1-4 \beta^2)\lambda \nonumber \\
&&
+\frac{\epsilon }{c^2}\left(2 (14 \beta^2+\nu -6)+\lambda(-20 \beta^2 \nu +92 \beta^2+5 \nu -15)\right),\nonumber \\
&&
e_t^2  - 1= 2(1-4 \beta^2) \lambda 
\nonumber \\
&&+\frac{\epsilon}{c^2}(4 \beta^2 (-5 \nu  \lambda +23 \lambda +7)+5 (\nu -3) \lambda +2 (\nu -6)),
\nonumber \\
&&
e_\theta^2 - 1 = \frac{4 \epsilon }{c^2} (\beta^2 (\nu +7)-3)\nonumber \\
&& + \lambda^2 (4 \beta^2 (\frac{\epsilon}{c^2} (\nu +23) -2)+\frac{\epsilon}{c^2} (\nu -15)+2), 
\end{eqnarray}
where $\lambda = \epsilon j^2/G^2 M^2$ is a dimensionless parameter.
Finally for the precession we have
\begin{equation}
\Delta\theta = \frac{\Phi}{2\pi} - 1 = \frac{6 \pi G M }{a c^2 (1-e^2)}\left(1 - \frac{2}{3} \beta^2 \right).
\end{equation}
 It is useful to use the relation
\begin{equation}
\frac{e_R}{e_t} = 1 + \frac{G M}{a_R c^2}\left[ 4-\frac{3}{2} \nu + \beta^2 \left(8-7 \nu \right)\right],
\end{equation}
{that connects the time eccentricity and the radial eccentricity. As expected, for the limit $\beta \rightarrow 0 $ the solutions from \cite{Damour:1988mr} are recovered.}
 
\subsection{Conformal and  disformal interactions}
To address the effect of the disformal coupling, we follow the approach of \cite{Damour:1999cr,Memmesheimer:2004cv}. First we attempt to solve the radial equation~(\ref{eq:fullEoM}). We can write the radial equation using  the Binet variable $s = 1/r$ 
\begin{equation}
\frac{\dot{s}}{s^4} = \left(\beta_0 + \beta_1 s + \beta_2 s^2 + \beta_3 s^3\right) \left(s - s_{-}\right)\left(s - s_{+}\right). 
\label{eq:radAsS}
\end{equation}
The full $\beta$ terms can be found  in Appendix \ref{appB}. This equation has two non-zero roots $s_{\pm}$. The eccentricity $e_R$ and semi major axis $a_R$ are defined using the periastron and the perihelion via $a_R = (s_{-} + s_{+})/(2s_{-}s_{+}) $ and $e_R = (s_{-} - s_{+})/(s_{-}+s_{+})$ leading to 
\begin{eqnarray}
&&
a_R = -\frac{G M }{4 \epsilon}\left(2+4 \beta^2-\frac{\epsilon }{c^2}(2 \beta^2 (\nu +1)+\nu -7)\right) \nonumber \\ &&  - \frac{32 \beta^2  M^3}{\Lambda^4 j^6} \left(\epsilon j^2+G^2 M^2\right),\nonumber \\
&&
e_R^2 - 1 = 2 (1-4 \beta^2)\lambda \nonumber \\
&& +\frac{\epsilon }{c^2}\left(2 (14 \beta^2+\nu -6)+\lambda(-20 \beta^2 \nu +92 \beta^2+5 \nu -15)\right)\nonumber \\
&& -\frac{64 \beta^2 \epsilon }{G \Lambda^4 j^6} \left(5 \epsilon^2 j^4+10 \epsilon G^2 j^2 M^2+4 G^4 M^4\right).
\end{eqnarray}
Using the Taylor expansion of $(1 + x)^{1/2}$, it is possible to solve for the time dependence by inverting  Eq.~(\ref{eq:radAsS})
\begin{equation}
t - t_0 = \int_{s}^{s_{-}} ds \, \frac{A_0 + A_1 s + A_2 s^2 +A_3 s^3}{\sqrt{(s-s_{-})(s_{+}-s)}s^2} .
\label{mott}
\end{equation}
with the $A$ coefficients given  in Appendix~(\ref{appB}). One can derive the modified $3^{rd}$ Keplerian law via the complete integral $\int_{s_{+}}^{s_{-}}$, that gives the period $P$. The frequency of the motion then reads
\begin{equation}
\begin{split}
\frac{n}{\sqrt{G M/a_R^3}} = 1 - \beta^2+ \frac{G M }{2 a_R c^2} (\nu - 9 + 2\beta^2 (2 \nu +1) \\-\frac{96 G \beta  M^2 }{ \Lambda^4 j^6} \left(\epsilon j^2+G^2 M^2\right).
\end{split}
\end{equation}
The conformal coupling changes the rate in the Keplerian $3^{rd}$ law  via the relativistic correction factor $G M/a_R c^2$. The angular equation~(\ref{eq:fullEoM}) becomes $d\theta/ds = d\dot{\theta}/d\dot{s} $ and can be expressed as
\begin{equation}
\theta - \theta_0 = \int_{s}^{s_{-}} ds \, \frac{C_0 + C_1 s + C_2 s^2 +C_3 s^3}{\sqrt{(s-s_{-})(s_{+}-s)}} .
\end{equation}
Next, one can calculate the advance of the perihelion $\Phi$ during a full orbit by taking the limits $s_{-}$ to $s_{+}$
\begin{equation}
\Phi = 2\int_{s_{+}}^{s_{-}} ds \, \frac{C_0 + C_1 s + C_2 s^2 +C_3 s^3}{\sqrt{(s-s_{-})(s_{+}-s)}} .
\end{equation}
The precession gives the modified relation
\begin{equation}
\begin{split}
k = \frac{3 G M }{a c^2 (1-e^2)}\left(1 - \frac{2}{3} \beta^2  + \frac{5 \beta^2 M }{6 \pi \Lambda^4 a^3 \left(1-e^2 \right)^3} \right), 
\end{split}
\end{equation}
where $2\pi k = \left(\Phi/2\pi - 1\right)$. This expression was already derived in \cite{Brax:2018bow,Brax:2019tcy} using the Binet equation. The above derivation uses  the quasi Keplerian parameterisation where we have reduced the description of the motion to a single integral (\ref{mott}) which generalises the description of the two-body motion in general relativity and appears to be akin to a 3PN parametrisation. This coincidence could become important as tests of GR become more and more precise, i.e. the scalar interactions could appear as systematic corrections to the expected result in GR at the 3PN order.

We will use  the periastron advance of binary pulsars defined as:
\begin{equation}
\begin{split}
 \dot{\omega} = \frac{k}{n} = \frac{\left(m T_\odot \right)^{2/3} n_b^{5/3}}{1 - e^2} \left[3 - 2\beta^2  + \frac{5\epsilon_\Lambda }{2 \pi T_{\odot}\Lambda^2}\right] ,   
\end{split}
\label{eq:prec}
\end{equation}
where $T_{\odot} = G M_{\odot}/c^2$ is the solar mass in time units and $P_b$ is the orbital period of the binary system. The effects of the conformal and disformal interaction appear as two additive corrections depending on $\beta^2$ and $\epsilon_\Lambda$ respectively.


\section{Radiation emission}
\label{sec:RadEm}
Dissipation in two-body problems is discussed from a phenomenological point of view in \cite{1994CeMDA..58..393M,1998MNRAS.299..237B,Brax:2019tcy}. The best modern timing model introduces the orbital period derivative
\begin{subequations}
\begin{equation}
\eta-e\sin \eta = 2\pi \left[ \left( {{t-t_0}\over{P_{\rm b}}} \right) - {{\dot P_{\rm b}}\over 2} \left( {{t-t_0}\over{P_{\rm b}}} \right)^2
  \right],
\end{equation}
\begin{equation} 
\theta - \theta_0 = \frac{\dot{\omega}}{n_b}
         \tilde{\nu}_e.  
\end{equation}
\end{subequations}
The first equation deforms  the relation between the mean anomaly and time by introducing a dependence on  the energy loss, and the second modifies the relation between the true and mean anomaly. Ref.~\cite{Brax:2019tcy} shows that such terms emerge from modified gravity due to the energy loss for binary objects and calculates in detail the scalar radiation emission. This section derives the corresponding PKP for the dissipating conformal and disformal dark energy by inferring the radiation reaction force in a manner akin to the standard GR treatment.

\subsection{Dissipation}
The scalar field radiates energy away from the binary system as shown in \cite{Brax:2019tcy}. This can be captured using an effective expansion of the interaction Lagrangian between the long wavelength field $\bar\phi$, i.e. the radiated field, once the short wavelength degrees of freedom corresponding to the scalar reaction to the motion of the two point masses has been integrated out. This reads explicitly
\begin{equation}
S_{\rm int}^{(\phi)} =  \frac{1}{m_\mathrm{Pl}} \int dt \left( I_\phi \bar{\phi} + I_\phi^i \partial_i \bar{\phi} + \frac{1}{2} I_\phi^{ij} \partial_i \partial_j \bar{\phi}  \right)
\end{equation}
where the multipole moments, characteristic of the binary system, are given by
\begin{equation}
\begin{split}
I_\phi  \equiv  \int d^3 x \left( J + \frac{1}{6}  \partial_t^2 J  x^2 \right) \;,\\ \quad I_\phi^i  \equiv \int d^3x J x^i \;, \quad  I_\phi^{ij} \equiv \int d^3x J \left(  x^i x^j - \frac{1}{3}x^2 \delta^{ij}\right) \; .    
\end{split}
\label{eq:scalar_dipole_quadrupole}
\end{equation}
From this expression, the power radiated into the scalar field reads
\begin{equation} \label{eq:radiated_power_scalar}
P_\phi = {2G}\left[  \big\langle \dot{I}_\phi^2  \big\rangle  + \frac{1}{3} \big\langle  \ddot{I}_\phi^i \ddot{I}_\phi^i  \big\rangle + \frac{1}{30} \big\langle \dddot{I}_\phi^{ij}  \dddot{I}_\phi^{ij}   \big\rangle \right]
\end{equation}
where the average is taken over many gravitational wave cycles. This is the scalar counterpart of the GR power radiated 
\begin{equation} \label{eq:quadrupole}
P_h = \frac{G}{5}  \left\langle \dddot{I}_h^{ij}\  \dddot{I}_h^{ij} \right\rangle
\end{equation}
where $I_h^{ij} \equiv \int d^3x T^{00} \left( x^i x^j - \frac{1}{3}x^2 \delta^{ij}\right)$ is the gravitational quadrupole of the source.
Working at leading order in the velocity expansion, 
the scalar current  $J$ is simply given by
\begin{equation} \label{eq:J_v0}
J_{v^0} = - \beta \left( m_A \delta^3(\vec{x}-\vec{x}_A) + m_B \delta^3(\vec{x}-\vec{x}_B) \right),
\end{equation}
i.e. this is nothing but the direct coupling between the point sources and the scalar field. 
The relativistic corrections to the current $J$ come by integrating out the short distance degrees of freedom and give for the conformal part
\begin{equation}
\begin{split}
J_{v^2} = \beta \left( m_A \frac{v_A^2}{2} \delta^3(\vec{x}-\vec{x}_A) + (A \leftrightarrow B) \right) \\+ \beta \frac{Gm_Am_B}{|\vec x_A - \vec x_B|} \left( \delta^3(\vec{x}-\vec{x}_A) + (A \leftrightarrow B) \right) \; .    
\end{split}
\end{equation}
and for the disformal part
\begin{equation}
J^\mathrm{disf} = 4 \beta \frac{Gm_Am_B}{\Lambda^2} \frac{d^2}{dt^2} \frac{\delta^3(\vec{x}-\vec{x}_A) + (A \leftrightarrow B)}{|\vec x_A - \vec x_B|} \; .
\end{equation}
The main contribution of the disformal current $J^\mathrm{disf}$ comes from the monopole at quadratic order in the velocities whilst higher order multipoles are suppressed by velocity powers . The  dipole and quadrupole emission terms are only dependent on  $J_{v^0}$ at this order
\begin{align}
\begin{split}
I_\phi^{ij} &= - \beta \left( m_A \left( x^i x^j - \frac{1}{3}x^2 \delta^{ij}\right) + (A \leftrightarrow B) \right) \\
I_\phi^{i} &= - \beta \left( m_A x_A^i + (A \leftrightarrow B) \right).
\end{split}
\end{align}
 As  the conformal coupling $\beta$ is universal, the dipole contribution to the radiated power vanishes, because its second derivative is zero by the centre-of-mass theorem at lowest order in the velocity. The lowest order contribution to the monopole coming from $J_{v^0}$ vanishes as $I_{\phi, \; v^0} = - \beta (m_A + m_B)$ is conserved. The next contribution to the monopole starts at the quadratic order in the velocities and reads
\begin{align}
\begin{split}
I_{\phi, \; v^2 + \mathrm{disf}} &= \frac{\beta}{6} (m_A v_A^2 + m_B v_B^2) + \frac{\beta}{3}(7+2\beta^2) \frac{Gm_1m_2}{|\vec x_A - \vec x_B|} \\
&+ 8 \beta \frac{Gm_Am_B}{\Lambda^2} \frac{d^2}{dt^2} \frac{1}{|\vec x_A - \vec x_B|} \; .
\end{split}
\end{align}
At this order the conservation of energy implies that  $m_A \frac{v_A^2}{2} + m_B \frac{v_B^2}{2} \simeq \frac{G(1+2\beta^2)m_Am_B}{|\vec x_A - \vec x_B|}$ and therefore
\begin{align} \label{eq:monopole}
\begin{split}
I_{\phi, \; v^2 + \mathrm{disf}} = 4 \beta G m_A m_B \\ \times \left( \frac{2+\beta^2}{3|\vec x_A - \vec x_B|} + \frac{2}{\Lambda^2} \frac{d^2}{dt^2} \frac{1}{|\vec x_A - \vec x_B|} \right).    
\end{split}
\end{align}
This can be used to deduce the radiation reaction force acting on the binary system.

\subsection{Radiation reaction force}

We are interested in deriving the secular effects of the radiated power on the  trajectories of the reduced two-body system in the centre of mass frame. As we are considering the non-relativistic regime and we shall focus on  the effect of energy loss at leading order, we can write
\be
\frac{d\epsilon}{dt}=-\vec{\mathcal{F}}_{\rm d}.\vec v
\ee
where $\vec{\mathcal{F}}_{\rm d}$ is the reduced force acting on the reduced particle in  the centre of mass frame. We can identify the dissipative force by integrating over a closed orbit of period $T$ 
\be 
\begin{split}
-\int_0^T dt \, \mu \mathcal{F}_{\rm d}.\vec v= {2G}\int_0^T dt \,[  \dot{I}_\phi^2    + \frac{1}{3}   \ddot{I}_\phi^i \ddot{I}_\phi^i\\ + \frac{1}{30}  \dddot{I}_\phi^{ij}  \dddot{I}_\phi^{ij} +\frac{1}{10}  \dddot{I}_\h^{ij}  \dddot{I}_\h^{ij} ].    
\end{split}
\ee
For universal couplings, the dipole radiation vanishes. Let us start with the monopole. It is convenient to rewrite
\be 
I_\phi=\beta\mu \left[ \frac{4}{3}  v^2 + \frac{8 G M}{\Lambda^2}\frac{d^2}{dt^2}\left(\frac{1}{r}\right) \right]
\label{Iphi}
\ee
where the conservation of energy at leading order has been used.
After one integration by parts we can identify
the drag force corresponding to an effective viscosity 
\be 
\vec F_{\rm drag}^{(0)}=\frac{8}{3} G\beta \ddot I_\phi \vec v.
\ee
This term involves a friction term which depends on the derivative of the acceleration.
There is also a reaction term
\be 
\vec F_{\rm re}^{(0)}=\frac{16 G^2 \beta   M}{\Lambda^2 r^2} {I}^{(3)}_\phi \hat{n}
\ee
which involves only the disformal term and the third time derivative of the scalar monopole. 

The quadrupole dissipation term is similar to the one in  GR, i.e. we find after two integrations by parts 
\be 
\vec F^{(2)}_{\rm d}= (1+ \frac{\beta^2}{3}) \vec F_{\rm GR}
\ee
where the  GR force contains only a radiation reaction components with 
\be 
  F_{\rm GR}^i= \frac{2G }{5} r^j Q_h^{(5) ij} 
\ee
involving the fifth time derivative of the quadrupole. 
We have used that in the centre of mass frame, the quadrupole moment reads
\be 
Q^{ij}=\mu (r^i r^j - \frac{\delta^{ij}}{3} r^2).
\ee
In conclusion, we find that dissipation due to the scalar field gives rise to an enhanced quadrudople radiation reaction and introduces two new forces from monopole radiation. One of them is characteristic of the disformal interaction and involves a modification of Newton's law proportional to the third time derivative of the monopole. All these interaction are of higher order and therefore potentially break causality. We will only consider them as perturbations to the GR trajectories. 

\begin{figure}
    \centering
\includegraphics[width=0.4\textwidth]{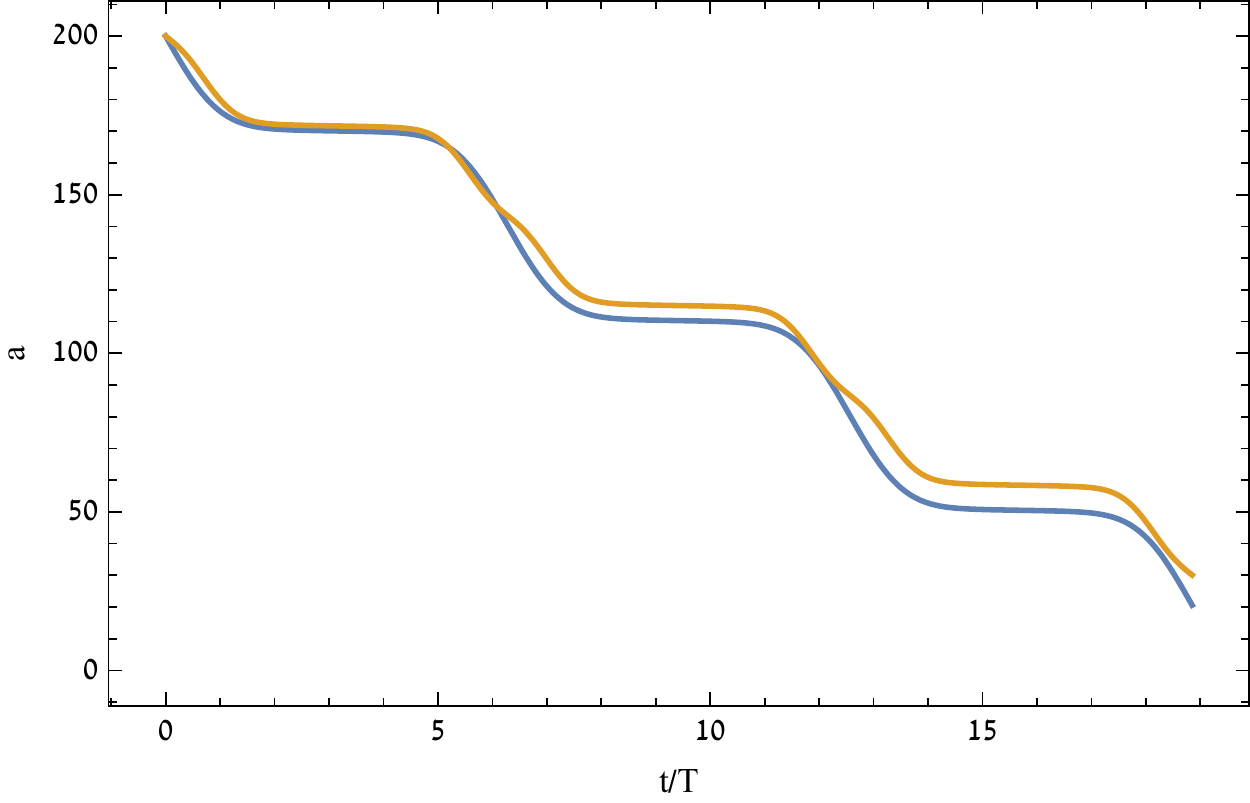}
\includegraphics[width=0.4\textwidth]{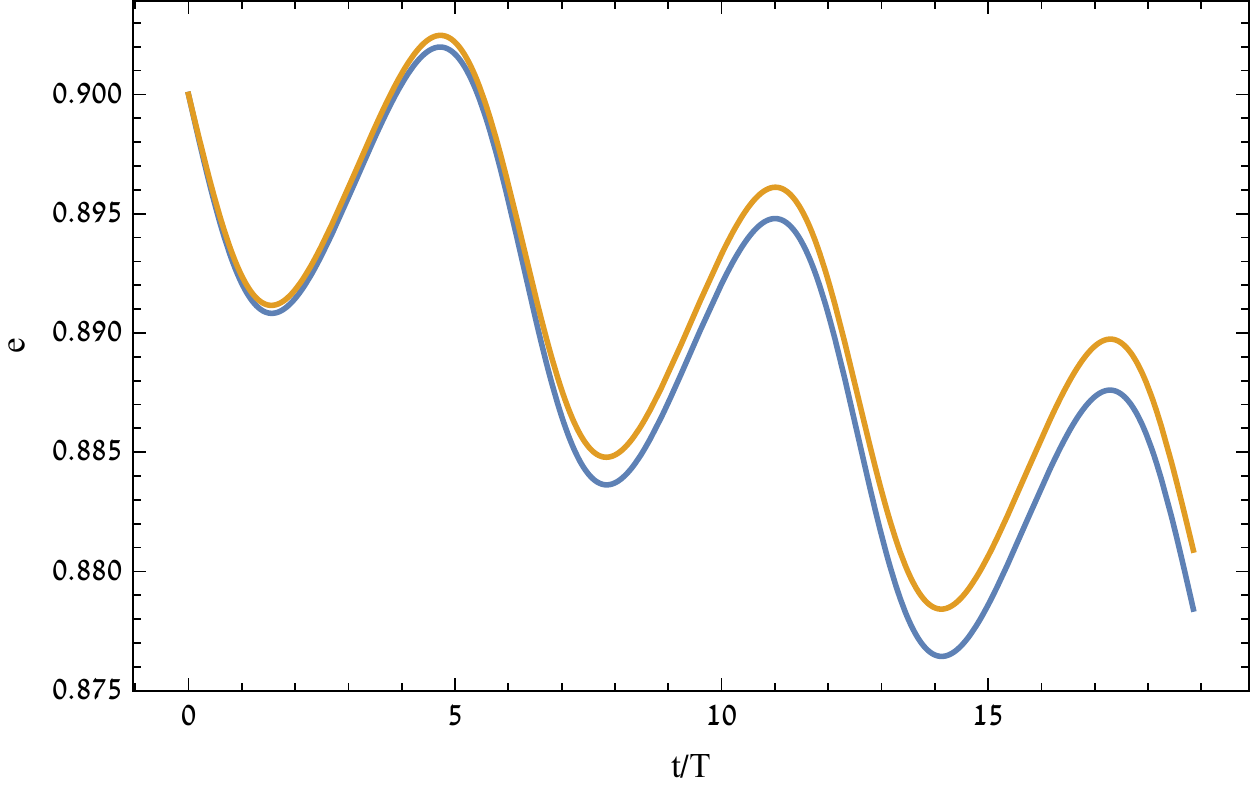}
\caption{\it{An illustration of the GR (blue) and disformal effects (yellow) for $\epsilon^{2} = 0.1$, $e = 0.9$ and $\beta^2 = 0.01$, $\epsilon_{\rm diff}^{(0)} = 0.06$. The upper plot shows $a$ and the lower one  $e$.}}
    \label{fig:effpe}
\end{figure}

\subsection{Reduced equations of motion}

Assuming that the main effect on the correction to the trajectories comes from the radiation reaction force, we can write in the centre of mass frame
\be 
\ddot{\vec{r}}  = -\frac{G M}{r^2} \hat{n} + \vec{\mathcal{F}}_d
\ee
where $\vec{\mathcal{F}}_d= \vec{\mathcal{F}}_d^{(0)} + \vec{\mathcal{F}}_d^{(2)}$. It is possible to separate the force $\vec {\mathcal{F}}_d$ into a radial $\mathcal{R}$ and into a tangential parts $\mathcal{S}$:
\begin{equation}
\vec{\mathcal{F}}_d = \mathcal{R} \,\hat{n} + \mathcal{S} \, \hat{\theta}.
\end{equation}
We use the cylindrical coordinate system $\left(r,\theta,z\right)$. The conformal and the disformal interaction effects are in the plane $\left(r,\theta\right)$. The loss for the angular momentum reads
\be 
\begin{split}
\frac{d}{dt} r^2 \dot \theta= \left(\vec{r} \wedge \vec{\mathcal{F}}_d\right)\cdot\hat{z} = \\ \frac{4G\beta(2+\beta^2)}{3(1+2\beta^2)} \ddot I_\phi r^2 \dot \theta  + (1+\frac{\beta^2}{3})\frac{2G }{5}\epsilon^{zki} r^k r^j Q_h^{(5) ij}    
\end{split}
\ee
The last term is the rescaled effect of the quadrupole radiation on the angular momentum. The first term is simply the effect of the monopole scalar radiation. Notice that the monopole $I_\phi$ is given by (\ref{Iphi}). Moreover both the monopole and the quadrupole are proportional to the reduced mass $\mu$. As a result, the effect of radiation will be proportional to $\nu$.

Since the forces include high derivatives of the separation and the anomaly, one can use perturbation theory in  the Newtonian solution to reduce the higher derivatives. More explicitly, we use the relations in the Appendix \ref{appsim}. For instance, the relation $\ddot{\theta} = -2 \dot{r} \dot{\theta}/r$ emerges from the conservation of the angular momentum $l = r^2 \dot{\theta}$, at $1^{st}$ order.  The relation~(\ref{Iphi}) becomes:
\begin{equation}
I_\phi  = \frac{2}{3} \beta  \mu  \left(\dot{r}^2+r^2 \dot{\theta}^2 + \frac{12 G M }{\Lambda^2 r^4} \left(G M+2 r \dot{r}^2-\dot{\theta}^2r^3\right) \right).
\end{equation}
Consequently, the modification for the force read
\begin{eqnarray}
&&
\mathcal{R}_{drag} =  \frac{32 \beta ^2 G^2 \mu  M \dot{r} }{9 r^4} \tilde{\alpha} + \frac{64 \beta ^2 G^2 \mu  M \dot{r} }{3 \Lambda^2 r^7} \tilde{\beta},
\nonumber \\
&&
\mathcal{S}_{drag} = \frac{32 \beta ^2 G^2 \mu  M \dot{\theta} }{9 r^3} \tilde{\alpha} + \frac{64 \beta ^2 G^2  \mu  M \dot{\theta} }{3 \Lambda^2 r^6} \tilde{\beta},
\nonumber \\
\end{eqnarray}
where
\begin{eqnarray}
&&
\tilde{\alpha} = G M+2 r \dot{r}^2 - r^3 \dot{\theta}^2,
\nonumber \\
&&
\tilde{\beta} = 8 G^2 M^2-r^3 \dot{\theta}^2 \left(17 G M+72 r \dot{r}^2\right)\nonumber \nonumber \\ && +58 G M r \dot{r}^2+24 r^2 \dot{r}^4+9 r^6 \dot{\theta}^4.   
\nonumber \\
\end{eqnarray}
together with
\begin{equation}
\begin{split}
\mathcal{R}_{\rm re} = -\frac{64 \beta ^2 G^3   \mu  M^2 \dot{r} }{3 \Lambda^2 r^7} \left(8 G M+6 r \dot{r}^2-9 r^3 \dot{\theta}^2\right), \quad  \mathcal{S}_{\rm re} = 0,
\end{split}
\end{equation}
and the quadrupolar expressions
\begin{eqnarray}
&&
\mathcal{R}_{d}^{(2)} = \left(1 + \frac{\beta ^2}{3}\right) \frac{16   G^2 M \mu \dot{r} }{15 c^5 r^4} \left(4 G M+3 r \dot{r}^2+3 r^3 \dot{\theta}^2\right),
\nonumber \\
&&\mathcal{S}_{d}^{(2)} = -\left(1 + \frac{\beta ^2}{3}\right) \frac{8  G^2 M \mu \dot{\theta} }{5 c^5 r^3} \left(3 G M+r \dot{r}^2+r^3 \dot{\theta}^2\right) .
\nonumber \\
\end{eqnarray}
With these relations we can write the system of equations for the orbital parameters for the  Keplerian trajectory where we define $p_0 = a(1-e^2)$. It is useful to introduce the small parameters which govern the evolution of the secular perturbations 
\be 
\epsilon^{(2)}=\frac{8}{5} \nu  \left(\frac{G M}{c^2 p_0}\right)^{5/2}.
\ee
This characterises the quadrupolar part of the evolution equation and  scales as $v^5/c^5$. Similarly for the scalar part we can distinguish the conformal part which depends on
\be 
\epsilon^{(0)}_{\rm conf}= \beta^2 \epsilon^{(2)}
\ee
and a disformal part 
\be 
\epsilon_{\rm diff}^{(0)}= \frac{8}{5} \nu  \left(\frac{GM}{p_0}\right)^{7/2} \frac{\beta^2}{p_0^2 \Lambda^2}
\ee
Notice that the ratio of the dimensionless parameters is the parameter $\epsilon_\Lambda$ that also appears in the precession term
\be 
\frac{\epsilon_{\rm diff}^{(0)}}{\epsilon^{(0)}} = \epsilon_\Lambda = \frac{\left(\beta n / \Lambda\right)^2}{\left(1-e^3\right)^3},
\ee
which is of order $v^7/c^7$. As a result the radial force reads
\begin{eqnarray}
&&
\mathcal{R}_{drag} = \epsilon^{(2)} \frac{2 e G M s }{3 p_0^2}  (c e+1)^3 \left(10 c e+3 e^2+7\right),    
\nonumber \\
&& \mathcal{R}_{\rm re} = -\beta^2 \epsilon^{(2)} \frac{2  e G M s  }{9 p_0^2} (c e+1)^3 \left(15 c_2 e^2-8 e^2-7\right),
\nonumber \\
&&
\mathcal{R}_{d}^{(2)} = \epsilon_{\text{diff}}^{(0)} \frac{5 e G M s}{3 p_0^2} (c e+1)^5\nonumber \\ && \times (5 e^2 \left(c_2 \left(40-12 e^2\right)+3 e \left(7 c_4 e+16 c_3\right)\right) \nonumber \\ &&-16 c \left(e^2-6\right) e+27 e^4+40 e^2+8) .   
\end{eqnarray}
Similarly, the tangential force becomes
\begin{eqnarray}
&&
\mathcal{S}_{drag} = -\epsilon^{(2)} \frac{G M}{p_0^2}  (c e+1)^4 \left(5 c e+e^2+4\right),    
\nonumber \\
&&
\mathcal{S}_{\rm re} = -\beta^2 \epsilon^{(2)} \frac{ G M }{9 p_0^2} (c e+1)^4 \left(5 e \left(6 c_2 e+7 c\right)-7 e^2+12\right),
\nonumber \\
&&
\mathcal{S}_{d}^{(2)} = \epsilon_{\text{diff}}^{(0)} \frac{5 e G M }{3 p_0^2}  (c e+1)^6 
\nonumber \\  && \times (c \left(8-58 e^2\right)+20 c_2 e \left(5-3 e^2\right)\nonumber \\ && +3 e \left(35 e \left(c_4 e+2 c_3\right)+9 e^2-4\right)).   
\nonumber \\
\end{eqnarray}
The secular variations for the orbital parameters becomes
\begin{eqnarray}
&&
\langle \dot{a} \rangle_{sec} = -\sqrt{\frac{G M}{a}} \frac{1}{72 \left(e^2-1\right)} \times \nonumber \\ && (2 \text{$\epsilon $}_2 \left(96 \left(\beta ^2+3\right)+\left(77 \beta ^2+111\right) e^4+\left(452 \beta ^2+876\right) e^2\right)\nonumber \\ && -45 e^2 \left(27 e^6+472 e^4+592 e^2+64\right) \epsilon_{\rm{diff}}^{(0)})  
\nonumber \\ &&
\langle \dot{p} \rangle_{sec} = - \left(1-e^2\right)^{3/2} \sqrt{\frac{G M}{p_0}} \left(\epsilon^{(2)} + \frac{1}{3}\epsilon^{(0)}_{\text{conf}} \right) \left(1+ \frac{7}{8} e^2\right)
\nonumber \\ &&
\frac{\langle \dot{e} \rangle_{sec}}{e} = - \frac{\left(1-e^2\right)^{3/2}}{144} \sqrt{\frac{G M}{p_0^3}}\times\nonumber \\ &&  (45 \epsilon_{\rm diff}^{(0)}  \left(27 e^6+472 e^4+592 e^2+64\right) \nonumber \\ && -2 \epsilon^{(2)} \left(464 \beta ^2+\left(161 \beta ^2+363\right) e^2+912\right)). 
\nonumber \\
\end{eqnarray}
Using astronomical units, the time drift of the eccentricity becomes
\begin{eqnarray}
&& \frac{\dot{e}}{e} = - \frac{304}{15} T_{\odot}^{5/3} \frac{m_p m_c}{m^{1/3}} \left( \frac{P_b}{2 \pi}\right)^{-8/3} \frac{1+\frac{121}{304} e^2}{\left( 1-e^2\right)^{5/2}}  
\nonumber \\ &&
\left[ 1 + \beta^2 \frac{161 e^2+464}{363 e^2+912}- \frac{15 \epsilon_\Lambda \left(27 e^6+472 e^4+592 e^2+64\right)}{2 \left(121 e^2+304\right)}\right].\nonumber \\
\end{eqnarray}
In the case of $\beta \rightarrow 0$ and $\epsilon_\Lambda \rightarrow 0$, these terms  reduce to the known terms in \cite{Peters:1964zz}. Fig. \ref{fig:effpe} illustrates the evolution of $p$ and $e$ vs. time $t$ using the true anomaly $\theta$ parameterization. The impact of the disformal coupling is such that it changes the evolution of $e$ and $p$ on average, i.e. in a secular way.

We note that the time variation in the eccentricity due to the scalar interactions is very different from the Kozai-Lidov effect discussed in \cite{Will_book} and in \cite{Randall:2019sab} where a third, distant, body distorts the eccentricity of the binaries potentially causing oscillations. In the Kozai-Lidov case, radiative effects are not taken into account, unlike in our work. Given the very different behaviour between the Kozai-Lidov effect and the time variation on the eccentricity induced by radiative loss, the two phenomena  should be distinguishable in future experiments which would be sensitive enough to observe such time variations.


\subsection{Emitted power and time variation of the period}

The total emitted power splits into the power lost into gravitons and the one lost into the scalar field. The emitted power  reads on average 
\begin{equation}
P = \langle \vec{\mathcal{F}} \cdot \vec{v}\rangle  = \langle \mathcal{R} \dot{r} + \mathcal{S} \dot{\theta} \, r\rangle . 
\end{equation}
For the graviton case the emitted power  for elliptic orbits is known from the Peter-Mathews formula~\cite{Peters:1963ux}:
\begin{align}
\begin{split}
P_h &= -\frac{195\pi T^{5/3}_{\odot}}{5 n^{5/3}}\frac{m_p m_c}{m^{1/3}}f_1 (e) \;,
\end{split}
\end{align}
where $m$ is the total mass of the system and:
\begin{equation}
f_1 (e) = \frac{1+\frac{73}{24}e^2+\frac{37}{96}e^4}{(1-e^2)^{7/2}}.
\end{equation}
The scalar quadrupole \eqref{eq:scalar_dipole_quadrupole} is proportional to  the gravitational quadrupole, i.e.
\begin{equation}
P_\phi^\mathrm{quad} = \frac{\beta^2}{3} P_h\; .
\end{equation}
The scalar dipole is zero as the coupling $\beta$ is universal. We are left  to calculate the monopole power starting from eq.~\eqref{eq:monopole}. 
As a consequence of the loss of power, the variation of the orbital period reads:
\begin{eqnarray}
&&
\dot{P} =  -\frac{195\pi T^{5/3}_{\odot}}{5 n^{5/3}}\frac{m_p m_c}{m^{1/3}}
\nonumber \\ &&\times\left[(1+\frac{\beta^2}{3})f_1(e)+\frac{10}{9} \beta^2 \, f_2(e) - \epsilon_{\Lambda} \frac{20}{3} f_3(e), \right],
\label{eq:pbdot}
\end{eqnarray}
with 
\begin{eqnarray}
&&
f_2(e) = \frac{e^2\left(1+\frac{1}{4} e^2\right) }{(1-e^2)^{7/2}} ,
\nonumber \\ 
&&
f_3(e) = \frac{e^2 \left(1+\frac{37}{4}e^2+\frac{59}{8}e^4 + \frac{27}{64}e^6 \right) }{(1-e^2)^{13/2}} .   
\end{eqnarray}
The expression for $\dot{P}$ was obtained originally in ~\cite{Brax:2019tcy} and is confirmed here from the averaged loss due to the radiation reaction force. 
\begin{figure}
    \centering
    \pgfmathsetlength{\imagewidth}{\linewidth}%
    \pgfmathsetlength{\imagescale}{\imagewidth/290}%
    \begin{tikzpicture}[x=\imagescale,y=-\imagescale]
        \node[anchor=north west] at (0,0) {\includegraphics[width=\imagewidth]{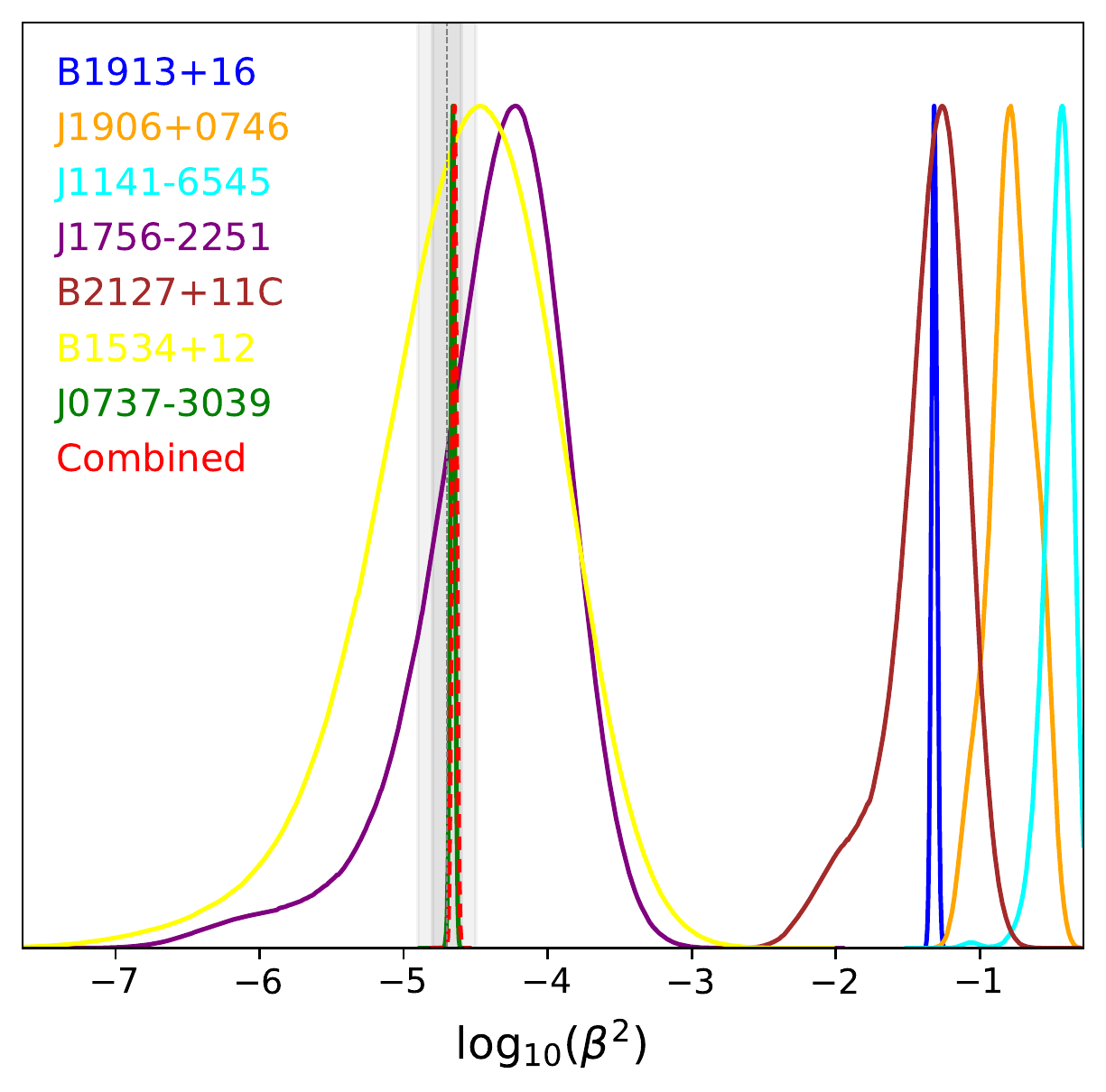}};
        \node[anchor=north west] at (10,130) {\includegraphics[width=0.35\imagewidth]{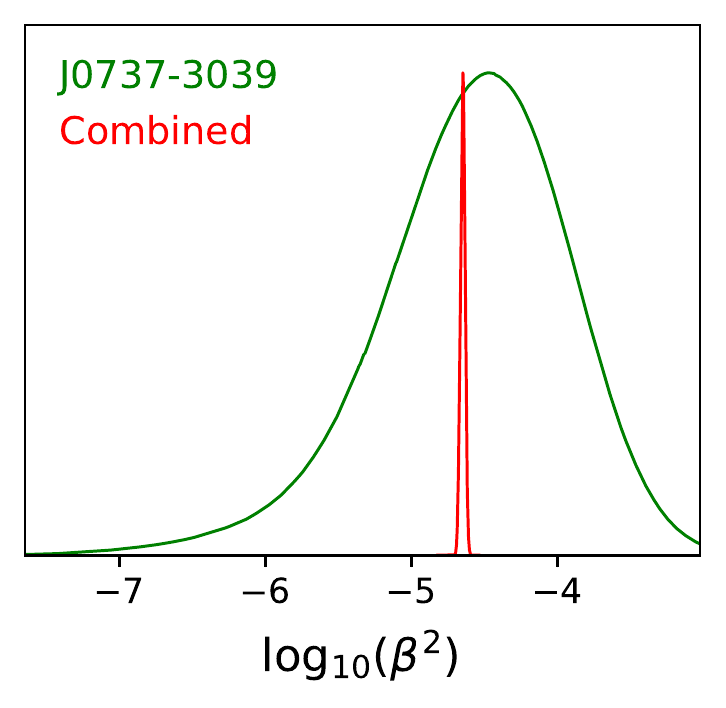}};
    \end{tikzpicture}
    \caption{\it{Conformal coupling constraint from different events in  comparison to the Cassini constraint (in gray). The subplot shows the double pulsar constraint (green) and the combined constraint (red) which is comparable to the Cassini upper bound.}}
\label{fig:conerplotsConformalContraints}
\end{figure}
\begin{figure}
    \centering
    \pgfmathsetlength{\imagewidth}{\linewidth}%
    \pgfmathsetlength{\imagescale}{\imagewidth/524}%
    \begin{tikzpicture}[x=\imagescale,y=-\imagescale]
        \node[anchor=north west] at (0,0) {\includegraphics[width=\imagewidth]{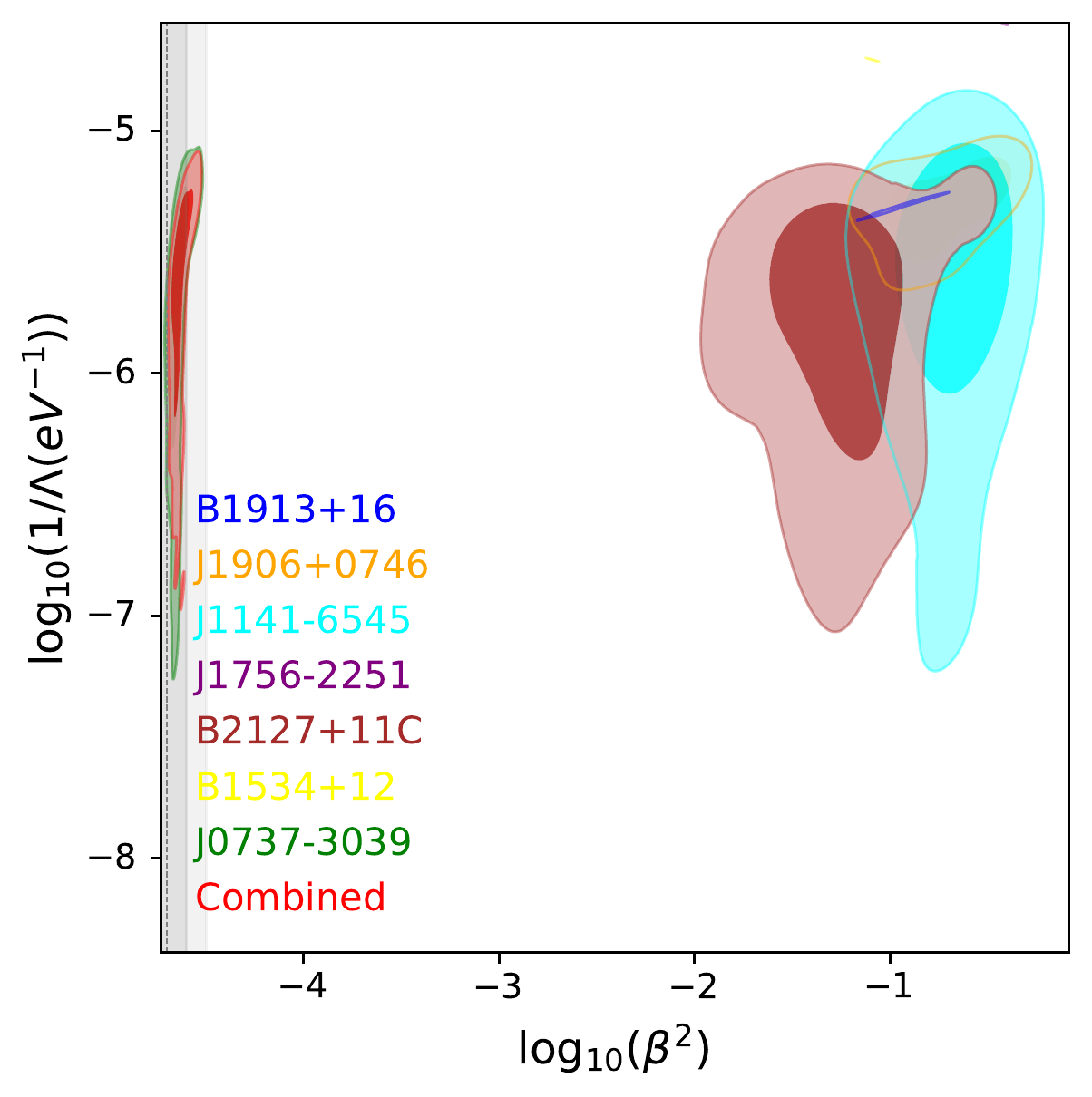}};
        \node[anchor=north west] at (90,10) {\includegraphics[width=0.45\imagewidth]{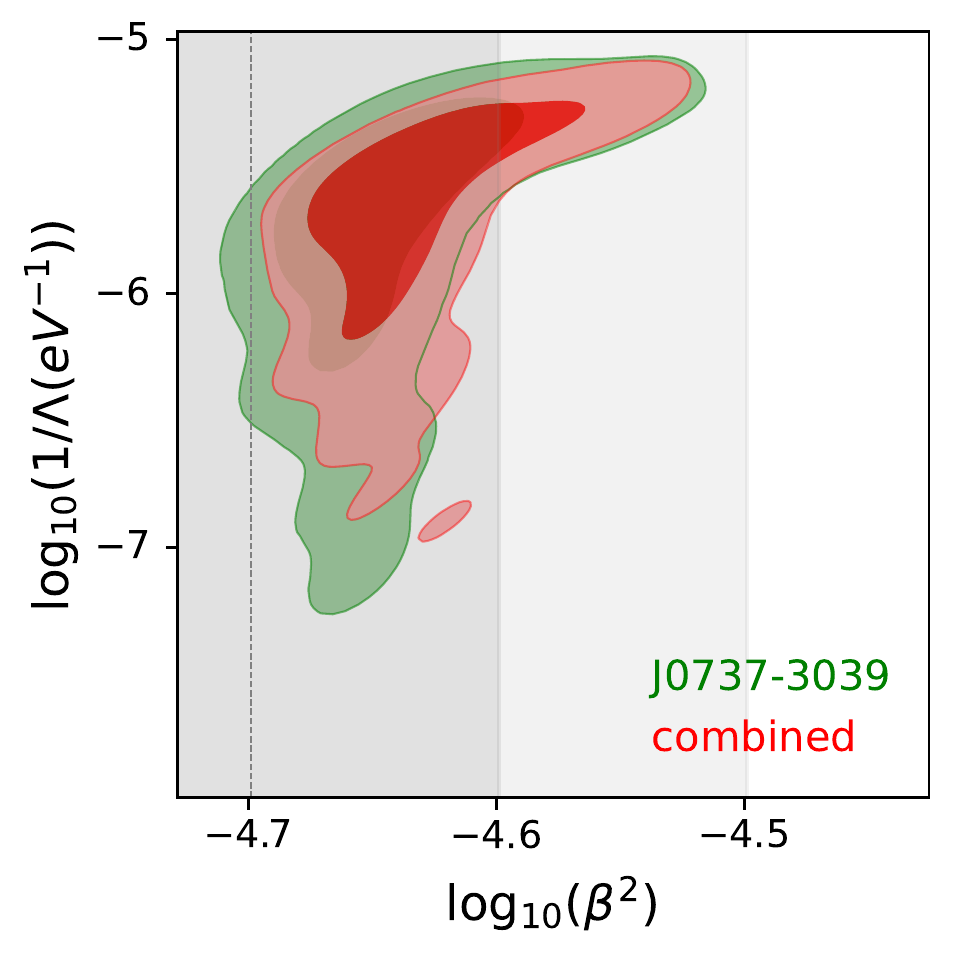}};
    \end{tikzpicture}
    \caption{\it{Conformal and disformal coupling constraints from different events in  comparison to the Cassini constraint (in gray). The subplot shows the double pulsar constraint (green) and the combined constraint (red) which is comparable to the Cassini upper bound.}}
\label{fig:conerplotsDisformalContraints}
\end{figure}

In GR, the PKP depend only on the two unknown masses of the pulsar $m_p$ and its companion $m_c$. When a light scalar is present, the PK parameters contain the conformal and the disformal interactions strengths. The full Time of Arrival (TOA) data allows  one to constrain these parameters efficiently. We use different events, i.e. PSR B1913+16~\cite{Weisberg:2016jye}, PSR J1906+0746~\cite{vanLeeuwen:2014sca}, PSR J1141-6545~\cite{VenkatramanKrishnan:2020pbi}, PSR B2127+11C~\cite{Ferdman:2014rna}, PSR B1534+12~  \cite{Fonseca:2014qla}, PSR J0737-3039A/B~\cite{Kramer:2021jcw}. We have chosen these events as they have the most  precise measurements of the PKP and the highest orbital frequencies, i.e. they provide the most stringent constraint on  the disformal coupling scale.

\section{Time delays}
\label{sec:timedelays}
In this section, we will describe the time delay between the signals emitted for instance by binary pulsars and the reception by detectors on Earth. For this, let us recall that the best available approximation  for an inertial reference frame is that of the Solar System barycentre. The required transformation between the observer's time  $\tau$, and the emission time $t$ from  a distant object such as a  pulsar is
\begin{equation}
t - \tau = -\frac{D}{f^2} + \Delta_{R\odot} + \Delta_{E\odot}
  -\Delta_{S\odot} - \Delta_R - \Delta_E - \Delta_S\,,
\label{equation:orbit}
\end{equation}
where $D/f^2$ accounts for the dispersive delay in seconds of the observed pulse relative to infinite frequency; the parameter $D$ is derived from the pulsar's dispersion measurements.  There are three time delays we take into account in this formula. First there is the Roemer delay, $\Delta_{R\odot}$, which  takes corresponds to the travel time across the Solar System based on the relative positions of the pulsar and the telescope. Then there is the Einstein delay, $\Delta_{E\odot}$, which accounts for the time dilation and  the gravitational redshifts due to the Sun and other masses in the Solar System. Finally, the Shapiro delay $\Delta_{S\odot}$  expresses the excess delay to the pulsar signal as it travels through the gravitational well of the Sun. The terms $\Delta_R$, $\Delta_E$, $\Delta_S$ account for similar delays within the pulsar binary system. The delay terms are:
\begin{eqnarray}
&&
\Delta_R = x \sin\omega (\cos \eta -e_r) + x
(1-e^2_\theta)^{1/2}\cos\omega \sin \eta,  
\nonumber \\
&&
\Delta_E = \gamma_E \sin \eta, \nonumber \\ &&
\exp \left[-\frac{\Delta_S}{2r}\right] = 1-e\cos \eta \nonumber \\ 
&&
- s \left[  \sin\omega (\cos \eta - e) + (1-e^2)^{1/2} \cos\omega \sin \eta \right] ,   
\nonumber \\
\end{eqnarray}
where $\gamma_E$ represents the combined time dilation and gravitational redshift due to the pulsar's orbit, and $r$ and $s$ are, respectively, the range and shape of the Shapiro delay.

In order to derive these Post-Keplerian-Parameters (PKP), one must  track the dynamics of photons compared to that of a light particle.  The trajectories of photons follow the null trajectories of the Jordan metric $g^J_{\mu\nu} dx^\mu dx^\nu\equiv 0$, where the Jordan metric is given by
\begin{eqnarray}
&& g^J_{00}= -(1-\frac{2G {m} (1+2 \beta^2)}{r})\nonumber \\
&& g^J_{ij}= (1+ \frac{2G {m} (1-2 \beta^2)}{r})\delta_{ij} + \frac{\beta^2 G}{\pi} \frac{ {m}^2 }{\Lambda^4 r^4}n_i n_j.
\nonumber \\
\label{effe2}
\end{eqnarray}
which involves the parallel velocity only. The study of the time delay of radio waves compared to its counterpart in GR is conveniently performed by introducing the  metric potential due to the presence of a point-like object
\begin{equation}
\Phi (r)= \Phi_N (r) + \beta \frac{\phi^{(0)}(r)}{m_{\rm Pl}} = -\frac{G_{\rm eff} m}{r}, 
\end{equation}
where the effective Newton constant is here $G_{\rm eff}= (1+2\beta^2) G$. This is the potential which appears in $g_{00}^J$. Since this potential is modified only by  the conformal coupling, the Einstein and the Shapiro delays are modified by replacing  $G \to G_{\text{eff}}$. Therefore the corresponding PKP ($\gamma, s, r$) become
\begin{eqnarray}
&&
\gamma_E =  e \, m_c \, \sqrt[3]{\frac{T_{\odot}^2}{ n m}  } \left( 1+ 2 \beta^2\right)^{2/3} \left(1 + \frac{m_c}{m}\right) ,
\nonumber \\
&&
s = \frac{x_p}{m_c} \sqrt[3]{\frac{n}{ 1+ 2 \beta^2} \frac{m^2}{T_{\odot}}},
\nonumber \\
&&
r = \left(1+ 2 \beta^2 \right) T_{\odot} m_c.
\nonumber \\
\end{eqnarray}
These parameters will constrain the conformal coupling especially as $\beta$ appears both in the numerator and the denominator of these expressions, making   the combined constraint  stronger. 

\section{Binary Pulsars Constraints}
\label{sec:constraints}
\subsection{Dataset}

\begin{figure}[t!]
\centering
\includegraphics[width=0.47\textwidth]{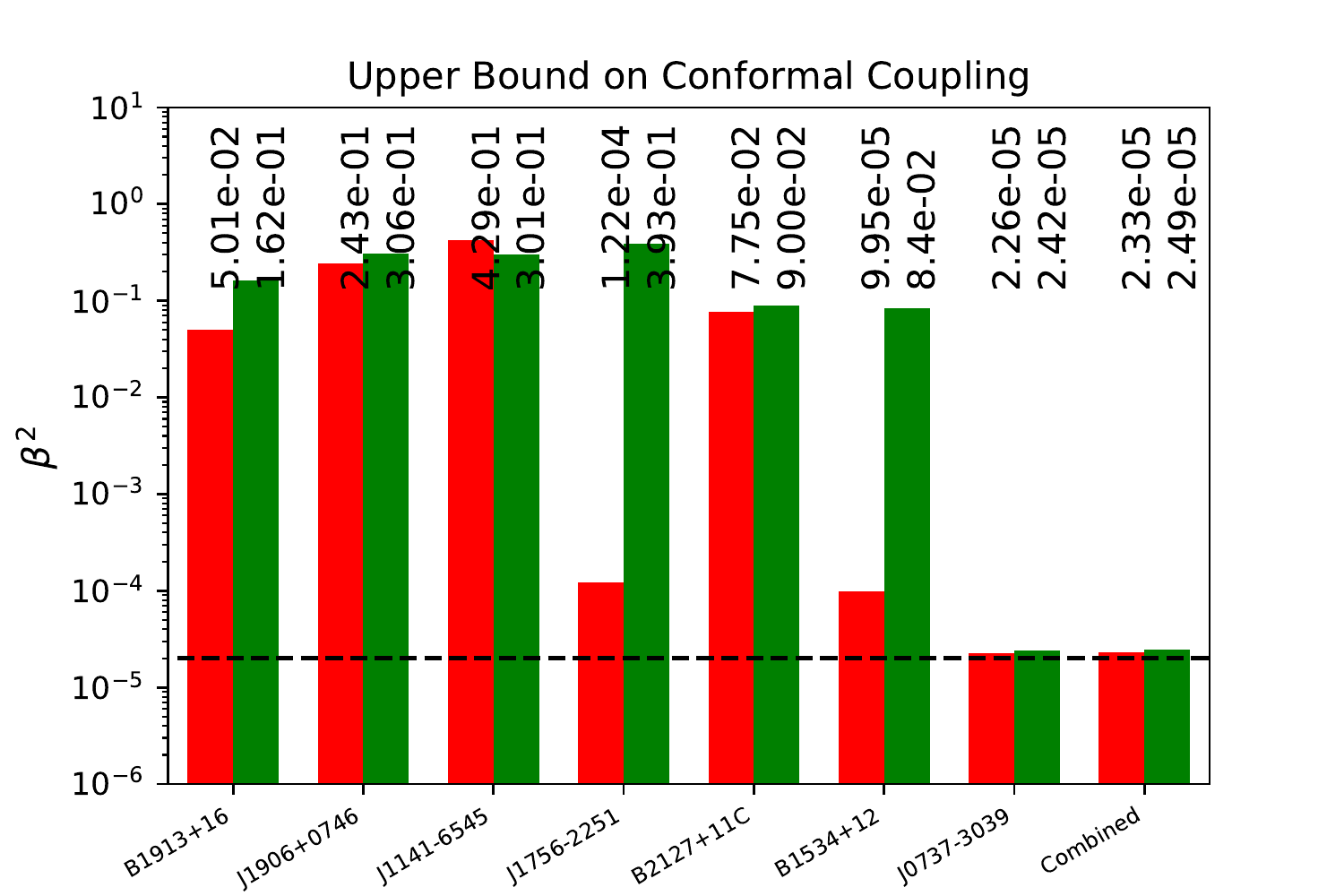}
\\
\includegraphics[width=0.47\textwidth]{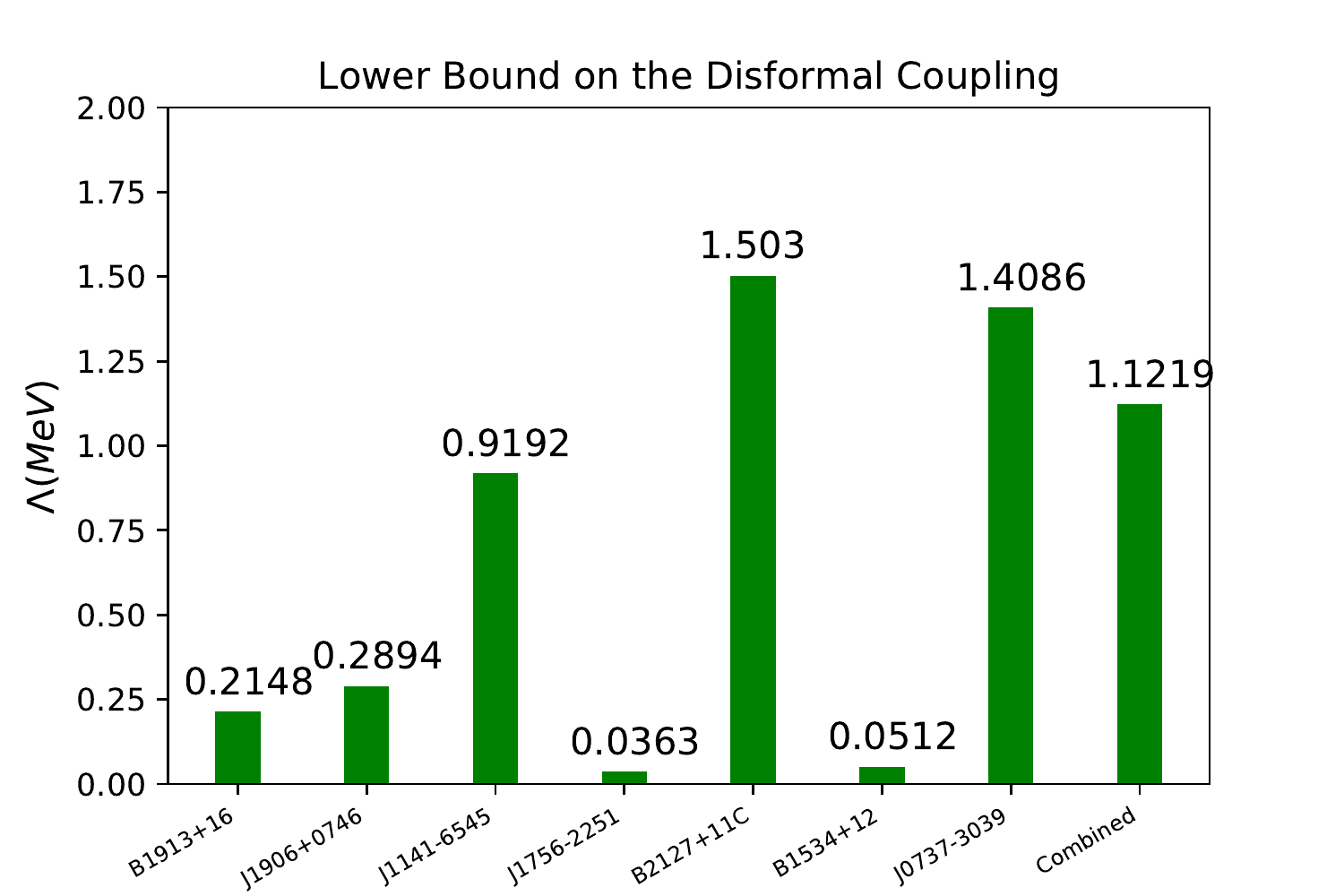}
\caption{\it{\textbf{Upper}: Upper bound on $\beta^2$ for a model with a conformal interaction (red) and for a model with a conformal and a disformal interaction (green). Cassini bound for $\beta^2$ is presented for a comparison with a dashed line. \textbf{Lower}: Lower bound on $\Lambda$ for a model with conformal and disformal interactions (green). }}
\label{fig:conformalConstraint}
\end{figure}

\begin{figure*}
    \centering
\includegraphics[width=0.48\textwidth]{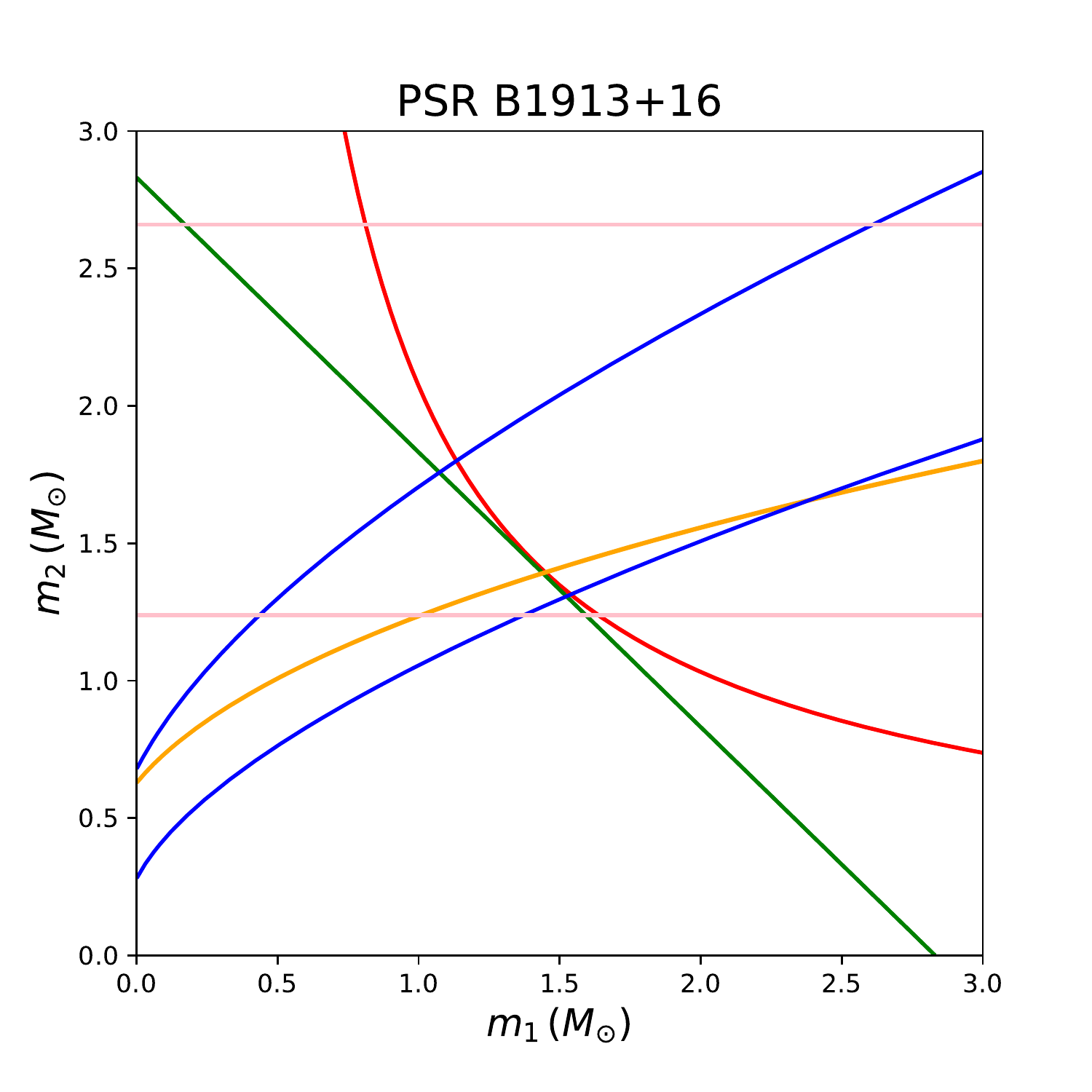}
\includegraphics[width=0.48\textwidth]{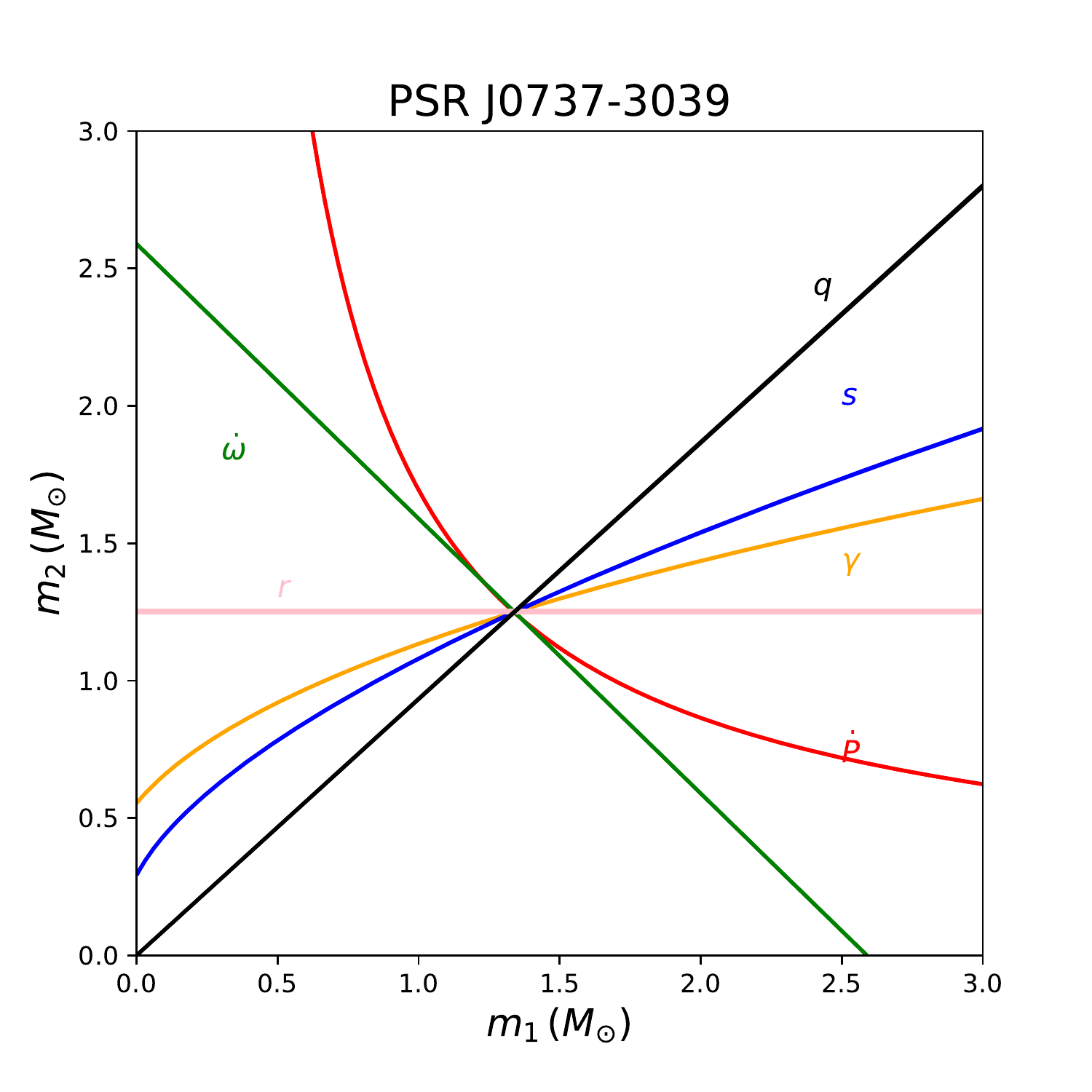}
\caption{\it{Mass-mass diagram for the pulsar events that gives the best constraints for the conformal and the disformal interactions: PSR B1913+16 and PSR J0737-3039. The contour describes the Post Keplerian Parameters and the width of each curve indicates the measurement uncertainty of the corresponding parameter.}}
    \label{fig:massdiagram}
\end{figure*}

There are two particularly relevant {{sets of observations}} to our study. First we will use the Hulse-Taylor Pulsar \cite{1981SciAm.245d..74W}, PSR B1913+16, as updated by a relativistic analysis of 9257 measurements  of times of arrival  acquired over the last 35 years  ~\cite{Weisberg:2016jye}. The updated analysis finds that the ratio of the observed orbital period decrease caused by gravitational wave damping to the general relativistic prediction is $0.9983 \pm 0.0016$ with very high precision. Then there is 
PSR J0737-3039A/B, which is the only known double pulsar with associated very high precision measurements \cite{Kramer:2006nb,Noutsos:2020uip,Piran:2004nn,Kramer:2021jcw}. The system has been studied continuously using a number of radio telescopes, with improved data acquisition systems and better sensitivity, resulting in much improved timing precision over time. The latest measurement of PSR J0737-3039A/B are published in ~\cite{Kramer:2021jcw} and include higher orders in the Post Newtonian expansion to guarantee a high precision on the determination of the PKP's. In our analysis, the contribution of the conformal and  the disformal couplings to the PKP will be  treated as perturbations compared to  the GR prediction.

The PKP's now contain  four unknown quantities $m_p, m_c, \beta, \Lambda$ which should be extracted from the observables $P_b, e, x_p, r, s, \dot{P}_b$. We use an affine-invariant Markov Chain Monte Carlo sampler \cite{Foreman-Mackey:2012any} for the minimisation of our likelihoods via the implementation of the open-source package $\text{Polychord}$ \cite{Handley:2015fda}. {{The likelihood reads}}
\begin{equation}
-2\ln \, \mathcal{L} \left(m_p,m_c,\beta,\Lambda \right) = \sum_{i = 1}^{N_{PSR}} \left(\frac{\xi(m_p,m_c,\beta,\Lambda) - \xi_{ob}}{\delta \xi_{ob}}\right)^2
\end{equation}
{{where $\xi$ is the corresponding PKP from $\xi \in [\dot{\omega}, \dot{P}, \gamma, r, s, q]$ with the error $\delta\xi$. $q$ is the ratio of the masses $q = m_p/m_c$.}} The prior we consider for the PKP's are Gaussian priors as reported in the original papers. { {For the masses we put a uniform prior of $[0,3] M_{\odot}$.}} For the conformal interaction we set a uniform prior of $\beta \in [0,1]$ and for the disformal coupling we set a uniform prior on $\Lambda^{-1} \in [0,n_b]$, where $n_b$ is the corresponding period of the system. Since the conformal interaction could be present without the disformal interaction, we test two different cases: only the conformal interaction and the the conformal with the disformal interaction.

\subsection{Results}
Fig.~\ref{fig:conerplotsConformalContraints} shows the posterior distribution for the conformal coupling for a model with the conformal interaction only. {The upper limit on $\beta^2$ are presented in the upper part of fig.~\ref{fig:conformalConstraint} in red. The double pulsar PSR J0737-3039 gives the strongest upper bound on $\beta^2$ ($<2.26 \cdot 10^{-5}$) which is similar to the Cassini bound. Since this constraint is very strong the combined constraint with the other pulsar events is also very similar.}

{Fig~\ref{fig:conerplotsDisformalContraints} shows the posterior distribution for the conformal and the disformal couplings for a model where both interactions are present. The upper limit on $\beta^2$ are presented in the upper part of fig.~\ref{fig:conformalConstraint} in red. The PSR 1913+16 timing gives a very strong bound on $\beta^2 \sim 10^{-2}$. The disformal lower limit is $0.21$ MeV  for that event. The double pulsar PSR J0838-3039 A/B gives a bound of $<2.26 \cdot 10^{-5}$ which is similar to the combined constraints  $<2.33 \cdot 10^{-5}$ due to the small errors for PSR J0838-3039 A/B. Correspondingly the lower bounds on $\Lambda $ are of the order $1.4 \, \rm{MeV}$ which is close to the GW 170817 constraint \cite{Sakstein:2017xjx}.}

For the completeness of our discussion, fig.~\ref{fig:massdiagram} introduces the mass-mass diagram for PSR B1913+16 and PSR J0737-3039 that give the best constraint on the light scalar interactions. It is possible to see that all of the parameters are intersecting at  the same point for both cases, and give a unique  mass for the pulsar and its companion. PSR B1913+16 includes larger errors for the $r$ curves and for the $s$ curves, but the double pulsar PSR J0737-3039 gives much smaller errors for the whole case. Since these observations are in good agreement with GR, the constraints on the light scalar interactions are the strongest constraints obtained from Pulsar Timing measurements. Our novel result is the tight constraint on the conformal and disformal interactions which are comparable to the Cassini constraint and to the GW-170817 constraint obtained from the Shapiro delay in the solar system and the speed of gravitational waves.

\section{Conclusions and future prospects}
\label{sec:Dis}

In this paper, we have described the effects of a massless scalar field coupled to matter on the motion of a two-body system. We have used the mean anomaly paramerisation of the two-body motion in the presence of  the conformal and disformal interactions to derive  exact analytical solutions for the trajectories of the two objects at the leading order in the parameters characterising both interactions, i.e. the conformal coupling strength $\beta$ and the coupling scale $\Lambda$ of the disformal interaction. We also derive the analytical corrections to the Keplerian $3^{rd}$ law and the precession of orbits. The formalism used to study the two-body motion in the presence of scalar interactions is similar to the one used at 3PN in GR.

The solutions to the binary motion  given  here in the presence of scalar interactions can be used to create  search templates for the detection of gravitational waves or for the improvement of the accuracy of the timing formula used for radio observations of relativistic binary pulsars. The steady improvement of sensitivity in observational astrophysics will most likely make these corrections more and more  relevant.   These correction terms to the two-body motion could be used in the analyses of future experimental data and could be seen as systematic deviations from GR. As such, they  should be taken into account in future high precision tests of  general relativity or  in the comparison of general relativity to alternative theories.

\begin{figure}[t!]
    \centering
    \includegraphics[width=0.48\textwidth]{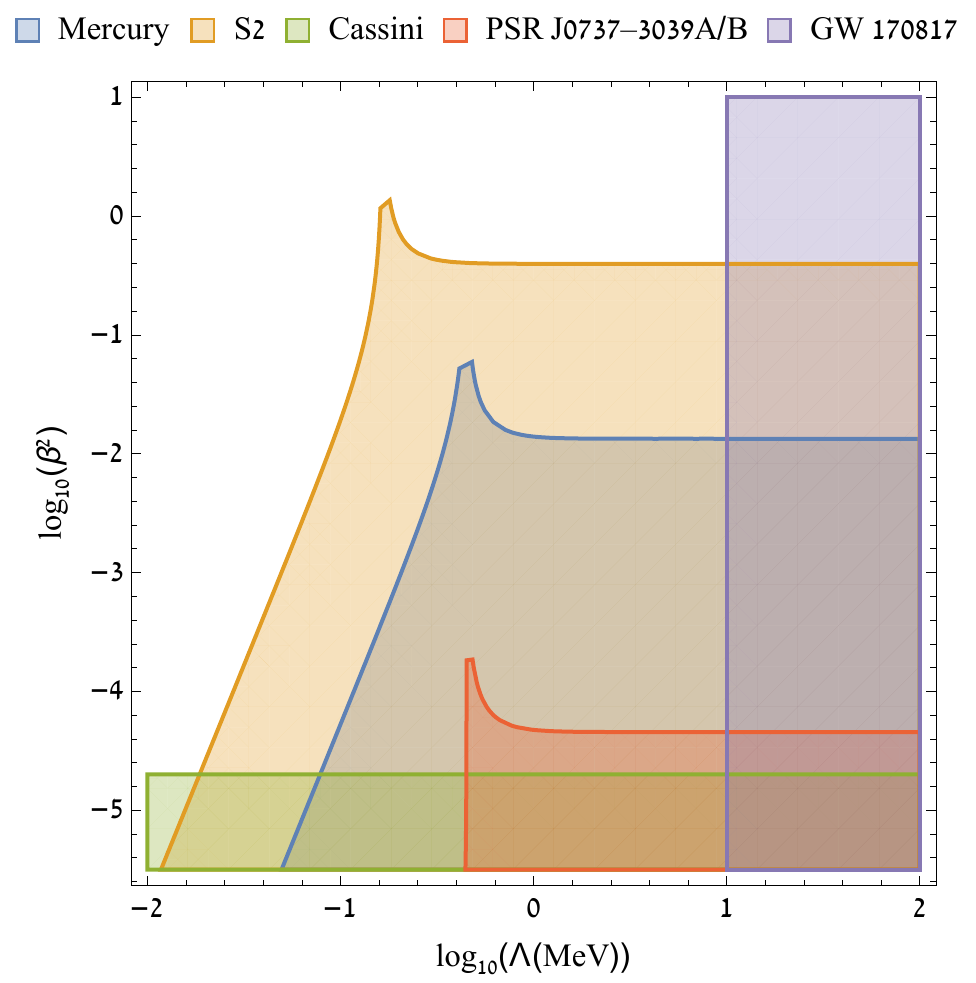}
\caption{\it{Qualitative comparison between different constraints on the conformal and the disformal couplings from different data sets. The strongest bounds are from Cassini together with the GW 170817 bound. However the analysis of PSR J0737-3039 A/B gives a comparable bound on the couplings only from one pulsar event. }}
    \label{fig:comparisonotherconstraints}
\end{figure}

Fig.~\ref{fig:comparisonotherconstraints} compares how different gravitational constraints impose different bounds on $\beta$ and $\Lambda$. The Cassini bound on $\beta^2$ gives $\lesssim 10^{-5}$ \cite{Bertotti:2003rm}. \cite{Brax:2018bow} gives a bound of $\Lambda >  10^{-4} \, {\rm MeV}$ for Mercury. \cite{Benisty:2021cmq} gives a bound of the suppression scale of the disformal interaction $\Lambda >  0.08 \ {\rm MeV}$. \cite{Sakstein:2017xjx} finds $\Lambda > 10 \ {\rm MeV}$ from the equivalence of the speed of gravity and the speed of light from the Neutron Star Merger GW-170817 \cite{LIGOScientific:2017vwq}. This is superseded by the  constraints coming from horizontal branch stars, which give $\Lambda > 100\ {\rm  MeV}$ \cite{Brax:2014vva}. Other limits on $\Lambda $ are discussed in \cite{Sakstein:2014isa}. The limit we obtain from the time drift of the revolution period of binary pulsars  are of the  order  $\lesssim 10^{-5}$ for  $\beta^2$ and $\Lambda \sim 1 {\rm MeV}$. The constraint from pulsar timing events is stronger than some of these constraints and comparable to Cassini bound for instance.  This should provide the possibility that future measurements will yield much stronger constraints from additional and more accurate measurements.

{ {Finally let us mention that the bounds on the disformal coupling scale $\Lambda$ obtained in the gravitational context such as pulsar observations are weaker and superseded by particle physics bounds, see \cite{Brax:2015hma}, where $\Lambda \gtrsim 650$ GeV can be obtained. Now of course, nothing guarantees that the models used to analyse pulsar data and gravitational phenomena are still valid at collider energy scales. In fact, it is quite likely that the low energy models used for gravitational phenomena need to be modified at higher energies and that there is no direct relationship between the gravitational and the particle physics bounds. In the absence of understanding of the UV completion of the low energy models leading to the screening of scalar effects in gravitational experiments and observations, we will refrain from stating strong conclusions from particle physics bound.}}

%

Let us comment on \cite{Liu:2014uka} which discusses  future constraints coming from  pulsar-black hole binary systems. Black holes go beyond our treatment as they have a horizon where relativistic effects cannot be neglected close to their horizon. On the other hand, when viewed from far enough away, where the black hole metric reduces to a nearly Minkowski metric, a Newtonian treatment can be considered.  It is then possible to see black holes as point particles  which interact with the scalar field despite no hair theorems. This happens in  the time dependent setting where  the  black holes acquire a scalar charge, see \cite{Wong:2022wni} and \cite{Sakstein:2014aca,Brax:2021wcv} for instance, i.e. in the models that we consider the usual no-hair theorems apply and no scalar charge is generated unless one of the assumptions of the theorems such as the absence of time dependence are violated. Extending our results to the black hole case would imply considering cosmologically induced scalar charge for instance. In this case, the results presented in this paper could help analysing black hole phenomena as long as the distance between stars and black holes is large and the point particle approximation remains valid. In this case, a scalar coupling can be assigned to black holes whose origin would be cosmological and whose magnitude can be left as a phenomenological parameter. Besides the higher precision and smaller errors for such events, some of the PKP's have stronger dependence on the black hole mass $m_{\bullet}$ since $m_p \ll m_{\bullet}$. Not only  $\dot{P}$ but   all  the PK parameters depend strongly on the black hole mass. Together with $\dot{P}$,  one could expect to  constrain both masses with a larger precision. However the constraint on the disformal coupling will not necessarily be stronger, since the dimensionless coupling $\epsilon_\Lambda$ depends on the orbital frequency. Indeed the total mass will be larger but   the semi-major axis is also crucial to obtain larger frequencies and could compensate the previous effect on the mass. Finally, the physics of conformal and disformal interactions will certainly benefit from improved measurements of the spin of binaries. Already known measurements from local experiments such as Gravity probe B~\cite{Brax:2020vgg} and future tests of the Lense-Thirring effects will certainly lead to interesting bounds. This is left for future work.

\acknowledgments
We thank Gilles Theureau, Salvatore Capozziello, Wyn Evans and Denitsa Staicova for useful discussion and comments. D.B gratefully acknowledge the supports of the Blavatnik and the Rothschild fellowships. D.B. acknowledges a Postdoctoral Research Associateship at the Queens' College, University of Cambridge. D.B. have received partial support from European COST actions CA15117 and CA18108 and the research grants KP-06-N58/5.

\bibliographystyle{apsrev4-1}
\bibliography{ref}

\begin{thebibliography}{135}%
\makeatletter
\providecommand \@ifxundefined [1]{%
 \@ifx{#1\undefined}
}%
\providecommand \@ifnum [1]{%
 \ifnum #1\expandafter \@firstoftwo
 \else \expandafter \@secondoftwo
 \fi
}%
\providecommand \@ifx [1]{%
 \ifx #1\expandafter \@firstoftwo
 \else \expandafter \@secondoftwo
 \fi
}%
\providecommand \natexlab [1]{#1}%
\providecommand \enquote  [1]{``#1''}%
\providecommand \bibnamefont  [1]{#1}%
\providecommand \bibfnamefont [1]{#1}%
\providecommand \citenamefont [1]{#1}%
\providecommand \href@noop [0]{\@secondoftwo}%
\providecommand \href [0]{\begingroup \@sanitize@url \@href}%
\providecommand \@href[1]{\@@startlink{#1}\@@href}%
\providecommand \@@href[1]{\endgroup#1\@@endlink}%
\providecommand \@sanitize@url [0]{\catcode `\\12\catcode `\$12\catcode
  `\&12\catcode `\#12\catcode `\^12\catcode `\_12\catcode `\%12\relax}%
\providecommand \@@startlink[1]{}%
\providecommand \@@endlink[0]{}%
\providecommand \url  [0]{\begingroup\@sanitize@url \@url }%
\providecommand \@url [1]{\endgroup\@href {#1}{\urlprefix }}%
\providecommand \urlprefix  [0]{URL }%
\providecommand \Eprint [0]{\href }%
\providecommand \doibase [0]{http://dx.doi.org/}%
\providecommand \selectlanguage [0]{\@gobble}%
\providecommand \bibinfo  [0]{\@secondoftwo}%
\providecommand \bibfield  [0]{\@secondoftwo}%
\providecommand \translation [1]{[#1]}%
\providecommand \BibitemOpen [0]{}%
\providecommand \bibitemStop [0]{}%
\providecommand \bibitemNoStop [0]{.\EOS\space}%
\providecommand \EOS [0]{\spacefactor3000\relax}%
\providecommand \BibitemShut  [1]{\csname bibitem#1\endcsname}%
\let\auto@bib@innerbib\@empty
\bibitem [{\citenamefont {Perlmutter}\ \emph {et~al.}(1999)\citenamefont
  {Perlmutter} \emph {et~al.}}]{SupernovaCosmologyProject:1998vns}%
  \BibitemOpen
  \bibfield  {author} {\bibinfo {author} {\bibfnamefont {S.}~\bibnamefont
  {Perlmutter}} \emph {et~al.} (\bibinfo {collaboration} {Supernova Cosmology
  Project}),\ }\href {\doibase 10.1086/307221} {\bibfield  {journal} {\bibinfo
  {journal} {Astrophys. J.}\ }\textbf {\bibinfo {volume} {517}},\ \bibinfo
  {pages} {565} (\bibinfo {year} {1999})},\ \Eprint
  {http://arxiv.org/abs/astro-ph/9812133} {arXiv:astro-ph/9812133} \BibitemShut
  {NoStop}%
\bibitem [{\citenamefont {Weinberg}(1989)}]{Weinberg:1988cp}%
  \BibitemOpen
  \bibfield  {author} {\bibinfo {author} {\bibfnamefont {S.}~\bibnamefont
  {Weinberg}},\ }\href {\doibase 10.1103/RevModPhys.61.1} {\bibfield  {journal}
  {\bibinfo  {journal} {Rev. Mod. Phys.}\ }\textbf {\bibinfo {volume} {61}},\
  \bibinfo {pages} {1} (\bibinfo {year} {1989})}\BibitemShut {NoStop}%
\bibitem [{\citenamefont {Lombriser}(2019)}]{Lombriser:2019jia}%
  \BibitemOpen
  \bibfield  {author} {\bibinfo {author} {\bibfnamefont {L.}~\bibnamefont
  {Lombriser}},\ }\href {\doibase 10.1016/j.physletb.2019.134804} {\bibfield
  {journal} {\bibinfo  {journal} {Phys. Lett. B}\ }\textbf {\bibinfo {volume}
  {797}},\ \bibinfo {pages} {134804} (\bibinfo {year} {2019})},\ \Eprint
  {http://arxiv.org/abs/1901.08588} {arXiv:1901.08588 [gr-qc]} \BibitemShut
  {NoStop}%
\bibitem [{\citenamefont {Copeland}\ \emph {et~al.}(2006)\citenamefont
  {Copeland}, \citenamefont {Sami},\ and\ \citenamefont
  {Tsujikawa}}]{Copeland:2006wr}%
  \BibitemOpen
  \bibfield  {author} {\bibinfo {author} {\bibfnamefont {E.~J.}\ \bibnamefont
  {Copeland}}, \bibinfo {author} {\bibfnamefont {M.}~\bibnamefont {Sami}}, \
  and\ \bibinfo {author} {\bibfnamefont {S.}~\bibnamefont {Tsujikawa}},\ }\href
  {\doibase 10.1142/S021827180600942X} {\bibfield  {journal} {\bibinfo
  {journal} {Int. J. Mod. Phys. D}\ }\textbf {\bibinfo {volume} {15}},\
  \bibinfo {pages} {1753} (\bibinfo {year} {2006})},\ \Eprint
  {http://arxiv.org/abs/hep-th/0603057} {arXiv:hep-th/0603057} \BibitemShut
  {NoStop}%
\bibitem [{\citenamefont {Frieman}\ \emph {et~al.}(2008)\citenamefont
  {Frieman}, \citenamefont {Turner},\ and\ \citenamefont
  {Huterer}}]{Frieman:2008sn}%
  \BibitemOpen
  \bibfield  {author} {\bibinfo {author} {\bibfnamefont {J.}~\bibnamefont
  {Frieman}}, \bibinfo {author} {\bibfnamefont {M.}~\bibnamefont {Turner}}, \
  and\ \bibinfo {author} {\bibfnamefont {D.}~\bibnamefont {Huterer}},\ }\href
  {\doibase 10.1146/annurev.astro.46.060407.145243} {\bibfield  {journal}
  {\bibinfo  {journal} {Ann. Rev. Astron. Astrophys.}\ }\textbf {\bibinfo
  {volume} {46}},\ \bibinfo {pages} {385} (\bibinfo {year} {2008})},\ \Eprint
  {http://arxiv.org/abs/0803.0982} {arXiv:0803.0982 [astro-ph]} \BibitemShut
  {NoStop}%
\bibitem [{\citenamefont {Riess}\ \emph {et~al.}(2019)\citenamefont {Riess},
  \citenamefont {Casertano}, \citenamefont {Yuan}, \citenamefont {Macri},\ and\
  \citenamefont {Scolnic}}]{Riess:2019cxk}%
  \BibitemOpen
  \bibfield  {author} {\bibinfo {author} {\bibfnamefont {A.~G.}\ \bibnamefont
  {Riess}}, \bibinfo {author} {\bibfnamefont {S.}~\bibnamefont {Casertano}},
  \bibinfo {author} {\bibfnamefont {W.}~\bibnamefont {Yuan}}, \bibinfo {author}
  {\bibfnamefont {L.~M.}\ \bibnamefont {Macri}}, \ and\ \bibinfo {author}
  {\bibfnamefont {D.}~\bibnamefont {Scolnic}},\ }\href {\doibase
  10.3847/1538-4357/ab1422} {\bibfield  {journal} {\bibinfo  {journal}
  {Astrophys. J.}\ }\textbf {\bibinfo {volume} {876}},\ \bibinfo {pages} {85}
  (\bibinfo {year} {2019})},\ \Eprint {http://arxiv.org/abs/1903.07603}
  {arXiv:1903.07603 [astro-ph.CO]} \BibitemShut {NoStop}%
\bibitem [{\citenamefont {Ooguri}\ \emph {et~al.}(2019)\citenamefont {Ooguri},
  \citenamefont {Palti}, \citenamefont {Shiu},\ and\ \citenamefont
  {Vafa}}]{Ooguri:2018wrx}%
  \BibitemOpen
  \bibfield  {author} {\bibinfo {author} {\bibfnamefont {H.}~\bibnamefont
  {Ooguri}}, \bibinfo {author} {\bibfnamefont {E.}~\bibnamefont {Palti}},
  \bibinfo {author} {\bibfnamefont {G.}~\bibnamefont {Shiu}}, \ and\ \bibinfo
  {author} {\bibfnamefont {C.}~\bibnamefont {Vafa}},\ }\href {\doibase
  10.1016/j.physletb.2018.11.018} {\bibfield  {journal} {\bibinfo  {journal}
  {Phys. Lett. B}\ }\textbf {\bibinfo {volume} {788}},\ \bibinfo {pages} {180}
  (\bibinfo {year} {2019})},\ \Eprint {http://arxiv.org/abs/1810.05506}
  {arXiv:1810.05506 [hep-th]} \BibitemShut {NoStop}%
\bibitem [{\citenamefont {Garg}\ and\ \citenamefont
  {Krishnan}(2019)}]{Garg:2018reu}%
  \BibitemOpen
  \bibfield  {author} {\bibinfo {author} {\bibfnamefont {S.~K.}\ \bibnamefont
  {Garg}}\ and\ \bibinfo {author} {\bibfnamefont {C.}~\bibnamefont
  {Krishnan}},\ }\href {\doibase 10.1007/JHEP11(2019)075} {\bibfield  {journal}
  {\bibinfo  {journal} {JHEP}\ }\textbf {\bibinfo {volume} {11}},\ \bibinfo
  {pages} {075} (\bibinfo {year} {2019})},\ \Eprint
  {http://arxiv.org/abs/1807.05193} {arXiv:1807.05193 [hep-th]} \BibitemShut
  {NoStop}%
\bibitem [{\citenamefont {Starobinsky}(1979)}]{Starobinsky:1979ty}%
  \BibitemOpen
  \bibfield  {author} {\bibinfo {author} {\bibfnamefont {A.~A.}\ \bibnamefont
  {Starobinsky}},\ }\href@noop {} {\bibfield  {journal} {\bibinfo  {journal}
  {JETP Lett.}\ }\textbf {\bibinfo {volume} {30}},\ \bibinfo {pages} {682}
  (\bibinfo {year} {1979})}\BibitemShut {NoStop}%
\bibitem [{\citenamefont {Starobinsky}(1980)}]{Starobinsky:1980te}%
  \BibitemOpen
  \bibfield  {author} {\bibinfo {author} {\bibfnamefont {A.~A.}\ \bibnamefont
  {Starobinsky}},\ }\href {\doibase 10.1016/0370-2693(80)90670-X} {\bibfield
  {journal} {\bibinfo  {journal} {Phys. Lett. B}\ }\textbf {\bibinfo {volume}
  {91}},\ \bibinfo {pages} {99} (\bibinfo {year} {1980})}\BibitemShut {NoStop}%
\bibitem [{\citenamefont {Guth}(1981)}]{Guth:1980zm}%
  \BibitemOpen
  \bibfield  {author} {\bibinfo {author} {\bibfnamefont {A.~H.}\ \bibnamefont
  {Guth}},\ }\href {\doibase 10.1103/PhysRevD.23.347} {\bibfield  {journal}
  {\bibinfo  {journal} {Phys. Rev. D}\ }\textbf {\bibinfo {volume} {23}},\
  \bibinfo {pages} {347} (\bibinfo {year} {1981})}\BibitemShut {NoStop}%
\bibitem [{\citenamefont {Albrecht}\ and\ \citenamefont
  {Steinhardt}(1982)}]{Albrecht:1982wi}%
  \BibitemOpen
  \bibfield  {author} {\bibinfo {author} {\bibfnamefont {A.}~\bibnamefont
  {Albrecht}}\ and\ \bibinfo {author} {\bibfnamefont {P.~J.}\ \bibnamefont
  {Steinhardt}},\ }\href {\doibase 10.1103/PhysRevLett.48.1220} {\bibfield
  {journal} {\bibinfo  {journal} {Phys. Rev. Lett.}\ }\textbf {\bibinfo
  {volume} {48}},\ \bibinfo {pages} {1220} (\bibinfo {year}
  {1982})}\BibitemShut {NoStop}%
\bibitem [{\citenamefont {Mukhanov}\ and\ \citenamefont
  {Chibisov}(1981)}]{Mukhanov:1981xt}%
  \BibitemOpen
  \bibfield  {author} {\bibinfo {author} {\bibfnamefont {V.~F.}\ \bibnamefont
  {Mukhanov}}\ and\ \bibinfo {author} {\bibfnamefont {G.~V.}\ \bibnamefont
  {Chibisov}},\ }\href@noop {} {\bibfield  {journal} {\bibinfo  {journal} {JETP
  Lett.}\ }\textbf {\bibinfo {volume} {33}},\ \bibinfo {pages} {532} (\bibinfo
  {year} {1981})}\BibitemShut {NoStop}%
\bibitem [{\citenamefont {Guth}\ and\ \citenamefont {Pi}(1982)}]{Guth:1982ec}%
  \BibitemOpen
  \bibfield  {author} {\bibinfo {author} {\bibfnamefont {A.~H.}\ \bibnamefont
  {Guth}}\ and\ \bibinfo {author} {\bibfnamefont {S.~Y.}\ \bibnamefont {Pi}},\
  }\href {\doibase 10.1103/PhysRevLett.49.1110} {\bibfield  {journal} {\bibinfo
   {journal} {Phys. Rev. Lett.}\ }\textbf {\bibinfo {volume} {49}},\ \bibinfo
  {pages} {1110} (\bibinfo {year} {1982})}\BibitemShut {NoStop}%
\bibitem [{\citenamefont {Linde}(1982)}]{Linde:1981mu}%
  \BibitemOpen
  \bibfield  {author} {\bibinfo {author} {\bibfnamefont {A.~D.}\ \bibnamefont
  {Linde}},\ }\href {\doibase 10.1016/0370-2693(82)91219-9} {\bibfield
  {journal} {\bibinfo  {journal} {Phys. Lett. B}\ }\textbf {\bibinfo {volume}
  {108}},\ \bibinfo {pages} {389} (\bibinfo {year} {1982})}\BibitemShut
  {NoStop}%
\bibitem [{\citenamefont {Barrow}\ and\ \citenamefont
  {Cotsakis}(1988)}]{Barrow:1988xh}%
  \BibitemOpen
  \bibfield  {author} {\bibinfo {author} {\bibfnamefont {J.~D.}\ \bibnamefont
  {Barrow}}\ and\ \bibinfo {author} {\bibfnamefont {S.}~\bibnamefont
  {Cotsakis}},\ }\href {\doibase 10.1016/0370-2693(88)90110-4} {\bibfield
  {journal} {\bibinfo  {journal} {Phys. Lett. B}\ }\textbf {\bibinfo {volume}
  {214}},\ \bibinfo {pages} {515} (\bibinfo {year} {1988})}\BibitemShut
  {NoStop}%
\bibitem [{\citenamefont {Barrow}(1988)}]{Barrow:1988xi}%
  \BibitemOpen
  \bibfield  {author} {\bibinfo {author} {\bibfnamefont {J.~D.}\ \bibnamefont
  {Barrow}},\ }\href {\doibase 10.1016/0550-3213(88)90040-5} {\bibfield
  {journal} {\bibinfo  {journal} {Nucl. Phys. B}\ }\textbf {\bibinfo {volume}
  {296}},\ \bibinfo {pages} {697} (\bibinfo {year} {1988})}\BibitemShut
  {NoStop}%
\bibitem [{\citenamefont {Elizalde}\ \emph {et~al.}(2008)\citenamefont
  {Elizalde}, \citenamefont {Nojiri}, \citenamefont {Odintsov}, \citenamefont
  {Saez-Gomez},\ and\ \citenamefont {Faraoni}}]{Elizalde:2008yf}%
  \BibitemOpen
  \bibfield  {author} {\bibinfo {author} {\bibfnamefont {E.}~\bibnamefont
  {Elizalde}}, \bibinfo {author} {\bibfnamefont {S.}~\bibnamefont {Nojiri}},
  \bibinfo {author} {\bibfnamefont {S.~D.}\ \bibnamefont {Odintsov}}, \bibinfo
  {author} {\bibfnamefont {D.}~\bibnamefont {Saez-Gomez}}, \ and\ \bibinfo
  {author} {\bibfnamefont {V.}~\bibnamefont {Faraoni}},\ }\href {\doibase
  10.1103/PhysRevD.77.106005} {\bibfield  {journal} {\bibinfo  {journal} {Phys.
  Rev. D}\ }\textbf {\bibinfo {volume} {77}},\ \bibinfo {pages} {106005}
  (\bibinfo {year} {2008})},\ \Eprint {http://arxiv.org/abs/0803.1311}
  {arXiv:0803.1311 [hep-th]} \BibitemShut {NoStop}%
\bibitem [{\citenamefont {Ratra}\ and\ \citenamefont
  {Peebles}(1988)}]{Ratra:1987rm}%
  \BibitemOpen
  \bibfield  {author} {\bibinfo {author} {\bibfnamefont {B.}~\bibnamefont
  {Ratra}}\ and\ \bibinfo {author} {\bibfnamefont {P.~J.~E.}\ \bibnamefont
  {Peebles}},\ }\href {\doibase 10.1103/PhysRevD.37.3406} {\bibfield  {journal}
  {\bibinfo  {journal} {Phys. Rev. D}\ }\textbf {\bibinfo {volume} {37}},\
  \bibinfo {pages} {3406} (\bibinfo {year} {1988})}\BibitemShut {NoStop}%
\bibitem [{\citenamefont {Caldwell}\ \emph {et~al.}(1998)\citenamefont
  {Caldwell}, \citenamefont {Dave},\ and\ \citenamefont
  {Steinhardt}}]{Caldwell:1997ii}%
  \BibitemOpen
  \bibfield  {author} {\bibinfo {author} {\bibfnamefont {R.~R.}\ \bibnamefont
  {Caldwell}}, \bibinfo {author} {\bibfnamefont {R.}~\bibnamefont {Dave}}, \
  and\ \bibinfo {author} {\bibfnamefont {P.~J.}\ \bibnamefont {Steinhardt}},\
  }\href {\doibase 10.1103/PhysRevLett.80.1582} {\bibfield  {journal} {\bibinfo
   {journal} {Phys. Rev. Lett.}\ }\textbf {\bibinfo {volume} {80}},\ \bibinfo
  {pages} {1582} (\bibinfo {year} {1998})},\ \Eprint
  {http://arxiv.org/abs/astro-ph/9708069} {arXiv:astro-ph/9708069} \BibitemShut
  {NoStop}%
\bibitem [{\citenamefont {Zlatev}\ \emph {et~al.}(1999)\citenamefont {Zlatev},
  \citenamefont {Wang},\ and\ \citenamefont {Steinhardt}}]{Zlatev:1998tr}%
  \BibitemOpen
  \bibfield  {author} {\bibinfo {author} {\bibfnamefont {I.}~\bibnamefont
  {Zlatev}}, \bibinfo {author} {\bibfnamefont {L.-M.}\ \bibnamefont {Wang}}, \
  and\ \bibinfo {author} {\bibfnamefont {P.~J.}\ \bibnamefont {Steinhardt}},\
  }\href {\doibase 10.1103/PhysRevLett.82.896} {\bibfield  {journal} {\bibinfo
  {journal} {Phys. Rev. Lett.}\ }\textbf {\bibinfo {volume} {82}},\ \bibinfo
  {pages} {896} (\bibinfo {year} {1999})},\ \Eprint
  {http://arxiv.org/abs/astro-ph/9807002} {arXiv:astro-ph/9807002} \BibitemShut
  {NoStop}%
\bibitem [{\citenamefont {Caldwell}(2002)}]{Caldwell:1999ew}%
  \BibitemOpen
  \bibfield  {author} {\bibinfo {author} {\bibfnamefont {R.~R.}\ \bibnamefont
  {Caldwell}},\ }\href {\doibase 10.1016/S0370-2693(02)02589-3} {\bibfield
  {journal} {\bibinfo  {journal} {Phys. Lett. B}\ }\textbf {\bibinfo {volume}
  {545}},\ \bibinfo {pages} {23} (\bibinfo {year} {2002})},\ \Eprint
  {http://arxiv.org/abs/astro-ph/9908168} {arXiv:astro-ph/9908168} \BibitemShut
  {NoStop}%
\bibitem [{\citenamefont {Chiba}\ \emph {et~al.}(2000)\citenamefont {Chiba},
  \citenamefont {Okabe},\ and\ \citenamefont {Yamaguchi}}]{Chiba:1999ka}%
  \BibitemOpen
  \bibfield  {author} {\bibinfo {author} {\bibfnamefont {T.}~\bibnamefont
  {Chiba}}, \bibinfo {author} {\bibfnamefont {T.}~\bibnamefont {Okabe}}, \ and\
  \bibinfo {author} {\bibfnamefont {M.}~\bibnamefont {Yamaguchi}},\ }\href
  {\doibase 10.1103/PhysRevD.62.023511} {\bibfield  {journal} {\bibinfo
  {journal} {Phys. Rev. D}\ }\textbf {\bibinfo {volume} {62}},\ \bibinfo
  {pages} {023511} (\bibinfo {year} {2000})},\ \Eprint
  {http://arxiv.org/abs/astro-ph/9912463} {arXiv:astro-ph/9912463} \BibitemShut
  {NoStop}%
\bibitem [{\citenamefont {Bento}\ \emph {et~al.}(2002)\citenamefont {Bento},
  \citenamefont {Bertolami},\ and\ \citenamefont {Sen}}]{Bento:2002ps}%
  \BibitemOpen
  \bibfield  {author} {\bibinfo {author} {\bibfnamefont {M.~C.}\ \bibnamefont
  {Bento}}, \bibinfo {author} {\bibfnamefont {O.}~\bibnamefont {Bertolami}}, \
  and\ \bibinfo {author} {\bibfnamefont {A.~A.}\ \bibnamefont {Sen}},\ }\href
  {\doibase 10.1103/PhysRevD.66.043507} {\bibfield  {journal} {\bibinfo
  {journal} {Phys. Rev. D}\ }\textbf {\bibinfo {volume} {66}},\ \bibinfo
  {pages} {043507} (\bibinfo {year} {2002})},\ \Eprint
  {http://arxiv.org/abs/gr-qc/0202064} {arXiv:gr-qc/0202064} \BibitemShut
  {NoStop}%
\bibitem [{\citenamefont {Tsujikawa}(2013)}]{Tsujikawa:2013fta}%
  \BibitemOpen
  \bibfield  {author} {\bibinfo {author} {\bibfnamefont {S.}~\bibnamefont
  {Tsujikawa}},\ }\href {\doibase 10.1088/0264-9381/30/21/214003} {\bibfield
  {journal} {\bibinfo  {journal} {Class. Quant. Grav.}\ }\textbf {\bibinfo
  {volume} {30}},\ \bibinfo {pages} {214003} (\bibinfo {year} {2013})},\
  \Eprint {http://arxiv.org/abs/1304.1961} {arXiv:1304.1961 [gr-qc]}
  \BibitemShut {NoStop}%
\bibitem [{\citenamefont {Peebles}\ and\ \citenamefont
  {Ratra}(1988)}]{Peebles:1987ek}%
  \BibitemOpen
  \bibfield  {author} {\bibinfo {author} {\bibfnamefont {P.~J.~E.}\
  \bibnamefont {Peebles}}\ and\ \bibinfo {author} {\bibfnamefont
  {B.}~\bibnamefont {Ratra}},\ }\href {\doibase 10.1086/185100} {\bibfield
  {journal} {\bibinfo  {journal} {Astrophys. J. Lett.}\ }\textbf {\bibinfo
  {volume} {325}},\ \bibinfo {pages} {L17} (\bibinfo {year}
  {1988})}\BibitemShut {NoStop}%
\bibitem [{\citenamefont {Barreiro}\ \emph {et~al.}(2000)\citenamefont
  {Barreiro}, \citenamefont {Copeland},\ and\ \citenamefont
  {Nunes}}]{Barreiro:1999zs}%
  \BibitemOpen
  \bibfield  {author} {\bibinfo {author} {\bibfnamefont {T.}~\bibnamefont
  {Barreiro}}, \bibinfo {author} {\bibfnamefont {E.~J.}\ \bibnamefont
  {Copeland}}, \ and\ \bibinfo {author} {\bibfnamefont {N.~J.}\ \bibnamefont
  {Nunes}},\ }\href {\doibase 10.1103/PhysRevD.61.127301} {\bibfield  {journal}
  {\bibinfo  {journal} {Phys. Rev. D}\ }\textbf {\bibinfo {volume} {61}},\
  \bibinfo {pages} {127301} (\bibinfo {year} {2000})},\ \Eprint
  {http://arxiv.org/abs/astro-ph/9910214} {arXiv:astro-ph/9910214} \BibitemShut
  {NoStop}%
\bibitem [{\citenamefont {Carroll}(1998)}]{Carroll:1998zi}%
  \BibitemOpen
  \bibfield  {author} {\bibinfo {author} {\bibfnamefont {S.~M.}\ \bibnamefont
  {Carroll}},\ }\href {\doibase 10.1103/PhysRevLett.81.3067} {\bibfield
  {journal} {\bibinfo  {journal} {Phys. Rev. Lett.}\ }\textbf {\bibinfo
  {volume} {81}},\ \bibinfo {pages} {3067} (\bibinfo {year} {1998})},\ \Eprint
  {http://arxiv.org/abs/astro-ph/9806099} {arXiv:astro-ph/9806099} \BibitemShut
  {NoStop}%
\bibitem [{\citenamefont {Chiba}(1999)}]{Chiba:1999wt}%
  \BibitemOpen
  \bibfield  {author} {\bibinfo {author} {\bibfnamefont {T.}~\bibnamefont
  {Chiba}},\ }\href {\doibase 10.1103/PhysRevD.60.083508} {\bibfield  {journal}
  {\bibinfo  {journal} {Phys. Rev. D}\ }\textbf {\bibinfo {volume} {60}},\
  \bibinfo {pages} {083508} (\bibinfo {year} {1999})},\ \Eprint
  {http://arxiv.org/abs/gr-qc/9903094} {arXiv:gr-qc/9903094} \BibitemShut
  {NoStop}%
\bibitem [{\citenamefont {Sahni}\ and\ \citenamefont
  {Wang}(2000)}]{Sahni:1999qe}%
  \BibitemOpen
  \bibfield  {author} {\bibinfo {author} {\bibfnamefont {V.}~\bibnamefont
  {Sahni}}\ and\ \bibinfo {author} {\bibfnamefont {L.-M.}\ \bibnamefont
  {Wang}},\ }\href {\doibase 10.1103/PhysRevD.62.103517} {\bibfield  {journal}
  {\bibinfo  {journal} {Phys. Rev. D}\ }\textbf {\bibinfo {volume} {62}},\
  \bibinfo {pages} {103517} (\bibinfo {year} {2000})},\ \Eprint
  {http://arxiv.org/abs/astro-ph/9910097} {arXiv:astro-ph/9910097} \BibitemShut
  {NoStop}%
\bibitem [{\citenamefont {Koivisto}(2008)}]{Koivisto:2008ak}%
  \BibitemOpen
  \bibfield  {author} {\bibinfo {author} {\bibfnamefont {T.~S.}\ \bibnamefont
  {Koivisto}},\ }\href@noop {} {\  (\bibinfo {year} {2008})},\ \Eprint
  {http://arxiv.org/abs/0811.1957} {arXiv:0811.1957 [astro-ph]} \BibitemShut
  {NoStop}%
\bibitem [{\citenamefont {Zumalacarregui}\ \emph {et~al.}(2010)\citenamefont
  {Zumalacarregui}, \citenamefont {Koivisto}, \citenamefont {Mota},\ and\
  \citenamefont {Ruiz-Lapuente}}]{Zumalacarregui:2010wj}%
  \BibitemOpen
  \bibfield  {author} {\bibinfo {author} {\bibfnamefont {M.}~\bibnamefont
  {Zumalacarregui}}, \bibinfo {author} {\bibfnamefont {T.~S.}\ \bibnamefont
  {Koivisto}}, \bibinfo {author} {\bibfnamefont {D.~F.}\ \bibnamefont {Mota}},
  \ and\ \bibinfo {author} {\bibfnamefont {P.}~\bibnamefont {Ruiz-Lapuente}},\
  }\href {\doibase 10.1088/1475-7516/2010/05/038} {\bibfield  {journal}
  {\bibinfo  {journal} {JCAP}\ }\textbf {\bibinfo {volume} {05}},\ \bibinfo
  {pages} {038} (\bibinfo {year} {2010})},\ \Eprint
  {http://arxiv.org/abs/1004.2684} {arXiv:1004.2684 [astro-ph.CO]} \BibitemShut
  {NoStop}%
\bibitem [{\citenamefont {Koivisto}\ \emph {et~al.}(2012)\citenamefont
  {Koivisto}, \citenamefont {Mota},\ and\ \citenamefont
  {Zumalacarregui}}]{Koivisto:2012za}%
  \BibitemOpen
  \bibfield  {author} {\bibinfo {author} {\bibfnamefont {T.~S.}\ \bibnamefont
  {Koivisto}}, \bibinfo {author} {\bibfnamefont {D.~F.}\ \bibnamefont {Mota}},
  \ and\ \bibinfo {author} {\bibfnamefont {M.}~\bibnamefont {Zumalacarregui}},\
  }\href {\doibase 10.1103/PhysRevLett.109.241102} {\bibfield  {journal}
  {\bibinfo  {journal} {Phys. Rev. Lett.}\ }\textbf {\bibinfo {volume} {109}},\
  \bibinfo {pages} {241102} (\bibinfo {year} {2012})},\ \Eprint
  {http://arxiv.org/abs/1205.3167} {arXiv:1205.3167 [astro-ph.CO]} \BibitemShut
  {NoStop}%
\bibitem [{\citenamefont {van~de Bruck}\ \emph {et~al.}(2013)\citenamefont
  {van~de Bruck}, \citenamefont {Morrice},\ and\ \citenamefont
  {Vu}}]{vandeBruck:2013yxa}%
  \BibitemOpen
  \bibfield  {author} {\bibinfo {author} {\bibfnamefont {C.}~\bibnamefont
  {van~de Bruck}}, \bibinfo {author} {\bibfnamefont {J.}~\bibnamefont
  {Morrice}}, \ and\ \bibinfo {author} {\bibfnamefont {S.}~\bibnamefont {Vu}},\
  }\href {\doibase 10.1103/PhysRevLett.111.161302} {\bibfield  {journal}
  {\bibinfo  {journal} {Phys. Rev. Lett.}\ }\textbf {\bibinfo {volume} {111}},\
  \bibinfo {pages} {161302} (\bibinfo {year} {2013})},\ \Eprint
  {http://arxiv.org/abs/1303.1773} {arXiv:1303.1773 [astro-ph.CO]} \BibitemShut
  {NoStop}%
\bibitem [{\citenamefont {Brax}\ \emph {et~al.}(2013)\citenamefont {Brax},
  \citenamefont {Burrage}, \citenamefont {Davis},\ and\ \citenamefont
  {Gubitosi}}]{Brax:2013nsa}%
  \BibitemOpen
  \bibfield  {author} {\bibinfo {author} {\bibfnamefont {P.}~\bibnamefont
  {Brax}}, \bibinfo {author} {\bibfnamefont {C.}~\bibnamefont {Burrage}},
  \bibinfo {author} {\bibfnamefont {A.-C.}\ \bibnamefont {Davis}}, \ and\
  \bibinfo {author} {\bibfnamefont {G.}~\bibnamefont {Gubitosi}},\ }\href
  {\doibase 10.1088/1475-7516/2013/11/001} {\bibfield  {journal} {\bibinfo
  {journal} {JCAP}\ }\textbf {\bibinfo {volume} {11}},\ \bibinfo {pages} {001}
  (\bibinfo {year} {2013})},\ \Eprint {http://arxiv.org/abs/1306.4168}
  {arXiv:1306.4168 [astro-ph.CO]} \BibitemShut {NoStop}%
\bibitem [{\citenamefont {Neveu}\ \emph {et~al.}(2014)\citenamefont {Neveu},
  \citenamefont {Ruhlmann-Kleider}, \citenamefont {Astier}, \citenamefont
  {Besan\c{c}on}, \citenamefont {Conley}, \citenamefont {Guy}, \citenamefont
  {M\"oller}, \citenamefont {Palanque-Delabrouille},\ and\ \citenamefont
  {Babichev}}]{Neveu:2014vua}%
  \BibitemOpen
  \bibfield  {author} {\bibinfo {author} {\bibfnamefont {J.}~\bibnamefont
  {Neveu}}, \bibinfo {author} {\bibfnamefont {V.}~\bibnamefont
  {Ruhlmann-Kleider}}, \bibinfo {author} {\bibfnamefont {P.}~\bibnamefont
  {Astier}}, \bibinfo {author} {\bibfnamefont {M.}~\bibnamefont
  {Besan\c{c}on}}, \bibinfo {author} {\bibfnamefont {A.}~\bibnamefont
  {Conley}}, \bibinfo {author} {\bibfnamefont {J.}~\bibnamefont {Guy}},
  \bibinfo {author} {\bibfnamefont {A.}~\bibnamefont {M\"oller}}, \bibinfo
  {author} {\bibfnamefont {N.}~\bibnamefont {Palanque-Delabrouille}}, \ and\
  \bibinfo {author} {\bibfnamefont {E.}~\bibnamefont {Babichev}},\ }\href
  {\doibase 10.1051/0004-6361/201423758} {\bibfield  {journal} {\bibinfo
  {journal} {Astron. Astrophys.}\ }\textbf {\bibinfo {volume} {569}},\ \bibinfo
  {pages} {A90} (\bibinfo {year} {2014})},\ \Eprint
  {http://arxiv.org/abs/1403.0854} {arXiv:1403.0854 [gr-qc]} \BibitemShut
  {NoStop}%
\bibitem [{\citenamefont {Sakstein}(2014)}]{Sakstein:2014isa}%
  \BibitemOpen
  \bibfield  {author} {\bibinfo {author} {\bibfnamefont {J.}~\bibnamefont
  {Sakstein}},\ }\href {\doibase 10.1088/1475-7516/2014/12/012} {\bibfield
  {journal} {\bibinfo  {journal} {JCAP}\ }\textbf {\bibinfo {volume} {12}},\
  \bibinfo {pages} {012} (\bibinfo {year} {2014})},\ \Eprint
  {http://arxiv.org/abs/1409.1734} {arXiv:1409.1734 [astro-ph.CO]} \BibitemShut
  {NoStop}%
\bibitem [{\citenamefont {Sakstein}(2015)}]{Sakstein:2014aca}%
  \BibitemOpen
  \bibfield  {author} {\bibinfo {author} {\bibfnamefont {J.}~\bibnamefont
  {Sakstein}},\ }\href {\doibase 10.1103/PhysRevD.91.024036} {\bibfield
  {journal} {\bibinfo  {journal} {Phys. Rev. D}\ }\textbf {\bibinfo {volume}
  {91}},\ \bibinfo {pages} {024036} (\bibinfo {year} {2015})},\ \Eprint
  {http://arxiv.org/abs/1409.7296} {arXiv:1409.7296 [astro-ph.CO]} \BibitemShut
  {NoStop}%
\bibitem [{\citenamefont {Desmond}\ \emph {et~al.}(2019)\citenamefont
  {Desmond}, \citenamefont {Jain},\ and\ \citenamefont
  {Sakstein}}]{Desmond:2019ygn}%
  \BibitemOpen
  \bibfield  {author} {\bibinfo {author} {\bibfnamefont {H.}~\bibnamefont
  {Desmond}}, \bibinfo {author} {\bibfnamefont {B.}~\bibnamefont {Jain}}, \
  and\ \bibinfo {author} {\bibfnamefont {J.}~\bibnamefont {Sakstein}},\ }\href
  {\doibase 10.1103/PhysRevD.100.043537} {\bibfield  {journal} {\bibinfo
  {journal} {Phys. Rev. D}\ }\textbf {\bibinfo {volume} {100}},\ \bibinfo
  {pages} {043537} (\bibinfo {year} {2019})},\ \bibinfo {note} {[Erratum:
  Phys.Rev.D 101, 069904 (2020), Erratum: Phys.Rev.D 101, 129901 (2020)]},\
  \Eprint {http://arxiv.org/abs/1907.03778} {arXiv:1907.03778 [astro-ph.CO]}
  \BibitemShut {NoStop}%
\bibitem [{\citenamefont {Brax}\ \emph {et~al.}(2008)\citenamefont {Brax},
  \citenamefont {van~de Bruck}, \citenamefont {Davis},\ and\ \citenamefont
  {Shaw}}]{Brax:2008hh}%
  \BibitemOpen
  \bibfield  {author} {\bibinfo {author} {\bibfnamefont {P.}~\bibnamefont
  {Brax}}, \bibinfo {author} {\bibfnamefont {C.}~\bibnamefont {van~de Bruck}},
  \bibinfo {author} {\bibfnamefont {A.-C.}\ \bibnamefont {Davis}}, \ and\
  \bibinfo {author} {\bibfnamefont {D.~J.}\ \bibnamefont {Shaw}},\ }\href
  {\doibase 10.1103/PhysRevD.78.104021} {\bibfield  {journal} {\bibinfo
  {journal} {Phys. Rev. D}\ }\textbf {\bibinfo {volume} {78}},\ \bibinfo
  {pages} {104021} (\bibinfo {year} {2008})},\ \Eprint
  {http://arxiv.org/abs/0806.3415} {arXiv:0806.3415 [astro-ph]} \BibitemShut
  {NoStop}%
\bibitem [{\citenamefont {Sotiriou}\ and\ \citenamefont
  {Faraoni}(2010)}]{Sotiriou:2008rp}%
  \BibitemOpen
  \bibfield  {author} {\bibinfo {author} {\bibfnamefont {T.~P.}\ \bibnamefont
  {Sotiriou}}\ and\ \bibinfo {author} {\bibfnamefont {V.}~\bibnamefont
  {Faraoni}},\ }\href {\doibase 10.1103/RevModPhys.82.451} {\bibfield
  {journal} {\bibinfo  {journal} {Rev. Mod. Phys.}\ }\textbf {\bibinfo {volume}
  {82}},\ \bibinfo {pages} {451} (\bibinfo {year} {2010})},\ \Eprint
  {http://arxiv.org/abs/0805.1726} {arXiv:0805.1726 [gr-qc]} \BibitemShut
  {NoStop}%
\bibitem [{\citenamefont {Bekenstein}(1993)}]{Bekenstein:1992pj}%
  \BibitemOpen
  \bibfield  {author} {\bibinfo {author} {\bibfnamefont {J.~D.}\ \bibnamefont
  {Bekenstein}},\ }\href {\doibase 10.1103/PhysRevD.48.3641} {\bibfield
  {journal} {\bibinfo  {journal} {Phys. Rev. D}\ }\textbf {\bibinfo {volume}
  {48}},\ \bibinfo {pages} {3641} (\bibinfo {year} {1993})},\ \Eprint
  {http://arxiv.org/abs/gr-qc/9211017} {arXiv:gr-qc/9211017} \BibitemShut
  {NoStop}%
\bibitem [{\citenamefont {van~de Bruck}\ and\ \citenamefont
  {Morrice}(2015)}]{vandeBruck:2015ida}%
  \BibitemOpen
  \bibfield  {author} {\bibinfo {author} {\bibfnamefont {C.}~\bibnamefont
  {van~de Bruck}}\ and\ \bibinfo {author} {\bibfnamefont {J.}~\bibnamefont
  {Morrice}},\ }\href {\doibase 10.1088/1475-7516/2015/04/036} {\bibfield
  {journal} {\bibinfo  {journal} {JCAP}\ }\textbf {\bibinfo {volume} {04}},\
  \bibinfo {pages} {036} (\bibinfo {year} {2015})},\ \Eprint
  {http://arxiv.org/abs/1501.03073} {arXiv:1501.03073 [gr-qc]} \BibitemShut
  {NoStop}%
\bibitem [{\citenamefont {Karwal}\ \emph {et~al.}(2022)\citenamefont {Karwal},
  \citenamefont {Raveri}, \citenamefont {Jain}, \citenamefont {Khoury},\ and\
  \citenamefont {Trodden}}]{Karwal:2021vpk}%
  \BibitemOpen
  \bibfield  {author} {\bibinfo {author} {\bibfnamefont {T.}~\bibnamefont
  {Karwal}}, \bibinfo {author} {\bibfnamefont {M.}~\bibnamefont {Raveri}},
  \bibinfo {author} {\bibfnamefont {B.}~\bibnamefont {Jain}}, \bibinfo {author}
  {\bibfnamefont {J.}~\bibnamefont {Khoury}}, \ and\ \bibinfo {author}
  {\bibfnamefont {M.}~\bibnamefont {Trodden}},\ }\href {\doibase
  10.1103/PhysRevD.105.063535} {\bibfield  {journal} {\bibinfo  {journal}
  {Phys. Rev. D}\ }\textbf {\bibinfo {volume} {105}},\ \bibinfo {pages}
  {063535} (\bibinfo {year} {2022})},\ \Eprint
  {http://arxiv.org/abs/2106.13290} {arXiv:2106.13290 [astro-ph.CO]}
  \BibitemShut {NoStop}%
\bibitem [{\citenamefont {Hui}\ \emph {et~al.}(2017)\citenamefont {Hui},
  \citenamefont {Ostriker}, \citenamefont {Tremaine},\ and\ \citenamefont
  {Witten}}]{Hui:2016ltb}%
  \BibitemOpen
  \bibfield  {author} {\bibinfo {author} {\bibfnamefont {L.}~\bibnamefont
  {Hui}}, \bibinfo {author} {\bibfnamefont {J.~P.}\ \bibnamefont {Ostriker}},
  \bibinfo {author} {\bibfnamefont {S.}~\bibnamefont {Tremaine}}, \ and\
  \bibinfo {author} {\bibfnamefont {E.}~\bibnamefont {Witten}},\ }\href
  {\doibase 10.1103/PhysRevD.95.043541} {\bibfield  {journal} {\bibinfo
  {journal} {Phys. Rev. D}\ }\textbf {\bibinfo {volume} {95}},\ \bibinfo
  {pages} {043541} (\bibinfo {year} {2017})},\ \Eprint
  {http://arxiv.org/abs/1610.08297} {arXiv:1610.08297 [astro-ph.CO]}
  \BibitemShut {NoStop}%
\bibitem [{\citenamefont {Brax}\ \emph {et~al.}(2018)\citenamefont {Brax},
  \citenamefont {Fichet},\ and\ \citenamefont {Pignol}}]{Brax:2017xho}%
  \BibitemOpen
  \bibfield  {author} {\bibinfo {author} {\bibfnamefont {P.}~\bibnamefont
  {Brax}}, \bibinfo {author} {\bibfnamefont {S.}~\bibnamefont {Fichet}}, \ and\
  \bibinfo {author} {\bibfnamefont {G.}~\bibnamefont {Pignol}},\ }\href
  {\doibase 10.1103/PhysRevD.97.115034} {\bibfield  {journal} {\bibinfo
  {journal} {Phys. Rev. D}\ }\textbf {\bibinfo {volume} {97}},\ \bibinfo
  {pages} {115034} (\bibinfo {year} {2018})},\ \Eprint
  {http://arxiv.org/abs/1710.00850} {arXiv:1710.00850 [hep-ph]} \BibitemShut
  {NoStop}%
\bibitem [{\citenamefont {Trojanowski}\ \emph {et~al.}(2020)\citenamefont
  {Trojanowski}, \citenamefont {Brax},\ and\ \citenamefont {van~de
  Bruck}}]{Trojanowski:2020xza}%
  \BibitemOpen
  \bibfield  {author} {\bibinfo {author} {\bibfnamefont {S.}~\bibnamefont
  {Trojanowski}}, \bibinfo {author} {\bibfnamefont {P.}~\bibnamefont {Brax}}, \
  and\ \bibinfo {author} {\bibfnamefont {C.}~\bibnamefont {van~de Bruck}},\
  }\href {\doibase 10.1103/PhysRevD.102.023035} {\bibfield  {journal} {\bibinfo
   {journal} {Phys. Rev. D}\ }\textbf {\bibinfo {volume} {102}},\ \bibinfo
  {pages} {023035} (\bibinfo {year} {2020})},\ \Eprint
  {http://arxiv.org/abs/2006.01149} {arXiv:2006.01149 [hep-ph]} \BibitemShut
  {NoStop}%
\bibitem [{\citenamefont {Brax}\ \emph
  {et~al.}(2021{\natexlab{a}})\citenamefont {Brax}, \citenamefont {Kaneta},
  \citenamefont {Mambrini},\ and\ \citenamefont {Pierre}}]{Brax:2020gqg}%
  \BibitemOpen
  \bibfield  {author} {\bibinfo {author} {\bibfnamefont {P.}~\bibnamefont
  {Brax}}, \bibinfo {author} {\bibfnamefont {K.}~\bibnamefont {Kaneta}},
  \bibinfo {author} {\bibfnamefont {Y.}~\bibnamefont {Mambrini}}, \ and\
  \bibinfo {author} {\bibfnamefont {M.}~\bibnamefont {Pierre}},\ }\href
  {\doibase 10.1103/PhysRevD.103.015028} {\bibfield  {journal} {\bibinfo
  {journal} {Phys. Rev. D}\ }\textbf {\bibinfo {volume} {103}},\ \bibinfo
  {pages} {015028} (\bibinfo {year} {2021}{\natexlab{a}})},\ \Eprint
  {http://arxiv.org/abs/2011.11647} {arXiv:2011.11647 [hep-ph]} \BibitemShut
  {NoStop}%
\bibitem [{\citenamefont {Maheshwari}\ \emph {et~al.}(1981)\citenamefont
  {Maheshwari}, \citenamefont {Nissimov},\ and\ \citenamefont
  {Todorov}}]{Maheshwari:1980gmr}%
  \BibitemOpen
  \bibfield  {author} {\bibinfo {author} {\bibfnamefont {A.}~\bibnamefont
  {Maheshwari}}, \bibinfo {author} {\bibfnamefont {E.}~\bibnamefont
  {Nissimov}}, \ and\ \bibinfo {author} {\bibfnamefont {I.}~\bibnamefont
  {Todorov}},\ }\href {\doibase 10.1007/BF02285306} {\bibfield  {journal}
  {\bibinfo  {journal} {Lett. Math. Phys.}\ }\textbf {\bibinfo {volume} {5}},\
  \bibinfo {pages} {359} (\bibinfo {year} {1981})},\ \Eprint
  {http://arxiv.org/abs/1611.02943} {arXiv:1611.02943 [gr-qc]} \BibitemShut
  {NoStop}%
\bibitem [{\citenamefont {Damour}\ and\ \citenamefont
  {Taylor}(1992)}]{Damour:1991rd}%
  \BibitemOpen
  \bibfield  {author} {\bibinfo {author} {\bibfnamefont {T.}~\bibnamefont
  {Damour}}\ and\ \bibinfo {author} {\bibfnamefont {J.~H.}\ \bibnamefont
  {Taylor}},\ }\href {\doibase 10.1103/PhysRevD.45.1840} {\bibfield  {journal}
  {\bibinfo  {journal} {Phys. Rev. D}\ }\textbf {\bibinfo {volume} {45}},\
  \bibinfo {pages} {1840} (\bibinfo {year} {1992})}\BibitemShut {NoStop}%
\bibitem [{\citenamefont {Buonanno}\ and\ \citenamefont
  {Damour}(1999)}]{Buonanno:1998gg}%
  \BibitemOpen
  \bibfield  {author} {\bibinfo {author} {\bibfnamefont {A.}~\bibnamefont
  {Buonanno}}\ and\ \bibinfo {author} {\bibfnamefont {T.}~\bibnamefont
  {Damour}},\ }\href {\doibase 10.1103/PhysRevD.59.084006} {\bibfield
  {journal} {\bibinfo  {journal} {Phys. Rev. D}\ }\textbf {\bibinfo {volume}
  {59}},\ \bibinfo {pages} {084006} (\bibinfo {year} {1999})},\ \Eprint
  {http://arxiv.org/abs/gr-qc/9811091} {arXiv:gr-qc/9811091} \BibitemShut
  {NoStop}%
\bibitem [{\citenamefont {Damour}\ and\ \citenamefont
  {Schaefer}(1988)}]{Damour:1988mr}%
  \BibitemOpen
  \bibfield  {author} {\bibinfo {author} {\bibfnamefont {T.}~\bibnamefont
  {Damour}}\ and\ \bibinfo {author} {\bibfnamefont {G.}~\bibnamefont
  {Schaefer}},\ }\href {\doibase 10.1007/BF02828697} {\bibfield  {journal}
  {\bibinfo  {journal} {Nuovo Cim. B}\ }\textbf {\bibinfo {volume} {101}},\
  \bibinfo {pages} {127} (\bibinfo {year} {1988})}\BibitemShut {NoStop}%
\bibitem [{\citenamefont {Damour}\ \emph {et~al.}(2000)\citenamefont {Damour},
  \citenamefont {Jaranowski},\ and\ \citenamefont {Schaefer}}]{Damour:1999cr}%
  \BibitemOpen
  \bibfield  {author} {\bibinfo {author} {\bibfnamefont {T.}~\bibnamefont
  {Damour}}, \bibinfo {author} {\bibfnamefont {P.}~\bibnamefont {Jaranowski}},
  \ and\ \bibinfo {author} {\bibfnamefont {G.}~\bibnamefont {Schaefer}},\
  }\href {\doibase 10.1103/PhysRevD.62.044024} {\bibfield  {journal} {\bibinfo
  {journal} {Phys. Rev. D}\ }\textbf {\bibinfo {volume} {62}},\ \bibinfo
  {pages} {044024} (\bibinfo {year} {2000})},\ \Eprint
  {http://arxiv.org/abs/gr-qc/9912092} {arXiv:gr-qc/9912092} \BibitemShut
  {NoStop}%
\bibitem [{\citenamefont {Memmesheimer}\ \emph {et~al.}(2004)\citenamefont
  {Memmesheimer}, \citenamefont {Gopakumar},\ and\ \citenamefont
  {Schaefer}}]{Memmesheimer:2004cv}%
  \BibitemOpen
  \bibfield  {author} {\bibinfo {author} {\bibfnamefont {R.-M.}\ \bibnamefont
  {Memmesheimer}}, \bibinfo {author} {\bibfnamefont {A.}~\bibnamefont
  {Gopakumar}}, \ and\ \bibinfo {author} {\bibfnamefont {G.}~\bibnamefont
  {Schaefer}},\ }\href {\doibase 10.1103/PhysRevD.70.104011} {\bibfield
  {journal} {\bibinfo  {journal} {Phys. Rev. D}\ }\textbf {\bibinfo {volume}
  {70}},\ \bibinfo {pages} {104011} (\bibinfo {year} {2004})},\ \Eprint
  {http://arxiv.org/abs/gr-qc/0407049} {arXiv:gr-qc/0407049} \BibitemShut
  {NoStop}%
\bibitem [{\citenamefont {Iorio}(2016)}]{Iorio:2014yga}%
  \BibitemOpen
  \bibfield  {author} {\bibinfo {author} {\bibfnamefont {L.}~\bibnamefont
  {Iorio}},\ }\href {\doibase 10.1093/mnras/stw1155} {\bibfield  {journal}
  {\bibinfo  {journal} {Mon. Not. Roy. Astron. Soc.}\ }\textbf {\bibinfo
  {volume} {460}},\ \bibinfo {pages} {2445} (\bibinfo {year} {2016})},\ \Eprint
  {http://arxiv.org/abs/1407.5021} {arXiv:1407.5021 [gr-qc]} \BibitemShut
  {NoStop}%
\bibitem [{\citenamefont {Benisty}\ and\ \citenamefont
  {Davis}(2022)}]{Benisty:2021cmq}%
  \BibitemOpen
  \bibfield  {author} {\bibinfo {author} {\bibfnamefont {D.}~\bibnamefont
  {Benisty}}\ and\ \bibinfo {author} {\bibfnamefont {A.-C.}\ \bibnamefont
  {Davis}},\ }\href {\doibase 10.1103/PhysRevD.105.024052} {\bibfield
  {journal} {\bibinfo  {journal} {Phys. Rev. D}\ }\textbf {\bibinfo {volume}
  {105}},\ \bibinfo {pages} {024052} (\bibinfo {year} {2022})},\ \Eprint
  {http://arxiv.org/abs/2108.06286} {arXiv:2108.06286 [astro-ph.CO]}
  \BibitemShut {NoStop}%
\bibitem [{\citenamefont {Zhang}\ \emph {et~al.}(2016)\citenamefont {Zhang},
  \citenamefont {Zhao}, \citenamefont {Huang},\ and\ \citenamefont
  {Cai}}]{Zhang:2016njn}%
  \BibitemOpen
  \bibfield  {author} {\bibinfo {author} {\bibfnamefont {X.}~\bibnamefont
  {Zhang}}, \bibinfo {author} {\bibfnamefont {W.}~\bibnamefont {Zhao}},
  \bibinfo {author} {\bibfnamefont {H.}~\bibnamefont {Huang}}, \ and\ \bibinfo
  {author} {\bibfnamefont {Y.}~\bibnamefont {Cai}},\ }\href {\doibase
  10.1103/PhysRevD.93.124003} {\bibfield  {journal} {\bibinfo  {journal} {Phys.
  Rev. D}\ }\textbf {\bibinfo {volume} {93}},\ \bibinfo {pages} {124003}
  (\bibinfo {year} {2016})},\ \Eprint {http://arxiv.org/abs/1603.09450}
  {arXiv:1603.09450 [gr-qc]} \BibitemShut {NoStop}%
\bibitem [{\citenamefont {Zhang}\ \emph
  {et~al.}(2019{\natexlab{a}})\citenamefont {Zhang}, \citenamefont {Zhao},
  \citenamefont {Liu}, \citenamefont {Lin}, \citenamefont {Zhang},
  \citenamefont {Zhao}, \citenamefont {Zhang}, \citenamefont {Zhu},\ and\
  \citenamefont {Wang}}]{Zhang:2018prg}%
  \BibitemOpen
  \bibfield  {author} {\bibinfo {author} {\bibfnamefont {X.}~\bibnamefont
  {Zhang}}, \bibinfo {author} {\bibfnamefont {W.}~\bibnamefont {Zhao}},
  \bibinfo {author} {\bibfnamefont {T.}~\bibnamefont {Liu}}, \bibinfo {author}
  {\bibfnamefont {K.}~\bibnamefont {Lin}}, \bibinfo {author} {\bibfnamefont
  {C.}~\bibnamefont {Zhang}}, \bibinfo {author} {\bibfnamefont
  {X.}~\bibnamefont {Zhao}}, \bibinfo {author} {\bibfnamefont {S.}~\bibnamefont
  {Zhang}}, \bibinfo {author} {\bibfnamefont {T.}~\bibnamefont {Zhu}}, \ and\
  \bibinfo {author} {\bibfnamefont {A.}~\bibnamefont {Wang}},\ }\href {\doibase
  10.1088/1475-7516/2019/01/019} {\bibfield  {journal} {\bibinfo  {journal}
  {JCAP}\ }\textbf {\bibinfo {volume} {01}},\ \bibinfo {pages} {019} (\bibinfo
  {year} {2019}{\natexlab{a}})},\ \Eprint {http://arxiv.org/abs/1811.00339}
  {arXiv:1811.00339 [gr-qc]} \BibitemShut {NoStop}%
\bibitem [{\citenamefont {Zhang}\ \emph
  {et~al.}(2019{\natexlab{b}})\citenamefont {Zhang}, \citenamefont {Zhao},
  \citenamefont {Liu}, \citenamefont {Lin}, \citenamefont {Zhang},
  \citenamefont {Zhao}, \citenamefont {Zhang}, \citenamefont {Zhu},\ and\
  \citenamefont {Wang}}]{Zhang:2019ufz}%
  \BibitemOpen
  \bibfield  {author} {\bibinfo {author} {\bibfnamefont {X.}~\bibnamefont
  {Zhang}}, \bibinfo {author} {\bibfnamefont {W.}~\bibnamefont {Zhao}},
  \bibinfo {author} {\bibfnamefont {T.}~\bibnamefont {Liu}}, \bibinfo {author}
  {\bibfnamefont {K.}~\bibnamefont {Lin}}, \bibinfo {author} {\bibfnamefont
  {C.}~\bibnamefont {Zhang}}, \bibinfo {author} {\bibfnamefont
  {X.}~\bibnamefont {Zhao}}, \bibinfo {author} {\bibfnamefont {S.}~\bibnamefont
  {Zhang}}, \bibinfo {author} {\bibfnamefont {T.}~\bibnamefont {Zhu}}, \ and\
  \bibinfo {author} {\bibfnamefont {A.}~\bibnamefont {Wang}},\ }\href {\doibase
  10.3847/1538-4357/ab09f4} {\bibfield  {journal} {\bibinfo  {journal}
  {Astrophys. J.}\ }\textbf {\bibinfo {volume} {874}},\ \bibinfo {pages} {121}
  (\bibinfo {year} {2019}{\natexlab{b}})},\ \Eprint
  {http://arxiv.org/abs/1902.08374} {arXiv:1902.08374 [gr-qc]} \BibitemShut
  {NoStop}%
\bibitem [{\citenamefont {Benisty}(2022)}]{Benisty:2022txp}%
  \BibitemOpen
  \bibfield  {author} {\bibinfo {author} {\bibfnamefont {D.}~\bibnamefont
  {Benisty}},\ }\href {\doibase 10.1103/PhysRevD.106.043001} {\bibfield
  {journal} {\bibinfo  {journal} {Phys. Rev. D}\ }\textbf {\bibinfo {volume}
  {106}},\ \bibinfo {pages} {043001} (\bibinfo {year} {2022})},\ \Eprint
  {http://arxiv.org/abs/2207.08235} {arXiv:2207.08235 [gr-qc]} \BibitemShut
  {NoStop}%
\bibitem [{\citenamefont {Benisty}\ \emph
  {et~al.}(2022{\natexlab{a}})\citenamefont {Benisty}, \citenamefont {Evans},\
  and\ \citenamefont {Davis}}]{Benisty:2022idt}%
  \BibitemOpen
  \bibfield  {author} {\bibinfo {author} {\bibfnamefont {D.}~\bibnamefont
  {Benisty}}, \bibinfo {author} {\bibfnamefont {N.~W.}\ \bibnamefont {Evans}},
  \ and\ \bibinfo {author} {\bibfnamefont {A.-C.}\ \bibnamefont {Davis}},\
  }\href {\doibase 10.1093/mnrasl/slac134} {\  (\bibinfo {year}
  {2022}{\natexlab{a}}),\ 10.1093/mnrasl/slac134},\ \Eprint
  {http://arxiv.org/abs/2210.06429} {arXiv:2210.06429 [astro-ph.SR]}
  \BibitemShut {NoStop}%
\bibitem [{\citenamefont {Benisty}\ \emph
  {et~al.}(2022{\natexlab{b}})\citenamefont {Benisty}, \citenamefont
  {Vasiliev}, \citenamefont {Evans}, \citenamefont {Davis}, \citenamefont
  {Hartl},\ and\ \citenamefont {Strigari}}]{Benisty:2022ive}%
  \BibitemOpen
  \bibfield  {author} {\bibinfo {author} {\bibfnamefont {D.}~\bibnamefont
  {Benisty}}, \bibinfo {author} {\bibfnamefont {E.}~\bibnamefont {Vasiliev}},
  \bibinfo {author} {\bibfnamefont {N.~W.}\ \bibnamefont {Evans}}, \bibinfo
  {author} {\bibfnamefont {A.-C.}\ \bibnamefont {Davis}}, \bibinfo {author}
  {\bibfnamefont {O.~V.}\ \bibnamefont {Hartl}}, \ and\ \bibinfo {author}
  {\bibfnamefont {L.~E.}\ \bibnamefont {Strigari}},\ }\href {\doibase
  10.3847/2041-8213/ac5c42} {\bibfield  {journal} {\bibinfo  {journal}
  {Astrophys. J. Lett.}\ }\textbf {\bibinfo {volume} {928}},\ \bibinfo {pages}
  {L5} (\bibinfo {year} {2022}{\natexlab{b}})},\ \Eprint
  {http://arxiv.org/abs/2202.00033} {arXiv:2202.00033 [astro-ph.GA]}
  \BibitemShut {NoStop}%
\bibitem [{\citenamefont {Brax}\ \emph
  {et~al.}(2012{\natexlab{a}})\citenamefont {Brax}, \citenamefont {Burrage},\
  and\ \citenamefont {Davis}}]{Brax:2012ie}%
  \BibitemOpen
  \bibfield  {author} {\bibinfo {author} {\bibfnamefont {P.}~\bibnamefont
  {Brax}}, \bibinfo {author} {\bibfnamefont {C.}~\bibnamefont {Burrage}}, \
  and\ \bibinfo {author} {\bibfnamefont {A.-C.}\ \bibnamefont {Davis}},\ }\href
  {\doibase 10.1088/1475-7516/2012/10/016} {\bibfield  {journal} {\bibinfo
  {journal} {JCAP}\ }\textbf {\bibinfo {volume} {10}},\ \bibinfo {pages} {016}
  (\bibinfo {year} {2012}{\natexlab{a}})},\ \Eprint
  {http://arxiv.org/abs/1206.1809} {arXiv:1206.1809 [hep-th]} \BibitemShut
  {NoStop}%
\bibitem [{\citenamefont {Brax}\ \emph {et~al.}(2014)\citenamefont {Brax},
  \citenamefont {Davis},\ and\ \citenamefont {Sakstein}}]{Brax:2013uh}%
  \BibitemOpen
  \bibfield  {author} {\bibinfo {author} {\bibfnamefont {P.}~\bibnamefont
  {Brax}}, \bibinfo {author} {\bibfnamefont {A.-C.}\ \bibnamefont {Davis}}, \
  and\ \bibinfo {author} {\bibfnamefont {J.}~\bibnamefont {Sakstein}},\ }\href
  {\doibase 10.1088/0264-9381/31/22/225001} {\bibfield  {journal} {\bibinfo
  {journal} {Class. Quant. Grav.}\ }\textbf {\bibinfo {volume} {31}},\ \bibinfo
  {pages} {225001} (\bibinfo {year} {2014})},\ \Eprint
  {http://arxiv.org/abs/1301.5587} {arXiv:1301.5587 [gr-qc]} \BibitemShut
  {NoStop}%
\bibitem [{\citenamefont {Zhang}\ \emph {et~al.}(2017)\citenamefont {Zhang},
  \citenamefont {Liu},\ and\ \citenamefont {Zhao}}]{Zhang:2017srh}%
  \BibitemOpen
  \bibfield  {author} {\bibinfo {author} {\bibfnamefont {X.}~\bibnamefont
  {Zhang}}, \bibinfo {author} {\bibfnamefont {T.}~\bibnamefont {Liu}}, \ and\
  \bibinfo {author} {\bibfnamefont {W.}~\bibnamefont {Zhao}},\ }\href {\doibase
  10.1103/PhysRevD.95.104027} {\bibfield  {journal} {\bibinfo  {journal} {Phys.
  Rev. D}\ }\textbf {\bibinfo {volume} {95}},\ \bibinfo {pages} {104027}
  (\bibinfo {year} {2017})},\ \Eprint {http://arxiv.org/abs/1702.08752}
  {arXiv:1702.08752 [gr-qc]} \BibitemShut {NoStop}%
\bibitem [{\citenamefont {Brax}\ and\ \citenamefont
  {Davis}(2018)}]{Brax:2018bow}%
  \BibitemOpen
  \bibfield  {author} {\bibinfo {author} {\bibfnamefont {P.}~\bibnamefont
  {Brax}}\ and\ \bibinfo {author} {\bibfnamefont {A.-C.}\ \bibnamefont
  {Davis}},\ }\href {\doibase 10.1103/PhysRevD.98.063531} {\bibfield  {journal}
  {\bibinfo  {journal} {Phys. Rev. D}\ }\textbf {\bibinfo {volume} {98}},\
  \bibinfo {pages} {063531} (\bibinfo {year} {2018})},\ \Eprint
  {http://arxiv.org/abs/1809.09844} {arXiv:1809.09844 [gr-qc]} \BibitemShut
  {NoStop}%
\bibitem [{\citenamefont {Davis}\ and\ \citenamefont
  {Melville}(2020)}]{Davis:2019ltc}%
  \BibitemOpen
  \bibfield  {author} {\bibinfo {author} {\bibfnamefont {A.-C.}\ \bibnamefont
  {Davis}}\ and\ \bibinfo {author} {\bibfnamefont {S.}~\bibnamefont
  {Melville}},\ }\href {\doibase 10.1088/1475-7516/2020/09/013} {\bibfield
  {journal} {\bibinfo  {journal} {JCAP}\ }\textbf {\bibinfo {volume} {09}},\
  \bibinfo {pages} {013} (\bibinfo {year} {2020})},\ \Eprint
  {http://arxiv.org/abs/1910.08831} {arXiv:1910.08831 [gr-qc]} \BibitemShut
  {NoStop}%
\bibitem [{\citenamefont {Brax}\ \emph {et~al.}(2019)\citenamefont {Brax},
  \citenamefont {Davis},\ and\ \citenamefont {Kuntz}}]{Brax:2019tcy}%
  \BibitemOpen
  \bibfield  {author} {\bibinfo {author} {\bibfnamefont {P.}~\bibnamefont
  {Brax}}, \bibinfo {author} {\bibfnamefont {A.-C.}\ \bibnamefont {Davis}}, \
  and\ \bibinfo {author} {\bibfnamefont {A.}~\bibnamefont {Kuntz}},\ }\href
  {\doibase 10.1103/PhysRevD.99.124034} {\bibfield  {journal} {\bibinfo
  {journal} {Phys. Rev. D}\ }\textbf {\bibinfo {volume} {99}},\ \bibinfo
  {pages} {124034} (\bibinfo {year} {2019})},\ \Eprint
  {http://arxiv.org/abs/1903.03842} {arXiv:1903.03842 [gr-qc]} \BibitemShut
  {NoStop}%
\bibitem [{\citenamefont {Liu}\ \emph {et~al.}(2018)\citenamefont {Liu},
  \citenamefont {Zhang},\ and\ \citenamefont {Zhao}}]{Liu:2017xef}%
  \BibitemOpen
  \bibfield  {author} {\bibinfo {author} {\bibfnamefont {T.}~\bibnamefont
  {Liu}}, \bibinfo {author} {\bibfnamefont {X.}~\bibnamefont {Zhang}}, \ and\
  \bibinfo {author} {\bibfnamefont {W.}~\bibnamefont {Zhao}},\ }\href {\doibase
  10.1016/j.physletb.2017.12.051} {\bibfield  {journal} {\bibinfo  {journal}
  {Phys. Lett. B}\ }\textbf {\bibinfo {volume} {777}},\ \bibinfo {pages} {286}
  (\bibinfo {year} {2018})},\ \Eprint {http://arxiv.org/abs/1711.08991}
  {arXiv:1711.08991 [astro-ph.CO]} \BibitemShut {NoStop}%
\bibitem [{\citenamefont {Shibata}\ and\ \citenamefont
  {Traykova}(2022)}]{Shibata:2022gec}%
  \BibitemOpen
  \bibfield  {author} {\bibinfo {author} {\bibfnamefont {M.}~\bibnamefont
  {Shibata}}\ and\ \bibinfo {author} {\bibfnamefont {D.}~\bibnamefont
  {Traykova}},\ }\href@noop {} {\  (\bibinfo {year} {2022})},\ \Eprint
  {http://arxiv.org/abs/2210.12139} {arXiv:2210.12139 [gr-qc]} \BibitemShut
  {NoStop}%
\bibitem [{\citenamefont {Wong}\ \emph {et~al.}(2019)\citenamefont {Wong},
  \citenamefont {Davis},\ and\ \citenamefont {Gregory}}]{Wong:2019yoc}%
  \BibitemOpen
  \bibfield  {author} {\bibinfo {author} {\bibfnamefont {L.~K.}\ \bibnamefont
  {Wong}}, \bibinfo {author} {\bibfnamefont {A.-C.}\ \bibnamefont {Davis}}, \
  and\ \bibinfo {author} {\bibfnamefont {R.}~\bibnamefont {Gregory}},\ }\href
  {\doibase 10.1103/PhysRevD.100.024010} {\bibfield  {journal} {\bibinfo
  {journal} {Phys. Rev. D}\ }\textbf {\bibinfo {volume} {100}},\ \bibinfo
  {pages} {024010} (\bibinfo {year} {2019})},\ \Eprint
  {http://arxiv.org/abs/1903.07080} {arXiv:1903.07080 [hep-th]} \BibitemShut
  {NoStop}%
\bibitem [{\citenamefont {Bertotti}\ \emph {et~al.}(2003)\citenamefont
  {Bertotti}, \citenamefont {Iess},\ and\ \citenamefont
  {Tortora}}]{Bertotti:2003rm}%
  \BibitemOpen
  \bibfield  {author} {\bibinfo {author} {\bibfnamefont {B.}~\bibnamefont
  {Bertotti}}, \bibinfo {author} {\bibfnamefont {L.}~\bibnamefont {Iess}}, \
  and\ \bibinfo {author} {\bibfnamefont {P.}~\bibnamefont {Tortora}},\ }\href
  {\doibase 10.1038/nature01997} {\bibfield  {journal} {\bibinfo  {journal}
  {Nature}\ }\textbf {\bibinfo {volume} {425}},\ \bibinfo {pages} {374}
  (\bibinfo {year} {2003})}\BibitemShut {NoStop}%
\bibitem [{\citenamefont {Khoury}\ and\ \citenamefont
  {Weltman}(2004)}]{Khoury:2003aq}%
  \BibitemOpen
  \bibfield  {author} {\bibinfo {author} {\bibfnamefont {J.}~\bibnamefont
  {Khoury}}\ and\ \bibinfo {author} {\bibfnamefont {A.}~\bibnamefont
  {Weltman}},\ }\href {\doibase 10.1103/PhysRevLett.93.171104} {\bibfield
  {journal} {\bibinfo  {journal} {Phys. Rev. Lett.}\ }\textbf {\bibinfo
  {volume} {93}},\ \bibinfo {pages} {171104} (\bibinfo {year} {2004})},\
  \Eprint {http://arxiv.org/abs/astro-ph/0309300} {arXiv:astro-ph/0309300}
  \BibitemShut {NoStop}%
\bibitem [{\citenamefont {Brax}\ \emph {et~al.}(2004)\citenamefont {Brax},
  \citenamefont {van~de Bruck}, \citenamefont {Davis}, \citenamefont {Khoury},\
  and\ \citenamefont {Weltman}}]{Brax:2004qh}%
  \BibitemOpen
  \bibfield  {author} {\bibinfo {author} {\bibfnamefont {P.}~\bibnamefont
  {Brax}}, \bibinfo {author} {\bibfnamefont {C.}~\bibnamefont {van~de Bruck}},
  \bibinfo {author} {\bibfnamefont {A.-C.}\ \bibnamefont {Davis}}, \bibinfo
  {author} {\bibfnamefont {J.}~\bibnamefont {Khoury}}, \ and\ \bibinfo {author}
  {\bibfnamefont {A.}~\bibnamefont {Weltman}},\ }\href {\doibase
  10.1103/PhysRevD.70.123518} {\bibfield  {journal} {\bibinfo  {journal} {Phys.
  Rev. D}\ }\textbf {\bibinfo {volume} {70}},\ \bibinfo {pages} {123518}
  (\bibinfo {year} {2004})},\ \Eprint {http://arxiv.org/abs/astro-ph/0408415}
  {arXiv:astro-ph/0408415} \BibitemShut {NoStop}%
\bibitem [{\citenamefont {Babichev}\ \emph {et~al.}(2009)\citenamefont
  {Babichev}, \citenamefont {Deffayet},\ and\ \citenamefont
  {Ziour}}]{Babichev:2009ee}%
  \BibitemOpen
  \bibfield  {author} {\bibinfo {author} {\bibfnamefont {E.}~\bibnamefont
  {Babichev}}, \bibinfo {author} {\bibfnamefont {C.}~\bibnamefont {Deffayet}},
  \ and\ \bibinfo {author} {\bibfnamefont {R.}~\bibnamefont {Ziour}},\ }\href
  {\doibase 10.1142/S0218271809016107} {\bibfield  {journal} {\bibinfo
  {journal} {Int. J. Mod. Phys. D}\ }\textbf {\bibinfo {volume} {18}},\
  \bibinfo {pages} {2147} (\bibinfo {year} {2009})},\ \Eprint
  {http://arxiv.org/abs/0905.2943} {arXiv:0905.2943 [hep-th]} \BibitemShut
  {NoStop}%
\bibitem [{\citenamefont {Vainshtein}(1972)}]{Vainshtein:1972sx}%
  \BibitemOpen
  \bibfield  {author} {\bibinfo {author} {\bibfnamefont {A.~I.}\ \bibnamefont
  {Vainshtein}},\ }\href {\doibase 10.1016/0370-2693(72)90147-5} {\bibfield
  {journal} {\bibinfo  {journal} {Phys. Lett. B}\ }\textbf {\bibinfo {volume}
  {39}},\ \bibinfo {pages} {393} (\bibinfo {year} {1972})}\BibitemShut
  {NoStop}%
\bibitem [{\citenamefont {Damour}\ and\ \citenamefont
  {Polyakov}(1994)}]{Damour:1994zq}%
  \BibitemOpen
  \bibfield  {author} {\bibinfo {author} {\bibfnamefont {T.}~\bibnamefont
  {Damour}}\ and\ \bibinfo {author} {\bibfnamefont {A.~M.}\ \bibnamefont
  {Polyakov}},\ }\href {\doibase 10.1016/0550-3213(94)90143-0} {\bibfield
  {journal} {\bibinfo  {journal} {Nucl. Phys. B}\ }\textbf {\bibinfo {volume}
  {423}},\ \bibinfo {pages} {532} (\bibinfo {year} {1994})},\ \Eprint
  {http://arxiv.org/abs/hep-th/9401069} {arXiv:hep-th/9401069} \BibitemShut
  {NoStop}%
\bibitem [{\citenamefont {Olive}\ and\ \citenamefont
  {Pospelov}(2008)}]{Olive:2007aj}%
  \BibitemOpen
  \bibfield  {author} {\bibinfo {author} {\bibfnamefont {K.~A.}\ \bibnamefont
  {Olive}}\ and\ \bibinfo {author} {\bibfnamefont {M.}~\bibnamefont
  {Pospelov}},\ }\href {\doibase 10.1103/PhysRevD.77.043524} {\bibfield
  {journal} {\bibinfo  {journal} {Phys. Rev. D}\ }\textbf {\bibinfo {volume}
  {77}},\ \bibinfo {pages} {043524} (\bibinfo {year} {2008})},\ \Eprint
  {http://arxiv.org/abs/0709.3825} {arXiv:0709.3825 [hep-ph]} \BibitemShut
  {NoStop}%
\bibitem [{\citenamefont {Brax}\ \emph {et~al.}(2010)\citenamefont {Brax},
  \citenamefont {van~de Bruck}, \citenamefont {Davis},\ and\ \citenamefont
  {Shaw}}]{Brax:2010gi}%
  \BibitemOpen
  \bibfield  {author} {\bibinfo {author} {\bibfnamefont {P.}~\bibnamefont
  {Brax}}, \bibinfo {author} {\bibfnamefont {C.}~\bibnamefont {van~de Bruck}},
  \bibinfo {author} {\bibfnamefont {A.-C.}\ \bibnamefont {Davis}}, \ and\
  \bibinfo {author} {\bibfnamefont {D.}~\bibnamefont {Shaw}},\ }\href {\doibase
  10.1103/PhysRevD.82.063519} {\bibfield  {journal} {\bibinfo  {journal} {Phys.
  Rev. D}\ }\textbf {\bibinfo {volume} {82}},\ \bibinfo {pages} {063519}
  (\bibinfo {year} {2010})},\ \Eprint {http://arxiv.org/abs/1005.3735}
  {arXiv:1005.3735 [astro-ph.CO]} \BibitemShut {NoStop}%
\bibitem [{\citenamefont {Hinterbichler}\ and\ \citenamefont
  {Khoury}(2010)}]{Hinterbichler:2010es}%
  \BibitemOpen
  \bibfield  {author} {\bibinfo {author} {\bibfnamefont {K.}~\bibnamefont
  {Hinterbichler}}\ and\ \bibinfo {author} {\bibfnamefont {J.}~\bibnamefont
  {Khoury}},\ }\href {\doibase 10.1103/PhysRevLett.104.231301} {\bibfield
  {journal} {\bibinfo  {journal} {Phys. Rev. Lett.}\ }\textbf {\bibinfo
  {volume} {104}},\ \bibinfo {pages} {231301} (\bibinfo {year} {2010})},\
  \Eprint {http://arxiv.org/abs/1001.4525} {arXiv:1001.4525 [hep-th]}
  \BibitemShut {NoStop}%
\bibitem [{\citenamefont {Damour}\ and\ \citenamefont
  {Esposito-Farese}(1992)}]{Damour:1992we}%
  \BibitemOpen
  \bibfield  {author} {\bibinfo {author} {\bibfnamefont {T.}~\bibnamefont
  {Damour}}\ and\ \bibinfo {author} {\bibfnamefont {G.}~\bibnamefont
  {Esposito-Farese}},\ }\href {\doibase 10.1088/0264-9381/9/9/015} {\bibfield
  {journal} {\bibinfo  {journal} {Class. Quant. Grav.}\ }\textbf {\bibinfo
  {volume} {9}},\ \bibinfo {pages} {2093} (\bibinfo {year} {1992})}\BibitemShut
  {NoStop}%
\bibitem [{\citenamefont {Juli\'e}\ and\ \citenamefont
  {Deruelle}(2017)}]{Julie:2017pkb}%
  \BibitemOpen
  \bibfield  {author} {\bibinfo {author} {\bibfnamefont {F.-L.}\ \bibnamefont
  {Juli\'e}}\ and\ \bibinfo {author} {\bibfnamefont {N.}~\bibnamefont
  {Deruelle}},\ }\href {\doibase 10.1103/PhysRevD.95.124054} {\bibfield
  {journal} {\bibinfo  {journal} {Phys. Rev. D}\ }\textbf {\bibinfo {volume}
  {95}},\ \bibinfo {pages} {124054} (\bibinfo {year} {2017})},\ \Eprint
  {http://arxiv.org/abs/1703.05360} {arXiv:1703.05360 [gr-qc]} \BibitemShut
  {NoStop}%
\bibitem [{\citenamefont {Williams}\ \emph {et~al.}(2004)\citenamefont
  {Williams}, \citenamefont {Turyshev},\ and\ \citenamefont
  {Boggs}}]{Williams:2004qba}%
  \BibitemOpen
  \bibfield  {author} {\bibinfo {author} {\bibfnamefont {J.~G.}\ \bibnamefont
  {Williams}}, \bibinfo {author} {\bibfnamefont {S.~G.}\ \bibnamefont
  {Turyshev}}, \ and\ \bibinfo {author} {\bibfnamefont {D.~H.}\ \bibnamefont
  {Boggs}},\ }\href {\doibase 10.1103/PhysRevLett.93.261101} {\bibfield
  {journal} {\bibinfo  {journal} {Phys. Rev. Lett.}\ }\textbf {\bibinfo
  {volume} {93}},\ \bibinfo {pages} {261101} (\bibinfo {year} {2004})},\
  \Eprint {http://arxiv.org/abs/gr-qc/0411113} {arXiv:gr-qc/0411113}
  \BibitemShut {NoStop}%
\bibitem [{\citenamefont {Brax}(2013)}]{Brax:2013ida}%
  \BibitemOpen
  \bibfield  {author} {\bibinfo {author} {\bibfnamefont {P.}~\bibnamefont
  {Brax}},\ }\href {\doibase 10.1088/0264-9381/30/21/214005} {\bibfield
  {journal} {\bibinfo  {journal} {Class. Quant. Grav.}\ }\textbf {\bibinfo
  {volume} {30}},\ \bibinfo {pages} {214005} (\bibinfo {year}
  {2013})}\BibitemShut {NoStop}%
\bibitem [{\citenamefont {Brax}\ \emph
  {et~al.}(2021{\natexlab{b}})\citenamefont {Brax}, \citenamefont {Casas},
  \citenamefont {Desmond},\ and\ \citenamefont {Elder}}]{Brax:2021wcv}%
  \BibitemOpen
  \bibfield  {author} {\bibinfo {author} {\bibfnamefont {P.}~\bibnamefont
  {Brax}}, \bibinfo {author} {\bibfnamefont {S.}~\bibnamefont {Casas}},
  \bibinfo {author} {\bibfnamefont {H.}~\bibnamefont {Desmond}}, \ and\
  \bibinfo {author} {\bibfnamefont {B.}~\bibnamefont {Elder}},\ }\href
  {\doibase 10.3390/universe8010011} {\bibfield  {journal} {\bibinfo  {journal}
  {Universe}\ }\textbf {\bibinfo {volume} {8}},\ \bibinfo {pages} {11}
  (\bibinfo {year} {2021}{\natexlab{b}})},\ \Eprint
  {http://arxiv.org/abs/2201.10817} {arXiv:2201.10817 [gr-qc]} \BibitemShut
  {NoStop}%
\bibitem [{\citenamefont {Damour}\ and\ \citenamefont
  {Esposito-Farese}(1993)}]{Damour:1993hw}%
  \BibitemOpen
  \bibfield  {author} {\bibinfo {author} {\bibfnamefont {T.}~\bibnamefont
  {Damour}}\ and\ \bibinfo {author} {\bibfnamefont {G.}~\bibnamefont
  {Esposito-Farese}},\ }\href {\doibase 10.1103/PhysRevLett.70.2220} {\bibfield
   {journal} {\bibinfo  {journal} {Phys. Rev. Lett.}\ }\textbf {\bibinfo
  {volume} {70}},\ \bibinfo {pages} {2220} (\bibinfo {year}
  {1993})}\BibitemShut {NoStop}%
\bibitem [{\citenamefont {Freire}\ \emph {et~al.}(2012)\citenamefont {Freire},
  \citenamefont {Wex}, \citenamefont {Esposito-Farese}, \citenamefont
  {Verbiest}, \citenamefont {Bailes}, \citenamefont {Jacoby}, \citenamefont
  {Kramer}, \citenamefont {Stairs}, \citenamefont {Antoniadis},\ and\
  \citenamefont {Janssen}}]{Freire:2012mg}%
  \BibitemOpen
  \bibfield  {author} {\bibinfo {author} {\bibfnamefont {P.~C.~C.}\
  \bibnamefont {Freire}}, \bibinfo {author} {\bibfnamefont {N.}~\bibnamefont
  {Wex}}, \bibinfo {author} {\bibfnamefont {G.}~\bibnamefont
  {Esposito-Farese}}, \bibinfo {author} {\bibfnamefont {J.~P.~W.}\ \bibnamefont
  {Verbiest}}, \bibinfo {author} {\bibfnamefont {M.}~\bibnamefont {Bailes}},
  \bibinfo {author} {\bibfnamefont {B.~A.}\ \bibnamefont {Jacoby}}, \bibinfo
  {author} {\bibfnamefont {M.}~\bibnamefont {Kramer}}, \bibinfo {author}
  {\bibfnamefont {I.~H.}\ \bibnamefont {Stairs}}, \bibinfo {author}
  {\bibfnamefont {J.}~\bibnamefont {Antoniadis}}, \ and\ \bibinfo {author}
  {\bibfnamefont {G.~H.}\ \bibnamefont {Janssen}},\ }\href {\doibase
  10.1111/j.1365-2966.2012.21253.x} {\bibfield  {journal} {\bibinfo  {journal}
  {Mon. Not. Roy. Astron. Soc.}\ }\textbf {\bibinfo {volume} {423}},\ \bibinfo
  {pages} {3328} (\bibinfo {year} {2012})},\ \Eprint
  {http://arxiv.org/abs/1205.1450} {arXiv:1205.1450 [astro-ph.GA]} \BibitemShut
  {NoStop}%
\bibitem [{\citenamefont {Doneva}\ and\ \citenamefont
  {Yazadjiev}(2016)}]{Doneva:2016xmf}%
  \BibitemOpen
  \bibfield  {author} {\bibinfo {author} {\bibfnamefont {D.~D.}\ \bibnamefont
  {Doneva}}\ and\ \bibinfo {author} {\bibfnamefont {S.~S.}\ \bibnamefont
  {Yazadjiev}},\ }\href {\doibase 10.1088/1475-7516/2016/11/019} {\bibfield
  {journal} {\bibinfo  {journal} {JCAP}\ }\textbf {\bibinfo {volume} {11}},\
  \bibinfo {pages} {019} (\bibinfo {year} {2016})},\ \Eprint
  {http://arxiv.org/abs/1607.03299} {arXiv:1607.03299 [gr-qc]} \BibitemShut
  {NoStop}%
\bibitem [{\citenamefont {Ramazano\u{g}lu}\ and\ \citenamefont
  {Pretorius}(2016)}]{Ramazanoglu:2016kul}%
  \BibitemOpen
  \bibfield  {author} {\bibinfo {author} {\bibfnamefont {F.~M.}\ \bibnamefont
  {Ramazano\u{g}lu}}\ and\ \bibinfo {author} {\bibfnamefont {F.}~\bibnamefont
  {Pretorius}},\ }\href {\doibase 10.1103/PhysRevD.93.064005} {\bibfield
  {journal} {\bibinfo  {journal} {Phys. Rev. D}\ }\textbf {\bibinfo {volume}
  {93}},\ \bibinfo {pages} {064005} (\bibinfo {year} {2016})},\ \Eprint
  {http://arxiv.org/abs/1601.07475} {arXiv:1601.07475 [gr-qc]} \BibitemShut
  {NoStop}%
\bibitem [{\citenamefont {Shao}\ \emph {et~al.}(2017)\citenamefont {Shao},
  \citenamefont {Sennett}, \citenamefont {Buonanno}, \citenamefont {Kramer},\
  and\ \citenamefont {Wex}}]{Shao:2017gwu}%
  \BibitemOpen
  \bibfield  {author} {\bibinfo {author} {\bibfnamefont {L.}~\bibnamefont
  {Shao}}, \bibinfo {author} {\bibfnamefont {N.}~\bibnamefont {Sennett}},
  \bibinfo {author} {\bibfnamefont {A.}~\bibnamefont {Buonanno}}, \bibinfo
  {author} {\bibfnamefont {M.}~\bibnamefont {Kramer}}, \ and\ \bibinfo {author}
  {\bibfnamefont {N.}~\bibnamefont {Wex}},\ }\href {\doibase
  10.1103/PhysRevX.7.041025} {\bibfield  {journal} {\bibinfo  {journal} {Phys.
  Rev. X}\ }\textbf {\bibinfo {volume} {7}},\ \bibinfo {pages} {041025}
  (\bibinfo {year} {2017})},\ \Eprint {http://arxiv.org/abs/1704.07561}
  {arXiv:1704.07561 [gr-qc]} \BibitemShut {NoStop}%
\bibitem [{\citenamefont {Zhao}\ \emph {et~al.}(2022)\citenamefont {Zhao},
  \citenamefont {Freire}, \citenamefont {Kramer}, \citenamefont {Shao},\ and\
  \citenamefont {Wex}}]{Zhao:2022vig}%
  \BibitemOpen
  \bibfield  {author} {\bibinfo {author} {\bibfnamefont {J.}~\bibnamefont
  {Zhao}}, \bibinfo {author} {\bibfnamefont {P.~C.~C.}\ \bibnamefont {Freire}},
  \bibinfo {author} {\bibfnamefont {M.}~\bibnamefont {Kramer}}, \bibinfo
  {author} {\bibfnamefont {L.}~\bibnamefont {Shao}}, \ and\ \bibinfo {author}
  {\bibfnamefont {N.}~\bibnamefont {Wex}},\ }\href {\doibase
  10.1088/1361-6382/ac69a3} {\bibfield  {journal} {\bibinfo  {journal} {Class.
  Quant. Grav.}\ }\textbf {\bibinfo {volume} {39}},\ \bibinfo {pages} {11LT01}
  (\bibinfo {year} {2022})},\ \Eprint {http://arxiv.org/abs/2201.03771}
  {arXiv:2201.03771 [astro-ph.HE]} \BibitemShut {NoStop}%
\bibitem [{\citenamefont {{Weisberg}}\ \emph {et~al.}(1981)\citenamefont
  {{Weisberg}}, \citenamefont {{Taylor}},\ and\ \citenamefont
  {{Fowler}}}]{1981SciAm.245d..74W}%
  \BibitemOpen
  \bibfield  {author} {\bibinfo {author} {\bibfnamefont {J.~M.}\ \bibnamefont
  {{Weisberg}}}, \bibinfo {author} {\bibfnamefont {J.~H.}\ \bibnamefont
  {{Taylor}}}, \ and\ \bibinfo {author} {\bibfnamefont {L.~A.}\ \bibnamefont
  {{Fowler}}},\ }\href {\doibase 10.1038/scientificamerican1081-74} {\bibfield
  {journal} {\bibinfo  {journal} {Scientific American}\ }\textbf {\bibinfo
  {volume} {245}},\ \bibinfo {pages} {74} (\bibinfo {year} {1981})}\BibitemShut
  {NoStop}%
\bibitem [{\citenamefont {Weisberg}\ and\ \citenamefont
  {Huang}(2016)}]{Weisberg:2016jye}%
  \BibitemOpen
  \bibfield  {author} {\bibinfo {author} {\bibfnamefont {J.~M.}\ \bibnamefont
  {Weisberg}}\ and\ \bibinfo {author} {\bibfnamefont {Y.}~\bibnamefont
  {Huang}},\ }\href {\doibase 10.3847/0004-637X/829/1/55} {\bibfield  {journal}
  {\bibinfo  {journal} {Astrophys. J.}\ }\textbf {\bibinfo {volume} {829}},\
  \bibinfo {pages} {55} (\bibinfo {year} {2016})},\ \Eprint
  {http://arxiv.org/abs/1606.02744} {arXiv:1606.02744 [astro-ph.HE]}
  \BibitemShut {NoStop}%
\bibitem [{\citenamefont {{Damour}}\ and\ \citenamefont
  {{Deruelle}}(1986)}]{1986AIHS...44..263D}%
  \BibitemOpen
  \bibfield  {author} {\bibinfo {author} {\bibfnamefont {T.}~\bibnamefont
  {{Damour}}}\ and\ \bibinfo {author} {\bibfnamefont {N.}~\bibnamefont
  {{Deruelle}}},\ }\href@noop {} {\bibfield  {journal} {\bibinfo  {journal}
  {Ann. Inst. Henri Poincar{\'e} Phys. Th{\'e}or}\ }\textbf {\bibinfo {volume}
  {44}},\ \bibinfo {pages} {263} (\bibinfo {year} {1986})}\BibitemShut
  {NoStop}%
\bibitem [{\citenamefont {Stairs}(2003)}]{Stairs:2003eg}%
  \BibitemOpen
  \bibfield  {author} {\bibinfo {author} {\bibfnamefont {I.~H.}\ \bibnamefont
  {Stairs}},\ }\href {\doibase 10.12942/lrr-2003-5} {\bibfield  {journal}
  {\bibinfo  {journal} {Living Rev. Rel.}\ }\textbf {\bibinfo {volume} {6}},\
  \bibinfo {pages} {5} (\bibinfo {year} {2003})},\ \Eprint
  {http://arxiv.org/abs/astro-ph/0307536} {arXiv:astro-ph/0307536} \BibitemShut
  {NoStop}%
\bibitem [{\citenamefont {De~Laurentis}\ and\ \citenamefont
  {Capozziello}(2011)}]{DeLaurentis:2011tp}%
  \BibitemOpen
  \bibfield  {author} {\bibinfo {author} {\bibfnamefont {M.}~\bibnamefont
  {De~Laurentis}}\ and\ \bibinfo {author} {\bibfnamefont {S.}~\bibnamefont
  {Capozziello}},\ }\href {\doibase 10.1016/j.astropartphys.2011.08.006}
  {\bibfield  {journal} {\bibinfo  {journal} {Astropart. Phys.}\ }\textbf
  {\bibinfo {volume} {35}},\ \bibinfo {pages} {257} (\bibinfo {year} {2011})},\
  \Eprint {http://arxiv.org/abs/1104.1942} {arXiv:1104.1942 [gr-qc]}
  \BibitemShut {NoStop}%
\bibitem [{\citenamefont {De~Laurentis}\ and\ \citenamefont
  {De~Martino}(2014)}]{DeLaurentis:2013zv}%
  \BibitemOpen
  \bibfield  {author} {\bibinfo {author} {\bibfnamefont {M.}~\bibnamefont
  {De~Laurentis}}\ and\ \bibinfo {author} {\bibfnamefont {I.}~\bibnamefont
  {De~Martino}},\ }\href {\doibase 10.1093/mnras/stt216} {\bibfield  {journal}
  {\bibinfo  {journal} {Mon. Not. Roy. Astron. Soc.}\ }\textbf {\bibinfo
  {volume} {431}},\ \bibinfo {pages} {741} (\bibinfo {year} {2014})},\ \Eprint
  {http://arxiv.org/abs/1302.0220} {arXiv:1302.0220 [gr-qc]} \BibitemShut
  {NoStop}%
\bibitem [{\citenamefont {De~Laurentis}\ and\ \citenamefont
  {De~Martino}(2015)}]{DeLaurentis:2013fra}%
  \BibitemOpen
  \bibfield  {author} {\bibinfo {author} {\bibfnamefont {M.}~\bibnamefont
  {De~Laurentis}}\ and\ \bibinfo {author} {\bibfnamefont {I.}~\bibnamefont
  {De~Martino}},\ }\href {\doibase 10.1142/S0219887815500401} {\bibfield
  {journal} {\bibinfo  {journal} {Int. J. Geom. Meth. Mod. Phys.}\ }\textbf
  {\bibinfo {volume} {12}},\ \bibinfo {pages} {1550040} (\bibinfo {year}
  {2015})},\ \Eprint {http://arxiv.org/abs/1310.0711} {arXiv:1310.0711 [gr-qc]}
  \BibitemShut {NoStop}%
\bibitem [{\citenamefont {Dyadina}\ \emph {et~al.}(2016)\citenamefont
  {Dyadina}, \citenamefont {Alexeyev}, \citenamefont {Capozziello},\ and\
  \citenamefont {De~Laurentis}}]{Dyadina:2016bzb}%
  \BibitemOpen
  \bibfield  {author} {\bibinfo {author} {\bibfnamefont {P.}~\bibnamefont
  {Dyadina}}, \bibinfo {author} {\bibfnamefont {S.}~\bibnamefont {Alexeyev}},
  \bibinfo {author} {\bibfnamefont {S.}~\bibnamefont {Capozziello}}, \ and\
  \bibinfo {author} {\bibfnamefont {M.}~\bibnamefont {De~Laurentis}},\ }\href
  {\doibase 10.1051/epjconf/201612503005} {\bibfield  {journal} {\bibinfo
  {journal} {EPJ Web Conf.}\ }\textbf {\bibinfo {volume} {125}},\ \bibinfo
  {pages} {03005} (\bibinfo {year} {2016})}\BibitemShut {NoStop}%
\bibitem [{\citenamefont {Narang}\ \emph {et~al.}(2022)\citenamefont {Narang},
  \citenamefont {Mohanty},\ and\ \citenamefont {Jana}}]{Narang:2022jkv}%
  \BibitemOpen
  \bibfield  {author} {\bibinfo {author} {\bibfnamefont {A.}~\bibnamefont
  {Narang}}, \bibinfo {author} {\bibfnamefont {S.}~\bibnamefont {Mohanty}}, \
  and\ \bibinfo {author} {\bibfnamefont {S.}~\bibnamefont {Jana}},\ }\href@noop
  {} {\  (\bibinfo {year} {2022})},\ \Eprint {http://arxiv.org/abs/2211.12947}
  {arXiv:2211.12947 [gr-qc]} \BibitemShut {NoStop}%
\bibitem [{\citenamefont {Brax}\ \emph
  {et~al.}(2021{\natexlab{c}})\citenamefont {Brax}, \citenamefont {Davis},
  \citenamefont {Melville},\ and\ \citenamefont {Wong}}]{Brax:2020vgg}%
  \BibitemOpen
  \bibfield  {author} {\bibinfo {author} {\bibfnamefont {P.}~\bibnamefont
  {Brax}}, \bibinfo {author} {\bibfnamefont {A.-C.}\ \bibnamefont {Davis}},
  \bibinfo {author} {\bibfnamefont {S.}~\bibnamefont {Melville}}, \ and\
  \bibinfo {author} {\bibfnamefont {L.~K.}\ \bibnamefont {Wong}},\ }\href
  {\doibase 10.1088/1475-7516/2021/03/001} {\bibfield  {journal} {\bibinfo
  {journal} {JCAP}\ }\textbf {\bibinfo {volume} {03}},\ \bibinfo {pages} {001}
  (\bibinfo {year} {2021}{\natexlab{c}})},\ \Eprint
  {http://arxiv.org/abs/2011.01213} {arXiv:2011.01213 [gr-qc]} \BibitemShut
  {NoStop}%
\bibitem [{\citenamefont {Brax}\ \emph
  {et~al.}(2021{\natexlab{d}})\citenamefont {Brax}, \citenamefont {Davis},
  \citenamefont {Melville},\ and\ \citenamefont {Wong}}]{Brax:2021qqo}%
  \BibitemOpen
  \bibfield  {author} {\bibinfo {author} {\bibfnamefont {P.}~\bibnamefont
  {Brax}}, \bibinfo {author} {\bibfnamefont {A.-C.}\ \bibnamefont {Davis}},
  \bibinfo {author} {\bibfnamefont {S.}~\bibnamefont {Melville}}, \ and\
  \bibinfo {author} {\bibfnamefont {L.~K.}\ \bibnamefont {Wong}},\ }\href
  {\doibase 10.1088/1475-7516/2021/10/075} {\bibfield  {journal} {\bibinfo
  {journal} {JCAP}\ }\textbf {\bibinfo {volume} {10}},\ \bibinfo {pages} {075}
  (\bibinfo {year} {2021}{\natexlab{d}})},\ \Eprint
  {http://arxiv.org/abs/2107.10841} {arXiv:2107.10841 [gr-qc]} \BibitemShut
  {NoStop}%
\bibitem [{\citenamefont {Brax}(2018)}]{Brax:2017idh}%
  \BibitemOpen
  \bibfield  {author} {\bibinfo {author} {\bibfnamefont {P.}~\bibnamefont
  {Brax}},\ }\href {\doibase 10.1088/1361-6633/aa8e64} {\bibfield  {journal}
  {\bibinfo  {journal} {Rept. Prog. Phys.}\ }\textbf {\bibinfo {volume} {81}},\
  \bibinfo {pages} {016902} (\bibinfo {year} {2018})}\BibitemShut {NoStop}%
\bibitem [{\citenamefont {Brax}\ \emph
  {et~al.}(2012{\natexlab{b}})\citenamefont {Brax}, \citenamefont {Davis},
  \citenamefont {Li},\ and\ \citenamefont {Winther}}]{Brax:2012gr}%
  \BibitemOpen
  \bibfield  {author} {\bibinfo {author} {\bibfnamefont {P.}~\bibnamefont
  {Brax}}, \bibinfo {author} {\bibfnamefont {A.-C.}\ \bibnamefont {Davis}},
  \bibinfo {author} {\bibfnamefont {B.}~\bibnamefont {Li}}, \ and\ \bibinfo
  {author} {\bibfnamefont {H.~A.}\ \bibnamefont {Winther}},\ }\href {\doibase
  10.1103/PhysRevD.86.044015} {\bibfield  {journal} {\bibinfo  {journal} {Phys.
  Rev. D}\ }\textbf {\bibinfo {volume} {86}},\ \bibinfo {pages} {044015}
  (\bibinfo {year} {2012}{\natexlab{b}})},\ \Eprint
  {http://arxiv.org/abs/1203.4812} {arXiv:1203.4812 [astro-ph.CO]} \BibitemShut
  {NoStop}%
\bibitem [{\citenamefont {Brax}\ \emph {et~al.}(2017)\citenamefont {Brax},
  \citenamefont {Davis},\ and\ \citenamefont {Jha}}]{Brax:2017wcj}%
  \BibitemOpen
  \bibfield  {author} {\bibinfo {author} {\bibfnamefont {P.}~\bibnamefont
  {Brax}}, \bibinfo {author} {\bibfnamefont {A.-C.}\ \bibnamefont {Davis}}, \
  and\ \bibinfo {author} {\bibfnamefont {R.}~\bibnamefont {Jha}},\ }\href
  {\doibase 10.1103/PhysRevD.95.083514} {\bibfield  {journal} {\bibinfo
  {journal} {Phys. Rev. D}\ }\textbf {\bibinfo {volume} {95}},\ \bibinfo
  {pages} {083514} (\bibinfo {year} {2017})},\ \Eprint
  {http://arxiv.org/abs/1702.02983} {arXiv:1702.02983 [gr-qc]} \BibitemShut
  {NoStop}%
\bibitem [{\citenamefont {{Eardley}}(1975)}]{1975ApJ19659E}%
  \BibitemOpen
  \bibfield  {author} {\bibinfo {author} {\bibfnamefont {D.~M.}\ \bibnamefont
  {{Eardley}}},\ }\href {\doibase 10.1086/181744} {\bibfield  {journal}
  {\bibinfo  {journal} {Astrophysical Journal Letters}\ }\textbf {\bibinfo
  {volume} {196}},\ \bibinfo {pages} {L59} (\bibinfo {year}
  {1975})}\BibitemShut {NoStop}%
\bibitem [{\citenamefont {Doneva}\ \emph {et~al.}(2022)\citenamefont {Doneva},
  \citenamefont {Ramazano\u{g}lu}, \citenamefont {Silva}, \citenamefont
  {Sotiriou},\ and\ \citenamefont {Yazadjiev}}]{Doneva:2022ewd}%
  \BibitemOpen
  \bibfield  {author} {\bibinfo {author} {\bibfnamefont {D.~D.}\ \bibnamefont
  {Doneva}}, \bibinfo {author} {\bibfnamefont {F.~M.}\ \bibnamefont
  {Ramazano\u{g}lu}}, \bibinfo {author} {\bibfnamefont {H.~O.}\ \bibnamefont
  {Silva}}, \bibinfo {author} {\bibfnamefont {T.~P.}\ \bibnamefont {Sotiriou}},
  \ and\ \bibinfo {author} {\bibfnamefont {S.~S.}\ \bibnamefont {Yazadjiev}},\
  }\href@noop {} {\  (\bibinfo {year} {2022})},\ \Eprint
  {http://arxiv.org/abs/2211.01766} {arXiv:2211.01766 [gr-qc]} \BibitemShut
  {NoStop}%
\bibitem [{\citenamefont {Damour}\ and\ \citenamefont
  {Deruelle}(1985)}]{AIHPA_1985__43_1_107_0}%
  \BibitemOpen
  \bibfield  {author} {\bibinfo {author} {\bibfnamefont {T.}~\bibnamefont
  {Damour}}\ and\ \bibinfo {author} {\bibfnamefont {N.}~\bibnamefont
  {Deruelle}},\ }\href {http://www.numdam.org/item/AIHPA_1985__43_1_107_0/}
  {\bibfield  {journal} {\bibinfo  {journal} {Annales de l'I.H.P. Physique
  th\'eorique}\ }\textbf {\bibinfo {volume} {43}},\ \bibinfo {pages} {107}
  (\bibinfo {year} {1985})}\BibitemShut {NoStop}%
\bibitem [{\citenamefont {Brouwer}\ \emph {et~al.}(1961)\citenamefont
  {Brouwer}, \citenamefont {Books},\ and\ \citenamefont
  {Clemence}}]{brouwer1961methods}%
  \BibitemOpen
  \bibfield  {author} {\bibinfo {author} {\bibfnamefont {D.}~\bibnamefont
  {Brouwer}}, \bibinfo {author} {\bibfnamefont {E.~S. .~T.}\ \bibnamefont
  {Books}}, \ and\ \bibinfo {author} {\bibfnamefont {G.}~\bibnamefont
  {Clemence}},\ }\href {https://books.google.co.il/books?id=eKgNAQAAIAAJ}
  {\emph {\bibinfo {title} {Methods of Celestial Mechanics}}}\ (\bibinfo
  {publisher} {Academic Press},\ \bibinfo {year} {1961})\BibitemShut {NoStop}%
\bibitem [{\citenamefont {{Mavraganis}}\ and\ \citenamefont
  {{Michalakis}}(1994)}]{1994CeMDA..58..393M}%
  \BibitemOpen
  \bibfield  {author} {\bibinfo {author} {\bibfnamefont {A.~G.}\ \bibnamefont
  {{Mavraganis}}}\ and\ \bibinfo {author} {\bibfnamefont {D.~G.}\ \bibnamefont
  {{Michalakis}}},\ }\href {\doibase 10.1007/BF00692013} {\bibfield  {journal}
  {\bibinfo  {journal} {Celestial Mechanics and Dynamical Astronomy}\ }\textbf
  {\bibinfo {volume} {58}},\ \bibinfo {pages} {393} (\bibinfo {year}
  {1994})}\BibitemShut {NoStop}%
\bibitem [{\citenamefont {{Breiter}}\ and\ \citenamefont
  {{Jackson}}(1998)}]{1998MNRAS.299..237B}%
  \BibitemOpen
  \bibfield  {author} {\bibinfo {author} {\bibfnamefont {S.}~\bibnamefont
  {{Breiter}}}\ and\ \bibinfo {author} {\bibfnamefont {A.~A.}\ \bibnamefont
  {{Jackson}}},\ }\href {\doibase 10.1046/j.1365-8711.1998.01768.x} {\bibfield
  {journal} {\bibinfo  {journal} {Monthly Notices of the Royal Astronomical
  Society}\ }\textbf {\bibinfo {volume} {299}},\ \bibinfo {pages} {237}
  (\bibinfo {year} {1998})}\BibitemShut {NoStop}%
\bibitem [{\citenamefont {Peters}(1964)}]{Peters:1964zz}%
  \BibitemOpen
  \bibfield  {author} {\bibinfo {author} {\bibfnamefont {P.~C.}\ \bibnamefont
  {Peters}},\ }\href {\doibase 10.1103/PhysRev.136.B1224} {\bibfield  {journal}
  {\bibinfo  {journal} {Phys. Rev.}\ }\textbf {\bibinfo {volume} {136}},\
  \bibinfo {pages} {B1224} (\bibinfo {year} {1964})}\BibitemShut {NoStop}%
\bibitem [{\citenamefont {{POISSON}}\ and\ \citenamefont
  {{WILL}}(2014)}]{Will_book}%
  \BibitemOpen
  \bibfield  {author} {\bibinfo {author} {\bibfnamefont {E.}~\bibnamefont
  {{POISSON}}}\ and\ \bibinfo {author} {\bibfnamefont {M.~C.}\ \bibnamefont
  {{WILL}}},\ }\href@noop {} {\emph {\bibinfo {title} {{Gravity: Newtonian,
  Post-Newtonian, Relativistic}}}}\ (\bibinfo {year} {2014})\BibitemShut
  {NoStop}%
\bibitem [{\citenamefont {Randall}\ and\ \citenamefont
  {Xianyu}(2019)}]{Randall:2019sab}%
  \BibitemOpen
  \bibfield  {author} {\bibinfo {author} {\bibfnamefont {L.}~\bibnamefont
  {Randall}}\ and\ \bibinfo {author} {\bibfnamefont {Z.-Z.}\ \bibnamefont
  {Xianyu}},\ }\href@noop {} {\  (\bibinfo {year} {2019})},\ \Eprint
  {http://arxiv.org/abs/1902.08604} {arXiv:1902.08604 [astro-ph.HE]}
  \BibitemShut {NoStop}%
\bibitem [{\citenamefont {Peters}\ and\ \citenamefont
  {Mathews}(1963)}]{Peters:1963ux}%
  \BibitemOpen
  \bibfield  {author} {\bibinfo {author} {\bibfnamefont {P.~C.}\ \bibnamefont
  {Peters}}\ and\ \bibinfo {author} {\bibfnamefont {J.}~\bibnamefont
  {Mathews}},\ }\href {\doibase 10.1103/PhysRev.131.435} {\bibfield  {journal}
  {\bibinfo  {journal} {Phys. Rev.}\ }\textbf {\bibinfo {volume} {131}},\
  \bibinfo {pages} {435} (\bibinfo {year} {1963})}\BibitemShut {NoStop}%
\bibitem [{\citenamefont {van Leeuwen}\ \emph {et~al.}(2015)\citenamefont {van
  Leeuwen} \emph {et~al.}}]{vanLeeuwen:2014sca}%
  \BibitemOpen
  \bibfield  {author} {\bibinfo {author} {\bibfnamefont {J.}~\bibnamefont {van
  Leeuwen}} \emph {et~al.},\ }\href {\doibase 10.1088/0004-637X/798/2/118}
  {\bibfield  {journal} {\bibinfo  {journal} {Astrophys. J.}\ }\textbf
  {\bibinfo {volume} {798}},\ \bibinfo {pages} {118} (\bibinfo {year}
  {2015})},\ \Eprint {http://arxiv.org/abs/1411.1518} {arXiv:1411.1518
  [astro-ph.SR]} \BibitemShut {NoStop}%
\bibitem [{\citenamefont {Venkatraman~Krishnan}\ \emph
  {et~al.}(2020)\citenamefont {Venkatraman~Krishnan} \emph
  {et~al.}}]{VenkatramanKrishnan:2020pbi}%
  \BibitemOpen
  \bibfield  {author} {\bibinfo {author} {\bibfnamefont {V.}~\bibnamefont
  {Venkatraman~Krishnan}} \emph {et~al.},\ }\href {\doibase
  10.1126/science.aax7007} {\bibfield  {journal} {\bibinfo  {journal}
  {Science}\ }\textbf {\bibinfo {volume} {367}},\ \bibinfo {pages} {577}
  (\bibinfo {year} {2020})},\ \Eprint {http://arxiv.org/abs/2001.11405}
  {arXiv:2001.11405 [astro-ph.HE]} \BibitemShut {NoStop}%
\bibitem [{\citenamefont {Ferdman}\ \emph {et~al.}(2014)\citenamefont {Ferdman}
  \emph {et~al.}}]{Ferdman:2014rna}%
  \BibitemOpen
  \bibfield  {author} {\bibinfo {author} {\bibfnamefont {R.~D.}\ \bibnamefont
  {Ferdman}} \emph {et~al.},\ }\href {\doibase 10.1093/mnras/stu1223}
  {\bibfield  {journal} {\bibinfo  {journal} {Mon. Not. Roy. Astron. Soc.}\
  }\textbf {\bibinfo {volume} {443}},\ \bibinfo {pages} {2183} (\bibinfo {year}
  {2014})},\ \Eprint {http://arxiv.org/abs/1406.5507} {arXiv:1406.5507
  [astro-ph.SR]} \BibitemShut {NoStop}%
\bibitem [{\citenamefont {Fonseca}\ \emph {et~al.}(2014)\citenamefont
  {Fonseca}, \citenamefont {Stairs},\ and\ \citenamefont
  {Thorsett}}]{Fonseca:2014qla}%
  \BibitemOpen
  \bibfield  {author} {\bibinfo {author} {\bibfnamefont {E.}~\bibnamefont
  {Fonseca}}, \bibinfo {author} {\bibfnamefont {I.~H.}\ \bibnamefont {Stairs}},
  \ and\ \bibinfo {author} {\bibfnamefont {S.~E.}\ \bibnamefont {Thorsett}},\
  }\href {\doibase 10.1088/0004-637X/787/1/82} {\bibfield  {journal} {\bibinfo
  {journal} {Astrophys. J.}\ }\textbf {\bibinfo {volume} {787}},\ \bibinfo
  {pages} {82} (\bibinfo {year} {2014})},\ \Eprint
  {http://arxiv.org/abs/1402.4836} {arXiv:1402.4836 [astro-ph.HE]} \BibitemShut
  {NoStop}%
\bibitem [{\citenamefont {Kramer}\ \emph {et~al.}(2021)\citenamefont {Kramer}
  \emph {et~al.}}]{Kramer:2021jcw}%
  \BibitemOpen
  \bibfield  {author} {\bibinfo {author} {\bibfnamefont {M.}~\bibnamefont
  {Kramer}} \emph {et~al.},\ }\href {\doibase 10.1103/PhysRevX.11.041050}
  {\bibfield  {journal} {\bibinfo  {journal} {Phys. Rev. X}\ }\textbf {\bibinfo
  {volume} {11}},\ \bibinfo {pages} {041050} (\bibinfo {year} {2021})},\
  \Eprint {http://arxiv.org/abs/2112.06795} {arXiv:2112.06795 [astro-ph.HE]}
  \BibitemShut {NoStop}%
\bibitem [{\citenamefont {Kramer}\ \emph {et~al.}(2006)\citenamefont {Kramer}
  \emph {et~al.}}]{Kramer:2006nb}%
  \BibitemOpen
  \bibfield  {author} {\bibinfo {author} {\bibfnamefont {M.}~\bibnamefont
  {Kramer}} \emph {et~al.},\ }\href {\doibase 10.1126/science.1132305}
  {\bibfield  {journal} {\bibinfo  {journal} {Science}\ }\textbf {\bibinfo
  {volume} {314}},\ \bibinfo {pages} {97} (\bibinfo {year} {2006})},\ \Eprint
  {http://arxiv.org/abs/astro-ph/0609417} {arXiv:astro-ph/0609417} \BibitemShut
  {NoStop}%
\bibitem [{\citenamefont {Noutsos}\ \emph {et~al.}(2020)\citenamefont {Noutsos}
  \emph {et~al.}}]{Noutsos:2020uip}%
  \BibitemOpen
  \bibfield  {author} {\bibinfo {author} {\bibfnamefont {A.}~\bibnamefont
  {Noutsos}} \emph {et~al.},\ }\href {\doibase 10.1051/0004-6361/202038566}
  {\bibfield  {journal} {\bibinfo  {journal} {Astron. Astrophys.}\ }\textbf
  {\bibinfo {volume} {643}},\ \bibinfo {pages} {A143} (\bibinfo {year}
  {2020})},\ \Eprint {http://arxiv.org/abs/2011.02357} {arXiv:2011.02357
  [astro-ph.HE]} \BibitemShut {NoStop}%
\bibitem [{\citenamefont {Piran}\ and\ \citenamefont
  {Shaviv}(2005)}]{Piran:2004nn}%
  \BibitemOpen
  \bibfield  {author} {\bibinfo {author} {\bibfnamefont {T.}~\bibnamefont
  {Piran}}\ and\ \bibinfo {author} {\bibfnamefont {N.~J.}\ \bibnamefont
  {Shaviv}},\ }\href {\doibase 10.1103/PhysRevLett.94.051102} {\bibfield
  {journal} {\bibinfo  {journal} {Phys. Rev. Lett.}\ }\textbf {\bibinfo
  {volume} {94}},\ \bibinfo {pages} {051102} (\bibinfo {year} {2005})},\
  \Eprint {http://arxiv.org/abs/astro-ph/0409651} {arXiv:astro-ph/0409651}
  \BibitemShut {NoStop}%
\bibitem [{\citenamefont {Foreman-Mackey}\ \emph {et~al.}(2013)\citenamefont
  {Foreman-Mackey}, \citenamefont {Hogg}, \citenamefont {Lang},\ and\
  \citenamefont {Goodman}}]{Foreman-Mackey:2012any}%
  \BibitemOpen
  \bibfield  {author} {\bibinfo {author} {\bibfnamefont {D.}~\bibnamefont
  {Foreman-Mackey}}, \bibinfo {author} {\bibfnamefont {D.~W.}\ \bibnamefont
  {Hogg}}, \bibinfo {author} {\bibfnamefont {D.}~\bibnamefont {Lang}}, \ and\
  \bibinfo {author} {\bibfnamefont {J.}~\bibnamefont {Goodman}},\ }\href
  {\doibase 10.1086/670067} {\bibfield  {journal} {\bibinfo  {journal} {Publ.
  Astron. Soc. Pac.}\ }\textbf {\bibinfo {volume} {125}},\ \bibinfo {pages}
  {306} (\bibinfo {year} {2013})},\ \Eprint {http://arxiv.org/abs/1202.3665}
  {arXiv:1202.3665 [astro-ph.IM]} \BibitemShut {NoStop}%
\bibitem [{\citenamefont {Handley}\ \emph {et~al.}(2015)\citenamefont
  {Handley}, \citenamefont {Hobson},\ and\ \citenamefont
  {Lasenby}}]{Handley:2015fda}%
  \BibitemOpen
  \bibfield  {author} {\bibinfo {author} {\bibfnamefont {W.~J.}\ \bibnamefont
  {Handley}}, \bibinfo {author} {\bibfnamefont {M.~P.}\ \bibnamefont {Hobson}},
  \ and\ \bibinfo {author} {\bibfnamefont {A.~N.}\ \bibnamefont {Lasenby}},\
  }\href {\doibase 10.1093/mnrasl/slv047} {\bibfield  {journal} {\bibinfo
  {journal} {Mon. Not. Roy. Astron. Soc.}\ }\textbf {\bibinfo {volume} {450}},\
  \bibinfo {pages} {L61} (\bibinfo {year} {2015})},\ \Eprint
  {http://arxiv.org/abs/1502.01856} {arXiv:1502.01856 [astro-ph.CO]}
  \BibitemShut {NoStop}%
\bibitem [{\citenamefont {Sakstein}\ and\ \citenamefont
  {Jain}(2017)}]{Sakstein:2017xjx}%
  \BibitemOpen
  \bibfield  {author} {\bibinfo {author} {\bibfnamefont {J.}~\bibnamefont
  {Sakstein}}\ and\ \bibinfo {author} {\bibfnamefont {B.}~\bibnamefont
  {Jain}},\ }\href {\doibase 10.1103/PhysRevLett.119.251303} {\bibfield
  {journal} {\bibinfo  {journal} {Phys. Rev. Lett.}\ }\textbf {\bibinfo
  {volume} {119}},\ \bibinfo {pages} {251303} (\bibinfo {year} {2017})},\
  \Eprint {http://arxiv.org/abs/1710.05893} {arXiv:1710.05893 [astro-ph.CO]}
  \BibitemShut {NoStop}%
\bibitem [{\citenamefont {Abbott}\ \emph {et~al.}(2017)\citenamefont {Abbott}
  \emph {et~al.}}]{LIGOScientific:2017vwq}%
  \BibitemOpen
  \bibfield  {author} {\bibinfo {author} {\bibfnamefont {B.~P.}\ \bibnamefont
  {Abbott}} \emph {et~al.} (\bibinfo {collaboration} {LIGO Scientific,
  Virgo}),\ }\href {\doibase 10.1103/PhysRevLett.119.161101} {\bibfield
  {journal} {\bibinfo  {journal} {Phys. Rev. Lett.}\ }\textbf {\bibinfo
  {volume} {119}},\ \bibinfo {pages} {161101} (\bibinfo {year} {2017})},\
  \Eprint {http://arxiv.org/abs/1710.05832} {arXiv:1710.05832 [gr-qc]}
  \BibitemShut {NoStop}%
\bibitem [{\citenamefont {Brax}\ and\ \citenamefont
  {Burrage}(2014)}]{Brax:2014vva}%
  \BibitemOpen
  \bibfield  {author} {\bibinfo {author} {\bibfnamefont {P.}~\bibnamefont
  {Brax}}\ and\ \bibinfo {author} {\bibfnamefont {C.}~\bibnamefont {Burrage}},\
  }\href {\doibase 10.1103/PhysRevD.90.104009} {\bibfield  {journal} {\bibinfo
  {journal} {Phys. Rev. D}\ }\textbf {\bibinfo {volume} {90}},\ \bibinfo
  {pages} {104009} (\bibinfo {year} {2014})},\ \Eprint
  {http://arxiv.org/abs/1407.1861} {arXiv:1407.1861 [astro-ph.CO]} \BibitemShut
  {NoStop}%
\bibitem [{\citenamefont {Brax}\ \emph {et~al.}(2015)\citenamefont {Brax},
  \citenamefont {Burrage},\ and\ \citenamefont {Englert}}]{Brax:2015hma}%
  \BibitemOpen
  \bibfield  {author} {\bibinfo {author} {\bibfnamefont {P.}~\bibnamefont
  {Brax}}, \bibinfo {author} {\bibfnamefont {C.}~\bibnamefont {Burrage}}, \
  and\ \bibinfo {author} {\bibfnamefont {C.}~\bibnamefont {Englert}},\ }\href
  {\doibase 10.1103/PhysRevD.92.044036} {\bibfield  {journal} {\bibinfo
  {journal} {Phys. Rev. D}\ }\textbf {\bibinfo {volume} {92}},\ \bibinfo
  {pages} {044036} (\bibinfo {year} {2015})},\ \Eprint
  {http://arxiv.org/abs/1506.04057} {arXiv:1506.04057 [hep-ph]} \BibitemShut
  {NoStop}%
\bibitem [{\citenamefont {Liu}\ \emph {et~al.}(2014)\citenamefont {Liu},
  \citenamefont {Eatough}, \citenamefont {Wex},\ and\ \citenamefont
  {Kramer}}]{Liu:2014uka}%
  \BibitemOpen
  \bibfield  {author} {\bibinfo {author} {\bibfnamefont {K.}~\bibnamefont
  {Liu}}, \bibinfo {author} {\bibfnamefont {R.~P.}\ \bibnamefont {Eatough}},
  \bibinfo {author} {\bibfnamefont {N.}~\bibnamefont {Wex}}, \ and\ \bibinfo
  {author} {\bibfnamefont {M.}~\bibnamefont {Kramer}},\ }\href {\doibase
  10.1093/mnras/stu1913} {\bibfield  {journal} {\bibinfo  {journal} {Mon. Not.
  Roy. Astron. Soc.}\ }\textbf {\bibinfo {volume} {445}},\ \bibinfo {pages}
  {3115} (\bibinfo {year} {2014})},\ \Eprint {http://arxiv.org/abs/1409.3882}
  {arXiv:1409.3882 [astro-ph.GA]} \BibitemShut {NoStop}%
\bibitem [{\citenamefont {Wong}\ \emph {et~al.}(2022)\citenamefont {Wong},
  \citenamefont {Herdeiro},\ and\ \citenamefont {Radu}}]{Wong:2022wni}%
  \BibitemOpen
  \bibfield  {author} {\bibinfo {author} {\bibfnamefont {L.~K.}\ \bibnamefont
  {Wong}}, \bibinfo {author} {\bibfnamefont {C.~A.~R.}\ \bibnamefont
  {Herdeiro}}, \ and\ \bibinfo {author} {\bibfnamefont {E.}~\bibnamefont
  {Radu}},\ }\href {\doibase 10.1103/PhysRevD.106.024008} {\bibfield  {journal}
  {\bibinfo  {journal} {Phys. Rev. D}\ }\textbf {\bibinfo {volume} {106}},\
  \bibinfo {pages} {024008} (\bibinfo {year} {2022})},\ \Eprint
  {http://arxiv.org/abs/2204.09038} {arXiv:2204.09038 [gr-qc]} \BibitemShut
  {NoStop}%
\bibitem [{\citenamefont {Brax}\ \emph {et~al.}(2007)\citenamefont {Brax},
  \citenamefont {van~de Bruck}, \citenamefont {Davis}, \citenamefont {Mota},\
  and\ \citenamefont {Shaw}}]{Brax:2007hi}%
  \BibitemOpen
  \bibfield  {author} {\bibinfo {author} {\bibfnamefont {P.}~\bibnamefont
  {Brax}}, \bibinfo {author} {\bibfnamefont {C.}~\bibnamefont {van~de Bruck}},
  \bibinfo {author} {\bibfnamefont {A.-C.}\ \bibnamefont {Davis}}, \bibinfo
  {author} {\bibfnamefont {D.~F.}\ \bibnamefont {Mota}}, \ and\ \bibinfo
  {author} {\bibfnamefont {D.~J.}\ \bibnamefont {Shaw}},\ }\href {\doibase
  10.1103/PhysRevD.76.085010} {\bibfield  {journal} {\bibinfo  {journal} {Phys.
  Rev. D}\ }\textbf {\bibinfo {volume} {76}},\ \bibinfo {pages} {085010}
  (\bibinfo {year} {2007})},\ \Eprint {http://arxiv.org/abs/0707.2801}
  {arXiv:0707.2801 [hep-ph]} \BibitemShut {NoStop}%
\bibitem [{\citenamefont {Upadhye}(2013)}]{Upadhye:2012rc}%
  \BibitemOpen
  \bibfield  {author} {\bibinfo {author} {\bibfnamefont {A.}~\bibnamefont
  {Upadhye}},\ }\href {\doibase 10.1103/PhysRevLett.110.031301} {\bibfield
  {journal} {\bibinfo  {journal} {Phys. Rev. Lett.}\ }\textbf {\bibinfo
  {volume} {110}},\ \bibinfo {pages} {031301} (\bibinfo {year} {2013})},\
  \Eprint {http://arxiv.org/abs/1210.7804} {arXiv:1210.7804 [hep-ph]}
  \BibitemShut {NoStop}%
\bibitem [{\citenamefont {Brax}\ and\ \citenamefont
  {Davis}(2015)}]{Brax:2014zta}%
  \BibitemOpen
  \bibfield  {author} {\bibinfo {author} {\bibfnamefont {P.}~\bibnamefont
  {Brax}}\ and\ \bibinfo {author} {\bibfnamefont {A.-C.}\ \bibnamefont
  {Davis}},\ }\href {\doibase 10.1103/PhysRevD.91.063503} {\bibfield  {journal}
  {\bibinfo  {journal} {Phys. Rev. D}\ }\textbf {\bibinfo {volume} {91}},\
  \bibinfo {pages} {063503} (\bibinfo {year} {2015})},\ \Eprint
  {http://arxiv.org/abs/1412.2080} {arXiv:1412.2080 [hep-ph]} \BibitemShut
  {NoStop}%
\bibitem [{\citenamefont {Brax}\ and\ \citenamefont
  {Pitschmann}(2018)}]{Brax:2017hna}%
  \BibitemOpen
  \bibfield  {author} {\bibinfo {author} {\bibfnamefont {P.}~\bibnamefont
  {Brax}}\ and\ \bibinfo {author} {\bibfnamefont {M.}~\bibnamefont
  {Pitschmann}},\ }\href {\doibase 10.1103/PhysRevD.97.064015} {\bibfield
  {journal} {\bibinfo  {journal} {Phys. Rev. D}\ }\textbf {\bibinfo {volume}
  {97}},\ \bibinfo {pages} {064015} (\bibinfo {year} {2018})},\ \Eprint
  {http://arxiv.org/abs/1712.09852} {arXiv:1712.09852 [gr-qc]} \BibitemShut
  {NoStop}%
\end{thebibliography}%

\begin{widetext}

\appendix

\section{Scalarisation in a sphere}
\label{app:sca}
The solution of the Klein-Gordon equation in the non-relativistic case with pressure-less matter and a flat metric 
for a sphere of radius $R$ and matter density $\rho_m$ is difficult to obtain analytically. We will use a method which has been successfully used for screened models in the case of cavity profiles \cite{Brax:2007hi,Brax:2012gr}.  The Klein-Gordon equation reads \cite{Khoury:2003aq}
\be 
\frac{d^2\phi}{dr^2} + \frac{2}{r} \frac{d\phi}{dr} = -\partial_\phi V_{\rm eff}
\ee
where
\be 
V(\phi)= \frac{1}{2} m^2 \phi^2 + (e^{-a\phi^2/2m^2_{\rm Pl}}-1) \rho_m.
\ee
We will find an approximate solution which captures the physics of scalarisation. Above the critical density 
\be 
\rho_c= \frac{m^2 m^2_{\rm Pl}}{a}
\ee
the effective potential has a minimum at 
\be 
\frac{\phi(\rho_m)}{m_{\rm Pl}}= \sqrt{\frac{2}{a}} \ln^{1/2}(\frac{\rho_m}{\rho_c})
\ee
where the effective mass is 
\be 
m^2(\rho_m)= 2m^2 \ln(\frac{\rho_m}{\rho_c}).
\ee
This is a weak dependence on the density.
Point-particles in a medium of density $\rho_m$ couple to the scalar with a strength $\beta(\phi)= m_{\rm Pl} \partial_\phi \ln A(\phi)= -a \frac{\phi}{m_{\rm Pl}}$ which coincides with
\be 
\beta (\rho_m)= - \sqrt{2a} \ln^{1/2}(\frac{\rho_m}{\rho_c})
\ee
and  increases with the density.

We will look for solutions when a sphere is embedded in vacuum. We will expand the solution inside the sphere by considering that the scalar field is a massive scalar of mass $m_0= m(\phi_0)$ and the field can be expanded around a fiducial value $\phi_0$ which is determined by a bootstrapping method , i.e.
\be
r\le R, \ \ \phi= \phi_0 +\delta \phi
\ee
where $\delta \phi$ satisfies
\be 
\delta \phi'' +\frac{2}{r}\delta\phi'= m^2_0 \delta \phi +V'_{\rm eff} (\phi_0).
\ee
Notice that $\phi_0$ is not assumed to be at the minimum of the effective potential hence $V'_{\rm eff} (\phi_0)$ acts as a source term for $\delta \phi$. 
Outside the sphere we have 
\be 
\phi'' +\frac{2}{r}\phi'= m^2 \phi
\ee
where $'=d/dr$. 
Explicitly we find that
\be 
m^2_0\equiv m^2(\phi_0)= m^2 + \frac{a\rho_m}{m^2_{\rm Pl}}(\frac{a\phi_0^2}{m_{\rm Pl}^2}-1) e^{-a\phi^2/2m^2_{\rm Pl}}
\ee
and 
\be 
V'_0\equiv V'_{\rm eff}(\phi_0)= \phi_0( m^2 - \frac{a\rho_m}{m^2_{\rm Pl}}e^{-a\phi^2/2m^2_{\rm Pl}}).
\ee
The solution inside the sphere reads
\be 
\delta\phi= \delta\phi_0 + A \frac{\sinh m_0r}{m_0r}
\ee
where
\be 
\delta\phi_0= -\frac{V'_0}{m_0^2}
\ee
and $\sinh m_0 r$ should be replaced by $\sin \vert m_0\vert r$ when
$m_0^2<0$ corresponding to $\rho_m\ge \rho_c$ and small $\phi_0$. Self-consistency requires that $\phi_0$ should be determined by the boostrap equation
\be 
A+ \delta\phi_0=0
\ee
guaranteeing that the field deep inside the object is $\phi_0$. 

After matching at $r=R$ and imposing that $\delta \phi'=0$ at the origin we have
\be 
A= - \frac{\phi_0 +\delta\phi_0}{\frac{mR}{1+mR}\frac{\sinh{m_0 R}}{m_0R}+ \cosh m_0 R}
\ee
leading to the bootstrap equation
\be 
(\frac{mR}{1+mR}\frac{\sinh{m_0 R}}{m_0R}+ \cosh m_0 R -1) \delta\phi_0= \phi_0
\ee
Given a solution to this equation, the solution inside the sphere is explicitly
\be 
r\le R, \ \ \phi(r)= (\delta \phi_0 +  \phi_0) (1- \frac{1}{\frac{mR}{1+mR}\frac{\sinh{m_0 R}}{m_0R}+ \cosh m_0 R}\frac{\sinh m_0 r}{m_0r})
\ee
whilst outside the sphere we have
\be 
\phi(r)= -\frac{\beta_{\rm eff} m_{\rm Pl}}{4\pi  r} \frac{ M_E}{r}e^{-m(r-R)}.
\ee
The Einstein frame mass is given by 
\be 
M_E= \alpha M
\ee
where
\be 
\alpha= \frac{4\pi}{V} \int_0^R dx x^2 A(\phi(x))
\ee
with $V= \frac{4\pi}{3} R^3$. The scalar charge is given by 
\be 
\frac{\beta_{\rm eff}}{m_{\rm Pl}}= -\frac{4\pi R\alpha }{(1+mR) M_E} \frac{1}{\frac{mR}{1+mR}\frac{\sinh{m_0 R}}{m_0R}+ \cosh m_0 R}(\cosh m_0 R -\frac{\sinh m_0 R}{m_0 R}) (\phi_0+\delta\phi_0)
\ee
The bootstrap equation has two branches of solutions.
Let us assume that $\ln \frac{\rho_m}{\rho_c}= {\cal O}(1)$ such that $m$ and $m(\rho_m)$ are of the same order of magnitude. Then taking first the $mR\gg 1$ limit,
the first branch corresponds to the equation
\be 
m^2 - \frac{a\rho_m}{m^2_{\rm Pl}}(1- C\frac{a\phi_0^2}{m_{\rm Pl}^2}) e^{-a\phi_0^2/2m^2_{\rm Pl}}
= 0 
\label{BT}
\ee
where 
\be 
C=  \left(\frac{mR}{1+mR}\frac{\sinh{m_0 R}}{m_0R}+ \cosh m_0 R +1\right)^{-1}
\ee
which implies that $\phi_0\simeq \phi(\rho_m)$ when $mR\gg 1$ as $C\propto e^{-m(\phi(\rho_m)) R}\ll 1$. As $mR$ decreases,  the $C$ coefficient increases and a transition occurs  when $C \phi^2 (\rho_m)>1$ implying that (\ref{BT}) does not have a solution anymore. This happens for $mR \le f(\rho_m/\rho_c)$ where $f(\rho_m/\rho_c)$ corresponds to the solution of $C \phi^2 (\rho_m)=1$. For low values of $mR$ below the threshold $f(\rho_m/\rho_c)$, the solution of the bootstrap equation is on the second branch which is simply
\be 
\phi_0=0
\ee
i.e. for small enough bodies the field does not respond to the presence of the sphere and remains uniformly vanishing in all space. This type of phase transition from a non-vanishing to a vanishing value for the field below a certain radius is common for symmetron models where it can be shown that when $mR\lesssim 1$ the value of the symmetron field inside a cavity exactly vanishes \cite{Upadhye:2012rc,Brax:2014zta,Brax:2017hna}. Notice that the  scalarisation behaviour is dual to the symmetron one, i.e. the phase transition happens for symmetron inside the vacuum of a cavity whilst for scalarisation it occurs inside matter of a sphere. This illustrates the fact that screening and scalarised models behave in opposite ways when coupled to  matter.

For such small objects, the coupling to the scalar vanishes exactly whilst for large bodies $mR\gg 1$ the field is nearly constant inside the body implying that $\alpha= A(\phi(\rho_m))= \frac{\rho_c}{\rho_m}$ and therefore
\be 
\beta_{\rm eff}\simeq \frac{4\pi m_{\rm Pl} A(\phi(\rho_m)) \phi(\rho_m)}{m M_E}= \sqrt{\frac{2}{a}} \frac{\rho_c}{\rho_m} \ln^{1/2}(\frac{\rho_m}{\rho_c})\frac{1}{mR} \frac{1}{2\Phi_N(R)}
\ee
where $\Phi_N(R)= G_N M_E/R$. This is suppressed by $mR$. This also depends on the body via its size $R$ and its density (or its mass). and  breaks the universality of couplings which could lead to a substantial dipolar emission in the case of binary systems.  

In the intermediate region where $mR$ is neither large nor small, the field profile will interpolate smoothly between the vanishing value in vacuum outside the body and a non-vanishing value $\phi_0$  inside the body. As a result, the mass $M_E$ will depend on the the scalar field profile via the $A(\phi)$ factor. This is also the case of the coupling to the scalar field.

\section{Solution for Post Newtonian case}
\label{appA}
In the absence of the disformal coupling, the $\dot{r}$ and the $\dot{\theta}$ have the same form as the first PN expansion of GR with extended polynomials. \cite{Damour:1988mr} gives the corresponding relations between the polynomials and the observables
\begin{eqnarray}
&&
n = \frac{\left(-\alpha_0\right)^{3/2}}{\alpha_1}, \quad e_t^2 = 1 - \frac{\alpha_0}{\alpha_1^2} \left(\alpha_2 - \frac{\alpha_1 \alpha_3}{\alpha_2^{(0)}}\right),
\nonumber \\ &&
a_R = -\frac{\alpha_1}{\alpha_0} + \frac{\alpha_3}{2 \alpha_2^{(0)}}, \quad e_R = e_t \left(1 + \frac{\alpha_0 \alpha_3}{2 \alpha_1 \alpha_2^{(0)}} \right)
\nonumber \\
&&
e_{\theta} = e_t \left(1+ \frac{\alpha_0 \alpha_3}{\alpha_1 \alpha_2^{(0)}} - \frac{\alpha_0 \gamma_1 }{\alpha_1 \gamma_0} \right)
\nonumber \\ &&
\frac{2 \pi}{\Phi} = \frac{n}{\gamma_0} (a_R - \gamma_1/2\gamma_0)^2 (1-e_\phi^2)^{1/2}
\nonumber \\
\end{eqnarray}
where $\alpha_2^{(0)} = - j^2$.

\section{Full $\beta$ terms}
\label{appB}
The parameterisation of the correction terms to the orbits used in the main text are given below. They enter in the relation between the orbital radius, the true anomaly and time. 
\begin{subequations}
\begin{equation}
\beta_0 = \frac{32 \beta^2 \epsilon^2 j^4 \Lambda^{-2}    M^2 \epsilon +192 \beta^2 \epsilon G^2 j^2 \Lambda^{-2}    M^4 \epsilon +j^8 (2 \epsilon (1-3 \nu ) \epsilon  +2)+128 \beta^2 G^4 \Lambda^{-2} M^6 \epsilon -G^2 j^6 M^2 \epsilon   (2 \beta^2 (2 \nu -5)+\nu -6)}{2 j^9}
\end{equation}
\begin{equation}
\beta_1 = \frac{G M \epsilon  \left(64 \beta^2 \epsilon j^2 M^2/\Lambda^2+64 \beta^2 G^2 M^4/\Lambda^2+j^6 (8-(8 \beta^2+3) \nu )\right)}{2  j^7},\quad \beta_2 = \frac{8 \beta^2 \epsilon  \left(\epsilon j^2 M^2+2 G^2 M^4\right)}{j^5\Lambda^2}, \quad \beta_ 3 = \frac{8 \beta^2 G  M^3 \epsilon }{j^3\Lambda^2}
\end{equation}
\end{subequations}

\begin{subequations}
\begin{equation}
A_0 = \frac{32 \beta^2 \epsilon^2 j^4 \Lambda^{-2}    M^2 \epsilon +192 \beta^2 \epsilon G^2 j^2 \Lambda^{-2}   M^4 \epsilon +j^8 (2 \epsilon (1-3 \nu ) \epsilon  +2)+128 \beta^2 G^4 \Lambda^{-2}  M^6 \epsilon -G^2 j^6 M^2 \epsilon   (2 \beta^2 (2 \nu -5)+\nu -6)}{2 j^9}
\end{equation}
\begin{equation}
A_1 = \frac{G M \epsilon  \left(64 \beta^2 \epsilon j^2 \Lambda^{-2}M^2+64 \beta^2 G^2 \Lambda^{-2}   M^4+j^6  (8-(8 \beta^2+3) \nu )\right)}{2 j^7}, \quad A_2 = \frac{8 \beta^2 }{j^5\Lambda^2}\left(\epsilon j^2 M^2+2 G^2 M^4\right), \quad A_3 = \frac{8 \beta^2 G   M^3 \epsilon }{j^3\Lambda^2}
\end{equation}
\end{subequations}

\section{Simplification terms}
\label{appsim}
The relevant simplification equations have been used in the main text. They arise as the radiation reaction for GR appears as a fifth time derivative and the monopole term from the scalar interaction is a third time derivative
\begin{eqnarray}
&&
\ddot{r} = r \dot{\theta}^2-\frac{G M}{r^2}  , \quad
\dddot{r} =  \frac{2 G M \dot{r}}{r^3}-6 \dot{r} \dot{\theta}^2,
\nonumber \\
&&
\ddddot{r} = \frac{r \dot{\theta}^2 \left(7 G M+30 r \dot{r}^2\right)-2 G M \left(G M+3 r \dot{r}^2\right)-5 r^2 \dot{\theta}^4}{r^5},
\nonumber \\
&&
\ddot{\theta} = -\frac{2 \dot{r} \dot{\theta}}{r}, \quad \dddot{\theta} = \frac{2 \dot{\theta} \left(G M+3 r \dot{r}^2\right)}{r^3}-2 \dot{\theta}^3.
\end{eqnarray}
With these identities the higher derivatives reduce to simpler forms.
\section{Orbital parameters}
The relevant  orbital parameters obey the following Gauss equations which are used in the main text.
\begin{subequations}
\begin{equation}
\frac{da}{dt}=\frac{2}{\sqrt{1-e^2}} \sqrt{\frac{a^3}{G M}} \left[e s \, \mathcal{R} + (1+ e c)\mathcal{S}\right],
\end{equation}
\begin{equation}
\frac{dp}{dt} =2\sqrt{\frac{p^3}{G M}} \frac{\mathcal{S} }{1 + e c},
\end{equation}
\begin{equation}
\frac{de}{dt} =\sqrt{\frac{p}{G M}} \left(\frac{e \left(c^2+1\right)+2 c }{e c+1} \mathcal{S} + s \, \mathcal{R}\right),
\end{equation}
\begin{equation}
\frac{d\omega}{dt}=\frac{1}{e}\sqrt{\frac{p}{G M}}\left(s \frac{2 + e c}{e c+1} \mathcal{S} - c \,\mathcal{R}\right).
\end{equation}
\end{subequations}
When expressed in terms of the true anomaly, they read
\begin{subequations}
\begin{equation}
\frac{d p}{d\theta} \simeq \frac{2 p^2}{GM}\frac{\cal S}{\left(1+e c \right)^3},    
\end{equation}
\begin{equation}
\frac{de}{d\theta} \simeq \frac{p^2}{GM}\left(\frac{s}{(1+e c)^2}{\cal R}+ \frac{e+ 2 c + e c^2}{(1+e c)^3}{\cal S}\right),  
\end{equation}
\begin{equation}
\frac{d\omega}{d\theta} \simeq \frac{p^2}{e G M}\left[-\frac{c}{(1+e c)^2}{\cal R}+ s \frac{2+ e c}{(1+e c)^3}{\cal S}\right]
\end{equation}
\begin{equation}
\begin{split}
\frac{dt}{d\theta}  \simeq\sqrt{\frac{p^3}{GM}} \frac{1}{(1+e c )^2} \left[1- \frac{p^2}{eGM}\left(\frac{c}{(1+e c)^2}{\cal R}- s\frac{ 2+ e c}{(1+e c)^3}{\cal S}\right)\right]    
\end{split}
\end{equation}
\end{subequations}

\end{widetext}
\end{document}